\documentclass[preprint,5p,times,twocolumn,2018]{elsarticle}

\usepackage{lineno,hyperref}
\modulolinenumbers[5]

\newcommand{\checkmark}{\ding{51}}%
\newcommand{\xmark}{\ding{53}}%

\def\BibTeX{{\rm B\kern-.05em{\sc i\kern-.025em b}\kern-.08em
    T\kern-.1667em\lower.7ex\hbox{E}\kern-.125emX}}

    \usepackage{amsmath,amssymb}
\usepackage[flushleft]{threeparttable}
\usepackage{array,booktabs,makecell}
\usepackage[font=small]{caption}
    
    \usepackage{adjustbox}

\usepackage{graphicx}
\usepackage[section]{placeins}
\usepackage{lipsum}
\usepackage{graphicx}
\usepackage{float}
\usepackage[figuresright]{rotating}

\usepackage{amsmath,amssymb,amsfonts}           
\usepackage{rotating}
\usepackage{pdflscape}
\usepackage{tabulary}                                                           
\usepackage{amsmath}
\usepackage{array}
\usepackage{booktabs}
\usepackage{textcomp}

\usepackage{subfloat}
\usepackage{multirow}
\usepackage{blindtext}

\usepackage{cite}
\usepackage{algorithmic}
\usepackage{pifont}

\usepackage{rotating}
\usepackage{pdflscape}
\usepackage{soul}

\usepackage[dvipsnames]{xcolor}

\usepackage{comment}

\usepackage{cite}

\usepackage{color,soul}

\usepackage[section]{placeins}

\usepackage{wrapfig}
\usepackage{lscape}
\usepackage{rotating}
\usepackage{epstopdf}

\usepackage{pdflscape}

\usepackage{changepage}
\usepackage{afterpage}
\usepackage{xcolor}

\usepackage{tikz}
\usepackage{array}
\newcolumntype{P}[1]{>{\centering\arraybackslash}p{#1}}

\usepackage{booktabs, caption, siunitx}

\usepackage{array}
\usepackage{booktabs}

\usepackage{caption}
\usepackage{subcaption}

\usepackage{hyperref}
\usepackage{color, soul}

\journal{Journal of \LaTeX\ Templates}

\begin{document}

\begin{frontmatter}

\title{A Taxonomy Study on Securing Blockchain-based Industrial Applications: An Overview, Application Perspectives, Requirements, Attacks, Countermeasures, and Open Issues}


\author[add1]{Khizar Hameed}
\ead{hameed.khizar@utas.edu.au}
\author[add1]{Mutaz Barika}
\ead{mutaz.barika@unisa.edu.au}
\author[add1]{Saurabh Garg}
\ead{saurabh.garg@utas.edu.au}
\author[add1]{Muhammad Bilal Amin}
\ead{bilal.amin@utas.edu.au}
\author[add1]{Byeong Kang}
\ead{byeong.kang@utas.edu.au}

\address[add1]{Discipline of ICT, School of Technology, Environments, and Design, University of Tasmania, Australia}

\begin{abstract}

Blockchain technology has taken on a leading role in today's industrial applications by providing salient features and showing significant performance since its beginning. \colorbox{white}{Blockchain began its journey} from the concept of cryptocurrency and \colorbox{white}{is now} part of a range of core applications \colorbox{white}{to achieve resilience and automation between various tasks.} With the integration of Blockchain technology into different industrial applications, many application designs, security and privacy challenges present themselves, posing serious threats to users and their data. Although several approaches have been proposed to address the specific security and privacy needs of targeted applications with functional parameters, there is still a need for a research study on the application, security and privacy challenges, and requirements of Blockchain-based industrial applications, along with possible security threats and countermeasures. This study presents a state-of-the-art survey of Blockchain-based Industry 4.0 applications, focusing on crucial application and security and privacy requirements, as well as corresponding attacks on Blockchain systems with potential countermeasures. We also analyse  and provide the classification of different security and privacy techniques used in these applications to enhance the advancement of security features. \colorbox{white}{Furthermore, we highlight} some open issues in industrial applications that help to design secure Blockchain-based applications as future directions.
\end{abstract}

\begin{keyword}
\texttt{Blockchain industrial applications, Application requirements, Security requirements, Privacy-preserving, Security attacks, Countermeasures}
\end{keyword}

\end{frontmatter}


\section{Introduction}

\textcolor{black}{The widespread adoption of the Internet of Things (IoT) and related network and communication technologies drives the modern industrial revolution known as Industry 4.0 \citep{oberer2018leadership}. Sensors, actuators, and embedded systems used in the IoT for sensing, computing, and communicating data for industrial automation have a significant impact on Industry 4.0 \citep{alladi2019blockchain}. As stated, Industry 4.0 is a series of cutting-edge technologies based on advanced knowledge and communication standards and industry guidelines applied to manufacturing to help manufacturers accomplish their goals more effectively \citep{lee2019blockchain}. With an emerging trend of new disruptive technologies being used in Industry 4.0, academics and researchers have focused their efforts on developing Industry 4.0-based applications for the benefit of society. This emerging trend provides an interconnected platform for exchanging large amounts of data used in different processes. However, as the number of users increases rapidly, the network often experiences bottlenecks, resulting in scalability and single point of failure issues. Furthermore, it is often vulnerable to different types of security and privacy threats \citep{bodkhe2020blockchain}. Additionally, due to the amount of data exchanged over such an unsecured network, ensuring the confidentiality, privacy, and integrity of data becomes a major concern in Industry 4.0. Blockchain seems to be an excellent solution for dealing with the aforementioned problems and issues \citep{mohamed2019applying}.}

Blockchain technology aims to eliminate the central third party between communication parties and to provide an equal opportunity to all network nodes for controlling and managing the operations over the network. In general, Blockchain technology stipulates a trusted P2P platform with an apparent motive to design the decentralised applications for performing secure computations on transactions using cryptography algorithms. In addition to secure computations, Blockchain technology also offers a promising solution for storing the verified transactions on a shared, immutable ledger. This immutable feature is an embellished concept of Blockchain technology which provides the irreversible guarantee of transactions stored at distributed databases \citep{swan2015blockchain}. After earning remarkable success in the field of digital cryptocurrencies, Blockchain technology has gained much momentum between different business communities, and even the interest of different industrial application domains such as IoT \citep{doi:10.1155/2016/3159805}, banking \citep{guo2016blockchain} and financial services \citep{fanning2016blockchain}, Smart Grid (SG) \citep{mengelkamp2018blockchain}, logistics \citep{dorri2017blockchain} and medical \citep{mettler2016blockchain}. Recent years are witnesses to the flexible nature of Blockchain technology utilised by many applications to provide ease of automation of different manufacturing tasks with the utilisation of inherent features of Blockchain, such as decentralised topology, distributed ledger, transparency, traceability and auditability of data. 

Considering the aforementioned core features of Blockchain that can be handy for different businesses and fit in Industry 4.0, this technology was characterised by a new revolution for the applications mentioned earlier. In the banking and financial sector, for example, a high level of security is required to keep the exchange of customers’ money, data and information secure. Practically, this needs many mediators to move the money and assets over a network infrastructure, making the transactions more expensive and prone to errors, fraud and misinterpretations \citep{cai2016fraud}. Blockchain holds the potential to transform and innovate the way of transferring transactions and assets securely without a trusted party. Thus, it streamlines the transactions, as well as reducing their complexities and associated costs with full transparency and accountability. However, many application domains are still hesitant to adopt Blockchain because of these aspects of security and privacy.

The focus of the previous surveys presented by Yang et al. \citep{yang2016survey} and Li et al. \citep{li2017survey} is on providing a high-level overview of security and privacy aspects without discussing the implementation scenario for different application domains. With the advancements of Bitcoin and the related cryptocurrency applications, the research and development communities moved their focus to investigate the security and privacy aspects of these applications. For example, Khalilov and Levi \citep{khalilov2018survey} targeted two security properties (that is,  anonymity and privacy) for Bitcoin-like Digital Cash Systems;  Conti et al. \citep{conti2018survey} identified the security and privacy needs in Bitcoin and their related cryptocurrency applications and Zhang et al. \citep{zhang2019security} covered the security and privacy needs, and requirements of the Bitcoin-like cryptocurrency systems. However, these studies are specific surveys for exploring the security and privacy of financial transactions in different models. In the same context, \citep{feng2018survey} conducted a state-of-the-art survey for studying the importance of anonymity and transactional privacy in finance-related applications. This survey only covered some of the security attacks and provided only limited cryptography solutions.

Considering the completion of a successful journey made by leading Blockchain versions, a new version of Blockchain, that is Blockchain 4.0, has been introduced to address the challenges and limitations of many real-world applications. The recent surveys covered security and privacy aspects with different application domains using Blockchain technology. Joshi et al. \citep{joshi2018survey} studied security and privacy issues in some Blockchain-based applications such as finance, healthcare, mobile, defence and IoT. Salman et al. \citep{salman2018security} investigated the importance of different security services such as confidentiality, authentication, access control in IoT, healthcare and some cloud computing applications. Dasgupta et al. \citep{dasgupta2019survey} outlined the different security services covering the numerous Blockchain-based applications such as big data, medical and social networks. Hassan at al. \citep{hassan2019privacy} presented a comprehensive survey that highlights the privacy issues which arise with the integration of IoT and Blockchain technology for the services available publicly. However, these surveys do not provide a comprehensive study for security and privacy requirements, or for the challenges and their mapping to corresponding attacks, with potential solutions for Blockchain-based industrial applications. Accordingly, there is a need to research the landscape of security and privacy issues related to Blockchain-based industrial applications in order to support the design setup of these applications and meet the needs of secure environment. Such research reduces any reluctance to embrace and adopt Blockchain technology in Industry 4.0.

\subsection{Our Contributions}

To assist in this topic and provide directions for both developers and research communities to implement secure industrial Blockchain-based systems that can meet security and privacy requirements in the industrial scenario, \colorbox{white}{we} present a state-of-the-art research survey that focuses specifically on security and privacy issues in Blockchain-based Industry 4.0 applications and then discusses potential security techniques and solutions used to address them. Our concrete contributions to this paper are as follows:

\begin{itemize}
    \item  A detailed comparison of existing state-of-the-art research studies focusing on design, security and privacy issues in different Blockchain-based Industry 4.0 applications drives the research enhancement guidelines for our survey study.
    \item We examine the need of developing secure Blockchain-based Industry 4.0 applications, focusing on the design requirements, measuring criteria, and security and privacy requirements.
    \item We provide a comprehensive discussion on Blockchain-based industrial applications to meet security and privacy requirements and we further elaborate on this to achieve these by utilising security enhancement solutions.
   \item We explore, discuss and analyse the various types of security attacks detected on Blockchain-based Industry 4.0 applications, in conjunction with attack categories, attackers’  objectives, vulnerabilities exploited and target applications.
   \item We identify some open issues of integrating Blockchain technology into Blockchain-based Industry 4.0 applications on a larger scale, which provide researchers with fuel to develop potential future solutions.

\end{itemize}
\subsection{Paper Organization}

The organisation of this paper is as follows. Section \ref{sec:2} provides a related work that includes a detailed comparison of existing published surveys on security and privacy for Blockchain-based applications and highlights the limitations in them. In section \ref{sec:3}, we provide an overview of Blockchain technology in its introduction, features, layers, types, evolution, storage structure and transaction models. Section \ref{sec:4} classifies the design, security and privacy requirements of Blockchain-based Industry 4.0 applications. A detailed discussion on security and privacy requirements for Blockchain-based Industry 4.0 applications is provided in section \ref{sec:5}. In section \ref{sec:6}, we illustrates and categorises the security and privacy enhancement techniques used in different Blockchain applications to fulfil the security and privacy objectives. Section \ref{sec:7} describes the different security and privacy attacks on Blockchain-based Industry 4.0 applications. Furthermore, section \ref{sec:8} highlights the open issues required to address the development of secure Blockchain applications. Finally, we conclude our paper and provide some future research directions in section \ref{sec:9}.

\section{Related Work}\label{sec:2}

This section compares the related surveys that explicitly focus on security and privacy issues in different Blockchain-enabled applications. We make a detailed comparison of existing state-of-the-art studies related to the security and privacy domain in Blockchain-based Industry 4.0 applications based on various properties, including published year, publisher, paper title, applications covered, problems addressed, existing threats and vulnerabilities, attacks detected, techniques and solutions proposed, and future directions. Table \ref{table:existing_surveys} shows a detailed comparison of these surveys.

\begin{sidewaystable*}
\footnotesize	    
\centering
\caption{Existing Surveys on Security and Privacy of Blockchain-based Industry 4.0 Applications}
    \label{table:existing_surveys}
    \resizebox{24cm}{!}{%
    \begin{tabular}{P{0.03\textwidth}|P{0.06\textwidth}|P{0.1\textwidth}|P{0.1\textwidth}|P{0.1\textwidth}|P{0.1\textwidth}|P{0.1\textwidth}|P{0.12\textwidth}|P{0.15\textwidth}}
    
      \toprule 
      
  \multicolumn{1}{c|}{\textbf{Ref}} & \multicolumn{1}{c|} {\textbf{\thead{Year \\ \space\space\space \space\space\space Published /Publisher}}}& \multicolumn{1}{c|}{\textbf{Paper Title}}&  \multicolumn{1}{c|}{\textbf{\thead{Applications \\ Covered}}}& \multicolumn{1}{c|}{\textbf{\thead{Problems \\ Addressed}}} & \multicolumn{1}{c|}{\textbf{\thead{Threats \\ /Vulnerabilities \\Discussed}}} &  \multicolumn{1}{c|}{\textbf{\thead{Attacks \\ Highlighted}}}& \multicolumn{1}{c|}{\textbf{\thead{Techniques \\ /Solutions \\ Discussed}}}& \multicolumn{1}{c}{\textbf{Future Directions}}\\ 
\midrule

\citep{yang2016survey}& 2016 R3-Zcash& Survey of confidentiality and privacy preserving technologies for Blockchain& Not defined& Confidentiality and privacy in Blockchain& Not defined& Denial of service attack, 51\% attack& Zero-knowledge proofs, Ring signatures, Mixing Pedersen commitments with range proofs, Hawk, Enigma& Not defined\\ \midrule

\citep{li2017survey}& 2017 Elsevier& A survey on the security of Blockchain systems &  Not defined & Security threats  to Blockchain &  51\% vulnerability, Privacy key security,  Criminal activity, Double spending, Transaction privacy leakage&  Selfish mining attack, DAO attack,  BGP hijacking attack, Eclipse attack, Liveness attack, Balance attack&  Smartpool, Quantitative framework,  OYENTE, Hawk, Town Crier& Develop efficient and less time-consuming consensus algorithms,  Design scalable and efficient  privacy preserving schemes for decentralised applications, Improve the data cleanup and detections method in smart contracts\\\midrule

 \citep{khalilov2018survey}& 2018 IEEE  Communications Surveys \& Tutorials& A survey on  anonymity and  privacy in  Bitcoin-like digital cash systems&  Bitcoin-like  digital cash  systems  &  Anonymity and privacy &  Discovering  Bitcoin  addresses and  identities,   Mapping Bitcoin addresses to IP  addresses, Linking Bitcoin  addresses and their mapping  to geo-locations &  Denial of service  attack,  Majority attack,  Re-identification attack, Fingerprinting attack,  Man-in-the  middle attack, &  Mixing,  Blind signatures,  Ring signatures,  Homomorphic  encryption, Zero knowledge proof &  Investigate more  effective methods  to improve  anonymity and privacy in Bitcoin, Design more secure  cryptography   protocols,  Improve scalability  in the existing Bitcoin related cash systems, Balance the trusted   and integrity relation-  ship with anonymity and privacy of users
\\ \midrule

\citep{joshi2018survey}& 2018  Mathematical  Foundations of  Computing& A survey on  security and   privacy issues   of Blockchain technology& Finance,  Healthcare, Mobile,  Defence, Automobile, IoT& Security and  privacy in  Blockchain & Privacy leakage, Selfish mining, Personally identifiable  information, Security& Denial of   service attack,  51\% attack & Traceability, cryptography  techniques& Design secure  Blockchain-based  applications in  the areas of  security and privacy \\ \midrule

\citep{conti2018survey}& 2018 IEEE  Communications Surveys \& Tutorials& A survey on  security and privacy Issues  of Bitcoin& Bitcoin & Security and   Privacy in  Bitcoin & 51\% vulnerability,  Sybil and  double-spending,  Mining pool, Client-side Security& Bitcoin system attacks,  Bitcoin network and entities attacks& CoinJoin,  CoinShuffle  Xim,  CoinShuffle++, DiceMix, ValueShuffle, Dandelion,  SecureCoin,  CoinParty,  MixCoin, BlindCoin, TumbleBit&  Propose game theory  and stability,  Design of cryptography  and keying protocols, Improve Blockchain consensus algorithms, Design of incentive mechanisms for miners, Generation of  privacy preserving  smart contracts \\ \midrule

 \citep{feng2018survey}& 2019 Elsevier & A survey on privacy  protection in  Blockchain system& Finance& Privacy  (Identity and  Transaction)& De-  anonymization, Transaction  pattern & DoS attack,  Sybil attack& Centralised and  decentralised mixing schemes, Ring signatures,  CryptoNote, NIZK& Design of scalable system,  Design of a strong  privacy scheme, Compatibility of transaction structure with  different privacy requirements,  Traceability and accountability of transactions \\

\bottomrule
\end{tabular}}
\end{sidewaystable*}

\begin{sidewaystable*}
\footnotesize	    
\centering
    
\resizebox{23cm}{!}{%
    \begin{tabular}{P{0.03\textwidth}|P{0.06\textwidth}|P{0.1\textwidth}|P{0.1\textwidth}|P{0.1\textwidth}|P{0.1\textwidth}|P{0.1\textwidth}|P{0.12\textwidth}|P{0.15\textwidth}}
    
     \toprule 
      
  \multicolumn{1}{c|}{\textbf{Ref}} &   \multicolumn{1}{c|} {\textbf{\thead{Year \\ \space\space\space \space\space\space Published /Publisher}}}& \multicolumn{1}{c|}{\textbf{Paper Title}}&  \multicolumn{1}{c|}{\textbf{\thead{Applications \\ Covered}}}& \multicolumn{1}{c|}{\textbf{\thead{Problems \\ Addressed}}} & \multicolumn{1}{c|}{\textbf{\thead{Threats \\ /Vulnerabilities \\Discussed}}} &  \multicolumn{1}{c|}{\textbf{\thead{Attacks \\ Highlighted}}}& \multicolumn{1}{c|}{\textbf{\thead{Techniques \\ /Solutions \\ Discussed}}}& \multicolumn{1}{c}{\textbf{Future Directions}}\\ 
\midrule

\citep{salman2018security}&  2019 IEEE  Communications Surveys \& Tutorials& Security  services using  Blockchain: A  state of the art survey& IoT,  Healthcare,  Cloud   computing& Explore security  challenges,  problems, and services, (Authentication, privacy, integrity, confidentiality, non-repudiation,  data provenance) in  existing  security architectures & Vulnerabilities in  traditional  centralised  architectures &  Man-in-the-  middle attack, Data theft attack & Presented the multiple Blockchain-  based architectures  to enhance and support security services&  Design the different  solutions covering  the large scale  applications and real- time environments \\ \midrule

 \citep{dasgupta2019survey}& 2019 Springer & A survey of  Blockchain from  security perspective & Big data,  Medical,  Social networks,  Sports, Shopping, Education, Entertainment, Finance& Security and   privacy in Blockchain & Keys,  Quantum,  Identity,  Reputation, Application, Manipulation, Service, Malware& Replay attack,  Impersonation  attack, Sybil attack,  Eclipse attack, Time jacking attack, Race attack, DDoS attack,  Double spending  attack,  Finney attack, Vector76 attack, Collusion attack& cryptography  operations &  Design     resilient   security   solutions  to overcome the cyber attacks, Propose energy efficient mining  algorithms,  Design query architecture for  Blockchain  \\ \midrule

\citep{hassan2019privacy}& 2019 Elsevier& Privacy   preservation in Blockchain -based IoT   systems: Integration issues, Prospects,  challenges, and future  research directions& IoT (Healthcare,  Energy, Intelligent  transportation, Finance)& Privacy in IoT & User identity  privacy,  Transaction  privacy& Address reuse  attack,  Deanonymisation  analysis using graphs attack, Wallet privacy leakage attack,  Sybil attack,  Message spoofing attack, Linking attack& Anonymisation, Encryption,  Private contract,  Mixing, Differential privacy&  Explore and   design  the further  Blockchain  -based IoT   areas such as   Industrial IoT,  Internet of farming, cities, Mobile  things, Smart  cities, Mobile  crowd sensing  \\ \midrule

\citep{zhang2019security}& 2019  ACM&  Security and  Privacy on  Blockchain& Financial transaction & Security and privacy issues  in Blockchain & Inconsistencies  between ledgers,  Falsifying or  forging the  certificates, Data unavailability,  Double  spending  problem,  Disclosure of information,& DoS attack,  DDoS attack,  Double spending   attack, 51\% consensus  attack, de-anonymisation  attack,& Mixing, Anonymous signatures,  Homomorphic encryption, Attribute-Based  Encryption (ABE),  Secure multi-party computation, Non-interactive  zero-knowledge  (NIZK) proof, The trusted  execution environment  (TEE)-based  smart contracts,  Game-based  smart contracts & Design efficient  consensus  algorithms,  Develop lightweight cryptography algorithms, User identity problem, Linkability of transactions \\
 
\bottomrule
\end{tabular}}
\end{sidewaystable*}

\begin{sidewaystable*}
\footnotesize	    
\centering
    
    \resizebox{23cm}{!}{%

    \begin{tabular}{P{0.03\textwidth}|P{0.06\textwidth}|P{0.1\textwidth}|P{0.1\textwidth}|P{0.1\textwidth}|P{0.1\textwidth}|P{0.1\textwidth}|P{0.12\textwidth}|P{0.15\textwidth}}
    
     \toprule 
      
  \multicolumn{1}{c|}{\textbf{Ref}} &   \multicolumn{1}{c|} {\textbf{\thead{Year \\ \space\space\space \space\space\space Published /Publisher}}}& \multicolumn{1}{c|}{\textbf{Paper Title}}&  \multicolumn{1}{c|}{\textbf{\thead{Applications \\ Covered}}}& \multicolumn{1}{c|}{\textbf{\thead{Problems \\ Addressed}}} & \multicolumn{1}{c|}{\textbf{\thead{Threats \\ /Vulnerabilities \\Discussed}}} &  \multicolumn{1}{c|}{\textbf{\thead{Attacks \\ Highlighted}}}& \multicolumn{1}{c|}{\textbf{\thead{Techniques \\ /Solutions \\ Discussed}}}& \multicolumn{1}{c}{\textbf{Future Directions}}\\ 
\midrule

\citep{casino2019systematic}& 2019  Elsevier &  A systematic  literature review of Blockchain-based   applications: current status classification and open issues&  Financial applications integrity verification Governance Public sector Voting  Internet of Things Healthcare Business Applications Education Data management construction and  real state  Banking and  Insurance Waste Management & Impact of Blockchain on different applications & Not Applicable  & Not Applicable  & Not Applicable & Suitability of  Blockchain for specific applications,  Explore the  latency and  scalability issues,  Explore sustainability  of mining  protocols  \\ \midrule

\citep{mohanta2019blockchain}& 2019  Elsevier&  Blockchain technology: a survey on applications and  security privacy challenges&  Healthcare Financial  Internet of Things Legal perspective Government  Power grid Transport Commercial CloudData Reputation system Education
 & Use of Blockchain in various applications  and their  linked challenges& Not Applicable  & Double spending   privacy leakage  private key security  Mining attack  Balanced attack & Not Applicable 
 &  Not Applicable   \\ \midrule

 \citep{fernandez2019review}& 2019  IEEE Access&  A Review on  the Application of Blockchain to the Next Generation of  Cybersecure Industry 4.0 Smart Factories&  Industrial  Internet of Things,  Vertical and, Horizontal Integration Systems,  Cyber Physical Production System, Industry Augmented and Virtual Reality, Big Data  and Data Analytical,   Autonomous Robots  and Vehicles,  Cloud and Edge   Computing,   Additive Manufacturing,   Cyber Security,  Simulation Software
 & Analysing benefits and challenges of Blockchain in Industry 4.0  applications& Not Applicable  & Not Applicable  & Not Applicable 
 &  Scalability  Consensus Mechanism  Privacy and Security  Energy Efficiency  Management of Chains   \\ \midrule

 \citep{akram2020adoption}& 2020  Wiley&
 Adoption of  Blockchain  technology in various realms:   Opportunities and challenges&  Energy
   Health Supply chain Internet of Things  Resource Monitoring & Blockchain-based security solutions  for Industry 4.0 applications    & Not Applicable  & Not Applicable

 & Not Applicable
 &   Interoperability  and governance,  Rules and  regulation for governance  \\

\bottomrule
\end{tabular}}
\end{sidewaystable*}

\begin{sidewaystable*}
   \footnotesize	    
\centering
    
    \resizebox{23cm}{!}{%

    \begin{tabular}{P{0.03\textwidth}|P{0.06\textwidth}|P{0.1\textwidth}|P{0.1\textwidth}|P{0.1\textwidth}|P{0.1\textwidth}|P{0.1\textwidth}|P{0.12\textwidth}|P{0.15\textwidth}}
    
      \toprule 
      
  \multicolumn{1}{c|}{\textbf{Ref}} &   \multicolumn{1}{c|} {\textbf{\thead{Year \\ \space\space\space \space\space\space Published /Publisher}}}& \multicolumn{1}{c|}{\textbf{Paper Title}}&  \multicolumn{1}{c|}{\textbf{\thead{Applications \\ Covered}}}& \multicolumn{1}{c|}{\textbf{\thead{Problems \\ Addressed}}} & \multicolumn{1}{c|}{\textbf{\thead{Threats \\ /Vulnerabilities \\Discussed}}} &  \multicolumn{1}{c|}{\textbf{\thead{Attacks \\ Highlighted}}}& \multicolumn{1}{c|}{\textbf{\thead{Techniques \\ /Solutions \\ Discussed}}}& \multicolumn{1}{c}{\textbf{Future Directions}}\\ 
\midrule

  \citep{maesa2020blockchain}& 2020  Elsevier&
 Blockchain 3.0  applications  survey&  E-voting
   Health care Record and Identity management  decentralised notary  Intellectual property  Supplychain management & Use of Blockchain in various  industrial  applications  & Not Applicable  & Not Applicable

 & Not Applicable
 &   Not Applicable  \\ \midrule

 \citep{perera2020blockchain}& 2020  Elsevier&  Blockchain technology: Is it hype or real in   construction  industry&  Finance Identity protection  Foreign aid Voting transportation  Food and agriculture Healthcare Logistics Management MultipleData Applications for construction 
 & Applicability of   Blockchain in various construction  applications  and their  feasibility & 50\% vulnerability,  code vulnerability, private key security  criminal activity  exposing identities  & Not Applicable  & Not Applicable 
 &  Not Applicable   \\ \midrule
 
 \citep{bodkhe2020blockchain}& 2020  IEEE Access&  Blockchain for Industry 4.0 A Comprehensive Review &  Supply chain  and Logistics,  Energy Domain, Digital Content Distribution,  Tourism and Hospitality Industry, Smart Healthcare, Smart City, Business,   IoT,  Manufacturing,  Agriculture, 
 & Blockchain-based solutions in various Industry 4.0  applications& Not Applicable  & Not Applicable  & Not Applicable 
 &  Not Applicable   \\

\bottomrule
\end{tabular}}
\end{sidewaystable*}


The R3-Zcash organisation presents an initial security and privacy survey on the Blockchain in a technical report. In this report, Yang et al. \citep{yang2016survey} addressed the security challenges in Blockchain by focusing on confidentiality and privacy in Blockchain applications. The authors also described the basic attacks on Blockchain services like denial of service and 51\% attacks, discussed some solutions and proposed approaches, such as Hawk and Enigma , to overcome these attacks. However, this survey only highlighted some of the security properties and objectives, and their solutions, without further discussion regarding the most recent vulnerabilities. Li et al. \citep{li2017survey} highlighted the security and privacy threats in Blockchain systems. In this survey, the authors discussed the different vulnerabilities and attacks in these systems. Similar to \citep{yang2016survey}, some of the security solutions, that is, smart pool, Oyente, Towncrier, and Hawk, are discussed to address the fundamental security weaknesses and privacy challenges in Blockchain systems. However, these solutions only deal with smart contract applications to enforce the security policies of the Blockchain systems.  

To address the anonymity and privacy challenges in Blockchain-based digital cash systems, Khalilov and Levi \citep{khalilov2018survey} provided a detailed survey to cover the given problems. Bitcoin and its further extension of digital cash systems aim to work with different community studies to resolve the various limitations of address mappings in digital cash systems. The authors also covered the multiple attacks on these systems with their prospective solutions. However, this research work is a specific study on the security and privacy of financial transactions with various models. Joshi et al. \citep{joshi2018survey} presented a survey to highlight the security and privacy required in some of the Blockchain-based applications such as finance, healthcare, mobile, defence and IoT. However, the drawback of this study is that the authors only addressed two forms of attacks, including denial of service and 51\% attacks, and proposed the specific cryptography primitives as a solution.

Conti et al. \citep{conti2018survey} outlined the security and privacy needs in Bitcoin and its related applications. This survey identified and mapped the Bitcoin’s system’s significant loopholes in their associated threats and categorised each threat with their own proposed solutions and techniques. Even though the study considered the Bitcoin system’s significant vulnerabilities in the literature part, it only focused on addressing the various security requirements and challenges of the financial system. To cover the privacy issues in Blockchain technology, Feng et al. \citep{feng2018survey} presented a survey study highlighting the importance of anonymity and transactional privacy in finance-related applications. Moreover, the limited cryptography solutions regarding denial of service and Sybil attacks were discussed in this study.

Salman et al. \citep{salman2018security} summarised the importance of different security services such as confidentiality, authentication, access control in IoT, healthcare and some cloud computing applications. This study’s major limitation is that the discussion provided security services and challenges intended for a limited number of applications in Blockchain areas. Dasgupta at al. \citep{dasgupta2019survey} outlined the different security services covering the numerous Blockchain-based applications such as big data, medical and social networks. However, this research only highlighted the security requirements and challenges for essential aspects of privacy, along with a limited number of cryptography solutions.

Hassan at al. \citep{hassan2019privacy} presented the privacy issues that arise with the integration of IoT and Blockchain technologies for making services publicly available. This survey considered basic privacy parameters to secure communication in different Blockchain-based applications. This survey is a preliminary study for privacy, preserving strategies of IoT-based applications with a limited target scope. The recent research work by Zhang et al. \citep{zhang2019security} covered the requirements of security and privacy for Bitcoin-like cryptocurrency systems. Different security attacks on Blockchain services, such as the denial of service and mining attacks with multiple security solutions, were discussed. The limitation of this research work is that it only explored the security and privacy requirements, focusing on different models for financial transactions.

In a further study, Wang et al. \citep{wang2020survey} discussed privacy issues related to user identity and transactional privacy in Blockchain systems. This research study covered the traditional security mechanisms used to protect privacy, such as zero-knowledge proof and ring signatures, channel protocol, encryption and coin mixing mechanisms (Mix coin, Blind coin, Coin join, and so on) based on Blockchain technology. However, this study only covered limited privacy protection solutions which were mainly based on Blockchain technology. Casino et al. \citep{casino2019systematic} underlined  the importance of Blockchain technology and its underlying features in various real-time applications, ranging from industrial to business perspectives. This study’s applications were health care, IoT, voting, supply chain and a few in the business sector, such as data management, banking and insurance. However, this study did not address security threats and privacy issues in these Blockchain applications, nor did it discuss further solutions to overcome them. Akram et al. \citep{akram2020adoption} presented a systematic review of existing security solutions specifically designed for Industry 4.0 applications. However, this study’s scope only covered a few Blockchain-based security approaches, described their merits and demerits, and discussed the challenges of interoperability and governance.

In another study, Maesa and Mori \citep{maesa2020blockchain} explored the use of Blockchain from an Industrial 3.0 perspective and its links to underlying applications. They then further discussed the problem and solution requirements of Blockchain adoption in Industry 3.0. However, this study’s limitation was that it focused only on the importance of Blockchain technology in Industry 3.0 and did not cover the needs of security and the privacy issues. Mohanta et al. \citep{mohanta2019blockchain} discussed the importance of Blockchain in various Blockchain applications, including healthcare, finance, IoT, cloud computing, power grids, smart transport and so on, and then highlighted the different security and privacy issues and challenges in those applications. The limitation found in this survey study was that it focused only on security and privacy challenges and did not discuss security solutions to overcome those challenges. In a recent survey, Perera et al. \citep{perera2020blockchain} explored the possibility of adopting Blockchain technology in Industry 3.0 application sectors, particularly in the construction sector, by demonstrating its relevance with different use-case perspectives. However, this study’s focus was solely on exploring and mapping the various aspects and features of Blockchain in the industrial sectors and did not cover in detail the security threats and issues related to these applications or possible countermeasures.

Fernandez-Carames and Fraga-Lamas \citep{fernandez2019review} presented a survey to analyse the advantages and disadvantages of using Blockchain and smart contracts to build Industry 4.0 applications. However, this study primarily focused on describing a general roadmap for Industry 4.0 researchers to illustrate how to use Blockchain for more cybersecure industries. Bodkhe et al. \citep{bodkhe2020blockchain} conducted a recent survey to investigate emerging Blockchain-based solutions and their applicability for various smart applications, especially in Industry 4.0. However, this study’s focus covered only the merit and demerits of available solutions with a few countermeasures. Furthermore, this study did not go into detail about security risks and privacy attacks in Blockchain applications.

The Industry 4.0 revolution has brought new paradigms to the manufacturing industry, for example Cyber-Physical Production Systems (CPPSs), which can provide many advantages and future opportunities, such as self-awareness, self-prediction and self-reconfiguration. CPPSs attempts to connect the virtual and physical production realms but an integrated computational platform is necessary to execute these systems in the real world. To achieve this, Lee et al. \citep{lee2019blockchain} investigated the possible consequences of introducing Blockchain in real-world cyber-physical systems for creation and implementation perspectives. Moreover, a three-tier Blockchain architecture was also provided to direct industrial researchers to clearly define the role of Blockchain technology in next-generation manufacturing processes. To achieve the security and privacy of the devices and networks in industrial manufacturing processes under a smart factory setup, Lin et al. \citep{Lin2018BSeInAB} presented a Blockchain-based secure mutual authentication system to enforce fine-grained access policies.

Business process management (BPM) integrates with Industry 4.0 and Blockchain technology features such as decentralisation, immutability and accountability to digitise and automate business process workflows and to support open inter-operations of service providers, in order to achieve asset trustworthiness. To accomplish this goal, Viriyasitavat et al. \citep{viriyasitavat2018blockchain} investigated a business process management method in the composition services  in which Blockchain technology is used to identify the best possible combinations and determine partner businesses’ trustworthiness, using automated process management solutions. Moreover, a middleware approach is provided in \citep{mohamed2019applying} for leveraging Blockchain tools and capabilities to allow for more stable and transparent autonomous smart manufacturing applications, enabling different parties to build trust in the manufacturing process.

\section{A Generalised Overview of Blockchain} \label{sec:3}

\begin{figure*}[t]
    \centering
  \includegraphics[width=18cm,height=12cm]{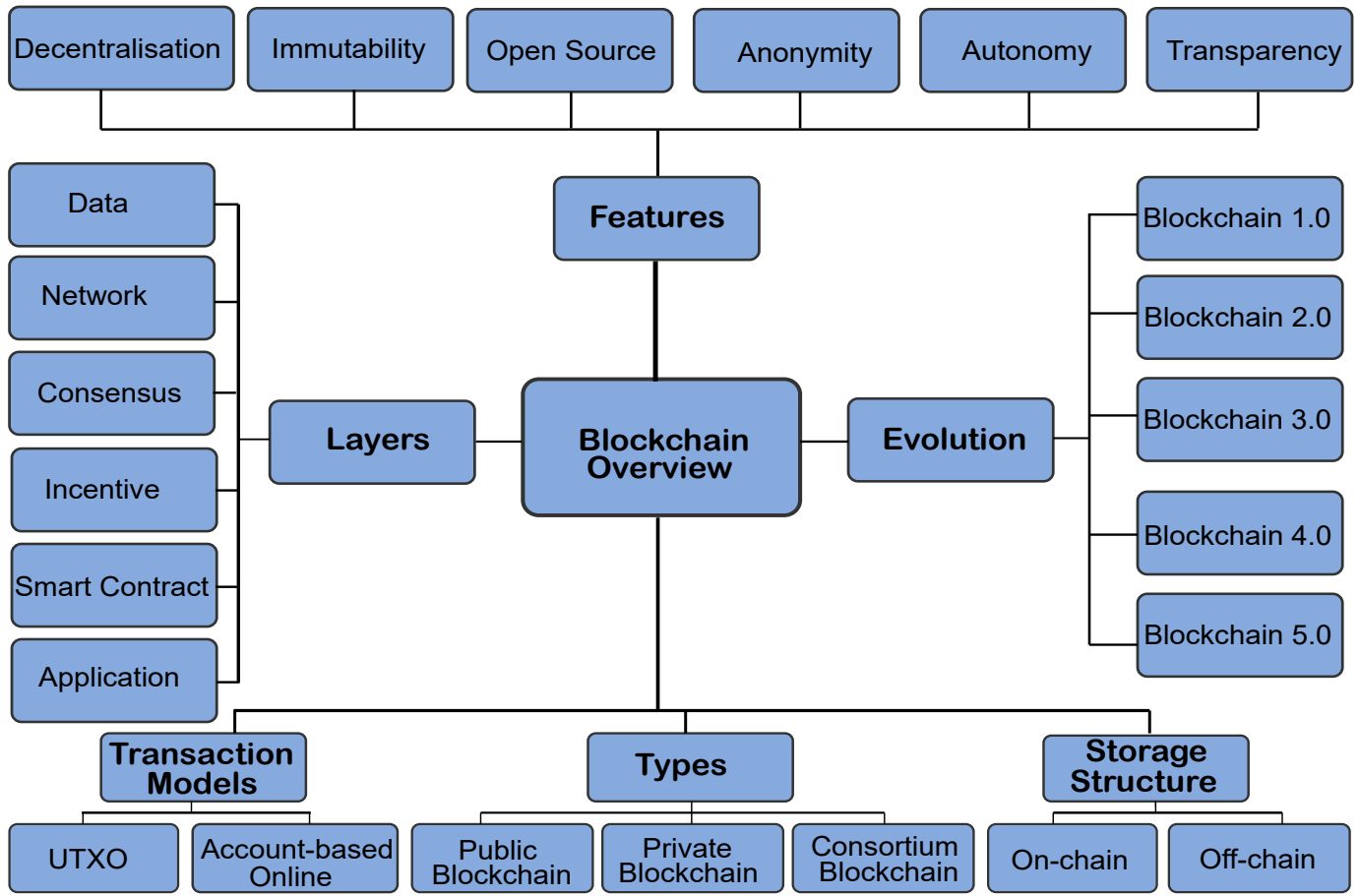}
  \caption{\colorbox{white}{A Generalised Overview of Blockchain}}
    \label{fig:GeneralizedBlockchainOverview}
\end{figure*}

Blockchain is a decentralised and distributed ledger technology that follows the peer-to-peer (P2P) network fashion in which participating nodes can interact and communicate with others, without having trusted third parties. The distributed ledger is a shared, timestamped, immutable and append-only database that keeps a record of transactions in a block structure. Each block is connected to its predecessor block by a cryptography hash stored in the block header to form a full chain called a Blockchain. Each block structure contains multiple information, such as timestamp, nonce and transaction-related, to a specific event. A timestamp indicates the time of creating each block, whereas nonce is a unique random number generated to each block and used in different cryptography operations. In a Blockchain, each block can contain multiple verified transactions stored as hash values that cannot be changed or modified regardless of the need for a lot of computing power \citep{swan2015blockchain, nofer2017blockchain}.

Blockchain allows the network’s participating nodes to interact and communicate with others without a significant third party to manage and provide verification services. Communication between network nodes is first validated and then stored as a transaction in a Blockchain database. Different cryptography primitives, such as digital signatures, are used in Blockchain to determine the level of trust for broadcasting transactions between nodes. Usually, there are two types of nodes involved in the Blockchain network which are responsible for creating and validating blocks. One is a simple node that can create  the account wallets and transactions in the network. Simultaneously, the others are full nodes (also called miner nodes) responsible for verifying or validating transactions before grouping and adding them to the Blockchain. Although both types of nodes can access all the blocks in the distributed ledger, no one has full control of the blocks and cannot modify them \citep{underwood2016blockchain}.

To ensure the reliability of data and transactions and to maintain trust between decentralised nodes, Blockchain systems follow the consensus concept, in which nodes do not accept any trusted third party’s services to manage their behaviour and interactions. Each interaction between the communicating nodes is cryptographically secured and recorded in the distributed ledger. By receiving broadcast transactions, full nodes or miner nodes on the Blockchain network can verify transactions using computational procedures. After verification, the miner nodes build a new block of validated transactions and add them to the Blockchain. To conclude, the complete process of validating and adding transactions to the Blockchain is called mining. followed by some decision-making or consensus mechanism. Each consensus mechanism is associated with miners’ rewards for their effort and computation \citep{crosby2016blockchain}.

Depending on the Blockchain systems and their types, several consensus mechanisms have been proposed. Nevertheless, the commonly used consensus mechanisms in most Blockchain systems are PoW (Proof of Work) \citep{pow}, PoS (Proof of Stake) \citep{postake}, PBFT (Practical Byzantine Fault Tolerance) \citep{castro1999practical} and DPoS (Delegated Proof of Stake) \citep{larimer2014delegated}. The PoW consensus mechanism is generally used by the Bitcoin cryptocurrency, while the Ethereum Blockchain systems use the PoS. Apart from these consensus mechanisms, several other consensus mechanisms have also been developed, such as PoA (Proof of Authentication) \citep{puthal2018proof}, PoET (Proof of Elapsed Time) \citep{chen2017security}, PoSpace (Proof of Space) \citep{pos} and PoI (Proof of Importance) \citep{poi}.

Blockchain technology can be classified into the following set of properties that may vary depending on the design perspectives of each application, ranging from single user level to business level. These properties include evolution, layered architecture, Blockchain types, storage structure and transaction models. A generalised overview of Blockchain, which illustrates its features, evolution, layers, types, storage structure and transaction models, is shown in Fig. \ref{fig:GeneralizedBlockchainOverview}.

\subsection{Features}

The overall Blockchain technology can be summarised with the following features: decentralisation, immutability, open source, anonymity, autonomy and transparency which is used to achieve a set of security features for different applications.

\subsubsection{Decentralisation}

Decentralisation feature allows a group of nodes to be organised in a P2P manner and is responsible for maintaining the network’s overall structure, rather than relying on a single governing authority to control and manage network-wide operations \citep{zheng2017overview}.
\subsubsection{Immutability}

Blockchain’s immutability feature relates to the distributed ledger, which means that the state of Blockchain remains unchanged. Since the data stored in the distributed ledger cannot be modified or changed once the majority of the nodes have been verified, immutability ensures the integrity and traceability of Blockchain data in a verifiable manner \citep{Zheng2018}.

\subsubsection{Open Source}

An open-source feature of Blockchain technology allows developers to build trust between network nodes and their data, using some of the available code features constructed. It can also provide a way to create new decentralised applications to govern the code and adopt a flexible approach \citep{muzammal2019renovating}. 

\subsubsection{Anonymity}

Anonymity applies to an entity’s status as being secret and unrevealed means that no one can access the users’ true identity from their behaviour or their transactions in the system \citep{conoscenti2016blockchain}. 
\subsubsection{Autonomy}

Autonomy can be defined as self-governing in any system capable of performing functions independently to achieve specific objectives. The anonymity feature of Blockchain enables users to participate in a self-organising system and gives them the freedom to verify transactions without involving any centralised third party \citep{lotti2016contemporary}.

\subsubsection{Transparency}
Transparency is one of the most appealing features of Blockchain technology as it allows any user to join the network and verify transactions before adding to the distributed ledger. In Bitcoin, transparency allows users to track the history of all transactions, for example, who created them and who verified them \citep{wang2020blockchain}.

\subsection{Evolution}
\begin{figure*}[!ht]
    \centering
\includegraphics[width=12cm,height=8cm]{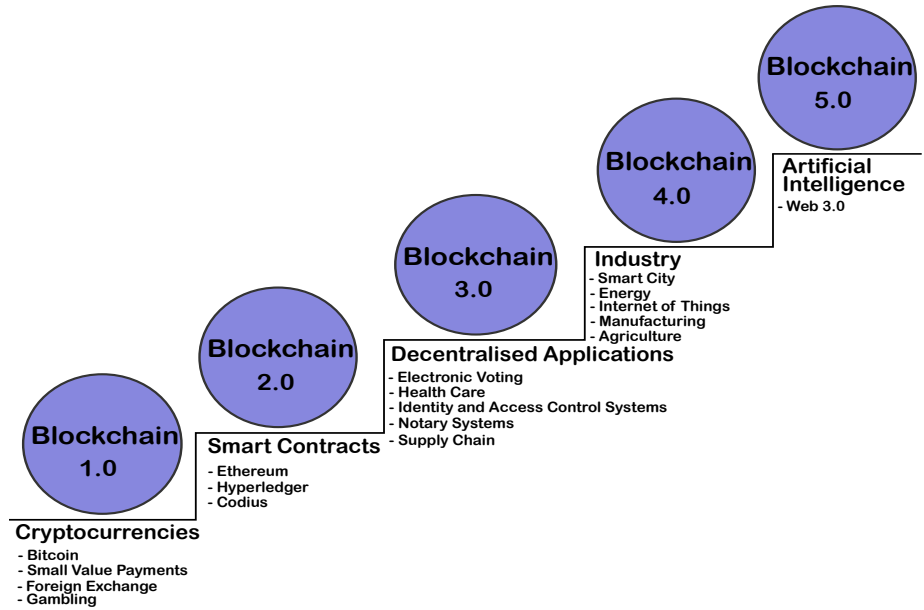}
  \caption{\colorbox{white}{Blockchain Evolution}}
   \label{fig:blockchainevaluation}
\end{figure*}

Blockchain technology continues to evolve its underlying architecture through a sequence of phases or evolution for developing a variety of applications, as illustrated in Fig. \ref{fig:blockchainevaluation}. In each phase, Blockchain technology identifies the various inherited challenges and has proposed splendid solutions to overcome them. To this end, the Blockchain evolution phases (1.0 to 4.0) are designed to provide a variety of lookouts, such as functionality, features, strengths, challenges, and security issues. Version 5.0 is currently under development, and research communities are working on it to improve its functionality for different business models. Table \ref{table:blockchaingenerations} summarises the different Blockchain generations (from 1.0 to 5.0) with respect to their applications, consensus mechanisms and features for each generation.

\subsubsection{Blockchain 1.0}

Following this, the first application of Blockchain technology was a very famous cryptocurrency named Bitcoin proposed by Satoshi Nakamoto in 2009 under the first evolution phase called Blockchain 1.0 \citep{nakamoto2008bitcoin}. The Bitcoin concept is very famous with the most commonly used terms on the internet being “Cryptocurrency” \citep{omohundro2014cryptocurrencies}, “Cash for the internet”  \citep{berentsen2016fallacy}, and “Internet of money” \citep{don2016technology}. Bitcoin used the concept of distributed ledger technology to transfer money without the need for a trusted third party. On the scene, this technology has become a fast and rapid growing digital payment system adopted by most of the financial organisations around the globe \citep{bohme2015bitcoin}. At present, Bitcoin is not just a currency system; it also changed the economic models and working structure of different organisations, for example, government sectors \citep{prisco2015estonian}, banking \citep{liebenau2016blockchain} and accounting. For security purposes, Bitcoin utilises the immutable feature of distributed ledger, to ensure the integrity of recorded transactions and to guarantee that no one can change or modify the transactions. In addition, advanced cryptography protocols, such as hashing algorithms and digital signatures, provide the authentication trust and privacy of users in the Blockchain environment \citep{zyskind2015decentralizing}. However, at present, in Blockchain 1.0, there are a few issues about computational cost, extended waiting times, lack of inter-operability and versatility which are recognised as major barriers to wider adoption.

\subsubsection{Blockchain 2.0}

Blockchain technology is considered a fast-growing technology that has been revolutionised by continuous improvements and rapid progression in the distributed ledger to develop smart applications for society and businesses. Blockchain version 2.0 comes with the concept of smart contracts, small executable user programs which run in the Blockchain environment called Ethereum Blockchain to carry out different automatic tasks and make valid decisions \citep{delmolino2016step}. The key features of such programs are that they execute automatically, based on defined logics and conditions in them, for example, time, performance, the decision and verification policies \citep{idelberger2016evaluation}. It is equally important to describe here that these small programs (or contracts) run with the autonomous identities of users to protect personal information in the Blockchain network \citep{watanabe2016blockchain}. The advantage of the smart contracts is that they can possibly reduce execution and verification times without requiring additional system resources to perform computation. Further, it can also allow the users to write smart contracts in a transparent way which prevents different fraud and hazard problems \citep{cai2016fraud}. To summarise, the Ethereum Blockchain \citep{wood2014ethereum} is the most prominent feature of Blockchain version 2.0 in which the users are allowed to write and execute smart contracts in a secure way.

\subsubsection{Blockchain 3.0}

The major limitations found in previous Blockchain versions (1.0 and 2.0) are that they mostly rely on the public Blockchain network and cannot store a massive amount of data in the distributed ledger of Blockchain technology. Bitcoin and Ethereum are open to everyone and the data are produced and recorded on the Blockchain daily. Therefore, the primary need is to store a large amount of data in different storage places, such as data servers and clouds \citep{khazraee2017asic}. For this purpose, a new version of the Blockchain has proposed a Blockchain 3.0 in which the decentralisation concept is utilised to store a huge amount of data and to legally support a wide variety of communication mediums  \citep{ali2018blockchain}. Indeed, the code in decentralised applications supports multiple servers to run and compile it; whereas a single server with limited storage only runs limited applications \citep{raval2016decentralized}. The advantage of Blockchain 3.0 is that it allows the developer to write the code of applications in any language since it requires system calls to communicate with the decentralised system for the execution of the program. Apart from the disadvantages, there are various security challenges faced by these decentralised networks such as authentication, authorisation and access control of users and their data. The privacy of users and their transactions in a decentralised network is also a challenging task, along with other security requirements \citep{aitzhan2016security}. To illustrate the concept of Blockchain 3.0, the developers of smart contracts introduced Genaro \citep{genaro}, a first Turing machine-based public Blockchain, which permits the users to write and deploy native smart contracts in decentralised storage systems with the support of different network modules in the one place.

\begin{sidewaystable*}
\footnotesize
\centering
\caption{Different Blockchain Generations along with their Applications, Consensus Mechanism and Unique Features}
\label{table:blockchaingenerations}

\resizebox{0.9\textwidth}{!}{%

    \begin{tabular}{p{0.08\textwidth}|p{0.35\textwidth}|p{0.3\textwidth}|p{0.3\textwidth}}
    
     \toprule

  \multicolumn{1}{c|}{\textbf{Blockchain Generations}} &  \multicolumn{1}{c|}{\textbf{Applications}}& \multicolumn{1}{c|}{\textbf{\thead{Consensus \\ Mechanism Used}}}&  \multicolumn{1}{c}{\textbf{\thead{Unique Features}}}\\ 
\midrule

\textbf{Blockchain 1.0} & 
\begin{itemize}
\item    Digital Currencies

        \begin{itemize}
            \item Bitcoin \citep{nakamoto2008bitcoin}, Bitcoin Cash \citep{bitcoincash}, Litecoin \citep{litecoin}, Ripple \citep{ripple}, etc.
        \end{itemize}

        \item    Small Value Payments \citep{xu2019systematic}
        \item    Foreign Exchange
        \item   Gambling
        \item   Money Laundering
        
\end{itemize}

  &  
  
  \begin{itemize}
    \item  PoW \citep{pow}
    \item  PoS
    \item Proof of Elapsed Time (PoET)
    \item Proof of Space 
    \item Federated Byzantine Agreement (Federated BA)
    \item Proof of Memory
\end{itemize}
  
   & \begin{itemize}
    \item  Mostly Designed for Cryptocurrencies
    \item Simple Ledgers
    \item Public Blockchain
\end{itemize} \\ \midrule

\textbf{Blockchain 2.0} & 
 
 \begin{itemize}
    \item  Ethereum \citep{wood2014ethereum}
    \item Hyperledger \citep{Hyperledger}
    \item Codius \citep{codius}
\end{itemize}

 & 
 
 \begin{itemize}
    \item PoS \citep{postake}
    \item Practical Byzantine Fault Tolerance (PBFT) \citep{castro1999practical}
    \item Byzantine Fault Tolerance (BFT) - BFT-SMaRt
\end{itemize}

 & \begin{itemize}
     \item Use of Smart Contracts
     \item Micro-transactions
     \item Digital Assets \citep{colomo2020critical}
     \item Privacy
     \item Decentralised Autonomous Organisations (DAOs)
     \item Decentralised Autonomous Corporations (DACs) \citep{swan2017anticipating}
     \item Has own Contact-Oriented Language (Solidity)
     \item Public Blockchain
 \end{itemize}  \\ \midrule

\textbf{Blockchain 3.0} &   Enterprise Blockchain Applications \citep{maesa2020blockchain}
 \begin{itemize}
     \item Electronic Voting
     \item Healthcare
     \item Identity and Access Control Systems
     \item Notary Systems
     \item Supply chain

 \end{itemize}

 &
 
  Only few of them listed here but not limited to \citep{bodkhe2020survey}
\begin{itemize}
    \item Tendermint
    \item  DPoS
    \item  Raft
    \item Casper
    \item Staller
\end{itemize}

 &  \begin{itemize}
     \item Instantaneous Transaction
     \item High Scalability
     \item Interoperability
     \item Sustainability
     \item Governance 
     \item Cloud Servicing
     \item Multi layer Middle-ware

 \end{itemize} \\\midrule
 
\textbf{Blockchain 4.0} & Industrial Perspectives \citep{lu2017industry} \begin{itemize}
    \item Cyber Physical Systems (CPS)
    \item Smart Manufacturing
    \item Industrial Internet of Things (IIoT)
    \item Agriculture
    \item Energy Trading
    \item Smart Product
    \item Smart City
    
\end{itemize}

& 
Only few of them listed here but not limited to \citep{bodkhe2020survey}

\begin{itemize}
    \item Hash DAG
    \item Proof of Importance (PoI)
    \item Proof of Burn (PoB)
    \item Proof of Value (PoV)
    \item Proof of Majority (PoM)
    \item Proof of History (PoH)

\end{itemize}

& \begin{itemize}
    \item Industry Consortium 
    \item Consensus Mechanism Efficiency
    \item Transparency
    \item Improve Scalability
    \item Energy Efficiency
    
\end{itemize}  \\ \midrule
\textbf{Blockchain 5.0} & Web 3.0 Applications

& Not Available& Combination of Artificial Intelligence and DLT  \begin{itemize}
    \item Data Privacy
    \item Security
    \item Interoperability
\end{itemize}    \\

\bottomrule
\end{tabular}}
\end{sidewaystable*}

\subsubsection{Blockchain 4.0}

With the completion of a successful journey made by leading Blockchain versions (from 1.0 to 3.0), the new version of Blockchain 4.0 is presented to address the industrial challenges and limitations of real-world applications. Blockchain 4.0 is a new generation or version of Blockchain technology that aims to introduce Blockchain into the industrial world and make it practical for developing and running real-world applications in a secure and decentralised way. The new version also enables us to propose new solutions and fills the gap between business and information technology industries \citep{chung2016internet}.

Furthermore, Blockchain 4.0 enables the industry and business sectors to transition their entire structure and processes (or parts of them) transparently, to stable, self-recording applications built on a decentralised, distributed and immutable ledger. As Industry 4.0 is known as a revolutionary technological wave for the interconnectivity  between people and machines, it provides substantial industry growth and productivity change that positively affects both the human quality of life and the environment \citep{bodkhe2020survey}.

The convergence of Industry 4.0 and the Blockchain 4.0 generation creates a joined paradigm based on trusted networks that eliminate the need for a third party. Individual manual processes are transformed into linked systems using automated, autonomous systems, which are also underpinned by Blockchain technology. This convergence is primarily centred on the use of Blockchain features such as public ledgers and distributed databases, as well as the implementation of smart contracts in industry processes to remove the need for paper-based contracts and to control the network through consensus \citep{fernandez2019review}. Moreover, introducing Blockchain version 4.0 into Industry 4.0 aims to achieve transparency in the industrial processes from planning to implementation, and to establish the relationship between industry policies and underlying Blockchain features \citep{badzar2016blockchain}.

There are a few examples of Industry 4.0 which have recently adopted this new version into their business processes: financial services \citep{guo2016blockchain}, IoT \citep{bahga2016blockchain}, Transport and Logistics \citep{abeyratne2016blockchain}, SG \citep{basden2017utilities,sharma2017block} and eHealth \citep{ekblaw2016case}.

\subsubsection{Blockchain 5.0}

Although Blockchain technology is relatively new, it has advanced dramatically. It is now used in a broad range of industrial sectors, including banking, healthcare, IoT and supply chain management. After achieving considerable success in earlier versions, Blockchain 5.0 is designed to serve the needs of the next generation business peoples’ by formalising and standardising digital lifelines.  Therefore, it is becoming extremely important to have Blockchain 5.0 in the today's world. The aim of Blockchain 5.0 is to concentrate on the integration of AI and DLT in order to develop the next generation of decentralised Web 3.0 applications to achieve data privacy, security, and interoperability.
By making this option, a project called "Relictum Pro" is well on its way to achieving success in the new age of Blockchain technology, which is characterised by Blockchain 5.0. The “Relictum Pro” Project has advanced technology to use Blockchain 5.0 to build virtual channels on this dedicated network. As a result, there is a significant increase in transfer rates and the introduction of a seamless system with smaller block sizes and faster transactions \citep{relictum}.

\begin{figure*}[!ht]
    \centering
    \includegraphics[width=16cm,height=10cm]{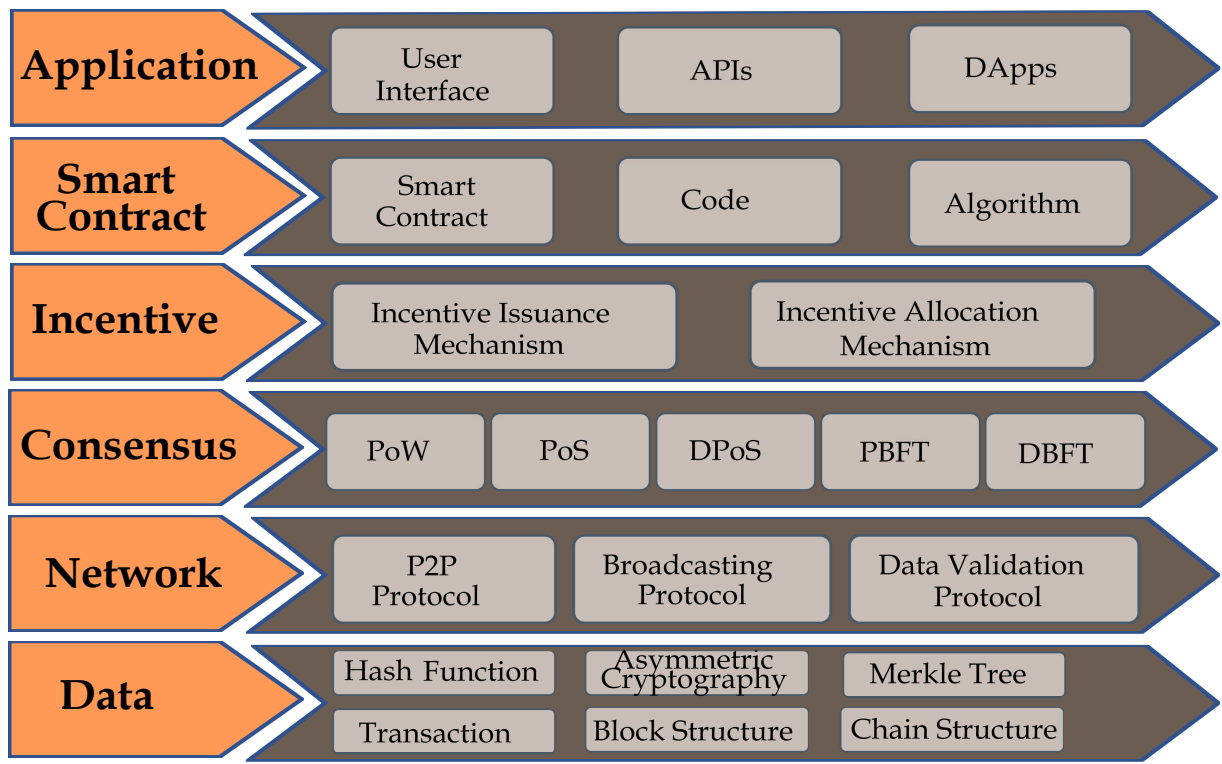}
    \caption{Layered Architecture of Blockchain}
    \label{fig:layerarchitecture}
\end{figure*}

\subsection{Layers}

The layered architecture of Blockchain can be divided into the following categories from top to bottom: application layer, smart contract layer, incentive layer, consensus layer, network layer and data layer \citep{wang2018overview, wen2021attacks}. Fig. \ref{fig:layerarchitecture} illustrates the layered architecture of Blockchain. 
\subsubsection{Application Layer}
The application layer is devoted to creating a wide range of Blockchain applications for use in many businesses and industrial sectors. This layer is an essential component of any architecture because it allows humans to communicate with the existing system and facilitates communication between an individual or a system over a network. The application layer comprises smart contracts, chain code, scripts, application program interface (APIs), user interfaces and frameworks. Further, it is also responsible for delivering specific user interface components and encompasses all that makes an application work, such as protocols and code.
\subsubsection{Smart Contract Layer}
The smart contract layer is the second layer of the Blockchain layer architecture, containing smart contract script and algorithmic logic for performing specific tasks inside the Blockchain application. In general, a smart contract script is a piece of code that is written and stored on the distributed ledger, and network nodes automatically execute it. Algorithmic logic is a set of rules and conditions that control how parties interact and communicate. When certain predefined conditions are met, the agreement is enforced and executed automatically.
\subsubsection{Incentive Layer}
The incentive layer is the third layer in the layered architecture of Blockchain. It is responsible for distributing incentives to nodes that contributed to the Blockchain by inserting valid blocks. The incentive method is primarily composed of two mechanisms: the issuance of incentives and the allocation of incentives. Aside from that, this layer enables nodes to participate in Blockchain verification by providing incentives. For instance, in Bitcoin, miners are rewarded with bitcoins, allowing additional users to join the network and mine the blocks. Similarly, ethers are used as mining incentives in Ethereum.
\subsubsection{Consensus Layer}

This layer is responsible for enforcing network rules that specify how nodes within the network should behave in order to achieve consensus on broadcasted transactions. It also ensures the integrity of records stored on the Blockchain as the fundamental layer of the Blockchain architecture. To accomplish this goal, the consensus layer incorporates several consensus protocols that enable Blockchain nodes to agree on the authenticity and legitimacy of newly created data blocks. The consensus layer contains specifications that define the rules for achieving consensus and how they can be applied depending on the consensus process. Various consensus mechanisms, such as PoW, PoS, DPoS, PBFT, DBFT, etc., have been proposed and used by various Blockchain-based applications. PoW algorithm was the first Blockchain algorithm to be implemented into the Blockchain network. PoS is a Blockchain consensus algorithm that allows miners to participate in the mining process by staking their coins. DPoS is a variant of PoS in which the stakeholders' problem is fully resolved, and any component on the network may act as a delegate. PBFT is primarily concerned with the state machine because it can replicate the system while avoiding the primary Byzantine general issue. DBFT is one of the most well-known consensus algorithms, and it is created to address the shortcomings of PBFT.

\subsubsection{Network Layer}

The network layer is the fifth layer in the Blockchain layer architecture, and it is primarily responsible for information exchange between Blockchain nodes. Although various components constitute the network layer and allow nodes to communicate on a Blockchain network, three primary components are considered primary components: the P2P network, the broadcasting protocol and the validation mechanism. In a P2P network, all nodes communicate using simple rules, and each node has an equal opportunity to create a new block in the Blockchain network. Following the generation of a block, each node broadcast the data to the P2P network for validation. All nodes do not have to receive the block data during broadcasting, but the primary node must accept it and connect it to the Blockchain to form a chain structure. The nodes in the validation mechanism obtain a new block containing information from other peer nodes and then verify the information before adding it to the Blockchain. The node agreed to add the new block to the Blockchain based on the validity of the information.

\subsubsection{Data Layer}
The data layer is responsible for handling and storing Blockchain data since it manages the data structure and physical storage space. As we know, Blockchain is based on distributed ledger technology; it enables the secure and efficient storage of data on a shared digital database. The ledger is constructed using a linked list of blocks, referred to as Merkle trees, that are encrypted using asymmetric encryption. The following components comprise the data layer: hash function, asymmetric cryptography, Merkle tree, transaction, block structure and chain structure. A hash function is used to convert the transactions into hash values since transactions are stored in the block in the form of hashes. Asymmetric encryption, such as public and private key pairs, is often used to secure the transfer of blocks through a network. The Merkle tree is used to arrange transactions as a tree and store them on the Blockchain. A transaction is any piece of data that is stored on the Blockchain. Blocks are primarily used as data structures, with the primary function of grouping all transactions and then distributing them to all nodes in the P2P network for verification. The transactions specified by the user are linked together in the chain structure by storing the hash of the previous block, in which each block stores the root hash.

\subsection{Types}

Starting with the types, there are three Blockchain types: public, private and consortium. These types are divided according to their assessment criteria and permission rules, all of which require access to the Blockchain network.

\subsubsection{Public Blockchain}

The public Blockchain is the most fundamental type of Blockchain network in which any user can participate, send and receive transactions, and validate (mine) the network’s transactions \citep{fromknecht2014decentralized}. The validation process is performed by specifically designated nodes called miners which run the consensus algorithm to verify the network’s transactions. The miners also add updated and validated blocks in the existing Blockchain \citep{garay2016Blockchain}. Indeed, the consensus algorithms such as PoW \citep{pow} and PoS \citep{postake} are mostly employed in a public Blockchain, in which a reward is given to the miners for their services (hashing or computations) in the network. The distribution of reward in the public network is directly proportional to the effort made by each miner; everyone has an equal opportunity to validate the blocks \citep{heires2016risks}. Moreover, different cryptography protocols are utilised in the public Blockchain to authenticate and secure users’ transactions \citep{kosba2016hawk}. For privacy purposes, the identity of each user remains anonymous in the public Blockchain. By confirming all the above features, Bitcoin \citep{nakamoto2008bitcoin}, Ethereum \citep{wood2014ethereum}, Litecoin \citep{litecoin} and Monero \citep{monero} are the most common and well-known examples of the public Blockchain network.

\subsubsection{Private Blockchain}

\begin{figure*}[!ht]
    \centering
    \includegraphics[width=12cm,height=8cm]{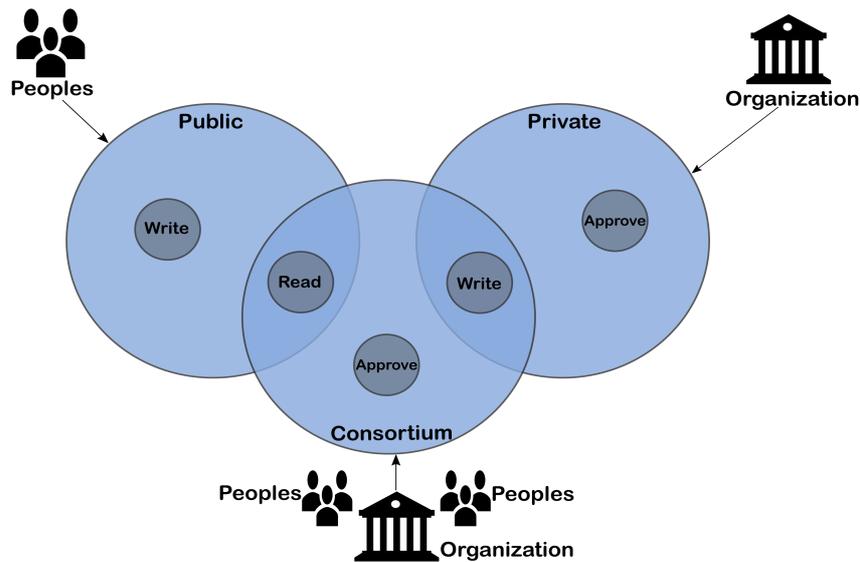}
    \caption{Consortium Blockchain with Different Access Policies}
    \label{fig:BCNETWORK}
\end{figure*}

In contrast to the public Blockchain, a private Blockchain is a permission-based Blockchain network that manages and specifies access for an organisation or group of people to read or write the blocks from the Blockchain \citep{gramoli2016danger}. Unlike a public Blockchain, a single authority is responsible for managing and updating the complete Blockchain setup by defining the rules and policies. In particular, a private Blockchain is designed for those organisations which want to keep their data safe within the given, defined boundaries, such as finance and audit companies \citep{li2017survey}. In a private Blockchain, the miners are the special nodes or trusted agents in which other nodes of the Blockchain network can blindly trust. The encrypted immutable ledger is shared with all other organisation members to keep the data safe \citep{meiklejohn2015privacy}. Generally speaking, a private Blockchain can solve challenging problems by providing secure solutions for different corporate sectors \citep{swan2015blockchain} and governmental organisations \citep{shin2017first}.

\subsubsection{Consortium Blockchain}

The hybrid form of the Blockchain network is referred to as a consortium Blockchain. In simpler terms, the consortium Blockchain combines different features and properties used in both public and private Blockchain networks. In most cases, the read request to access the specific block could be from a public Blockchain, whereas the write request is allowed only to private Blockchain nodes \citep{suberg2017we}. Fig. \ref{fig:BCNETWORK} depicts an example of various access policies (read, write and approve) on a consortium Blockchain as executed by other Blockchain types, such as public and private \citep{chen2020hyperbsa}. However, the consortium Blockchain’s consensus mechanism is run by those specific nodes that are initially defined to control and maintain the Blockchain setup \citep{maras2016r3}. For instance, an organisation consists of ten departments involved in different processes and activities; however, seven departments are the most authentic departments responsible for creating the organisation’s laws.

\subsection{Storage Structure}

Blockchain technology consists of different valuable and dominant features of cryptography to build real-time and networked applications based on decentralised and distributed databases. As the real-time applications gather and produce a large volume of data, it is essential to manage and secure the storage locations used either internally (that is, a local hard drive) or externally, such as a server or cloud. In general, there are two types of storage models utilised by Blockchain applications in order to store real-time data. The first model used in most of the applications is the on-chain storage model. The other is the off-chain model, which is employed by those applications which produced the data in a streaming fashion, such as IoT, SG, vehicular network and financial services applications.

\subsubsection{On-Chain Storage}

Built on distributed systems to manage the data across the network, the primary storage of Blockchain is called on-chain storage. In the Blockchain, the data is stored as confirmed transactions in the form of blocks. A block is linked to the previous block to form a complete chain. The special nodes (or miners) are responsible for undertaking the validation tasks, in order to add the existing Blockchain’s confirmed blocks \citep{eberhardt2017or}. For this purpose, the miners are rewarded with some financial benefits for their services provided to the Blockchain network. However, all these tasks, such as transaction execution and verification, distribution of reward and decentralised execution, can incur an extra storage overhead on the system. Moreover, in a public Blockchain, the transaction structure is open to everyone and, therefore, not completely anonymous. There are a few advantages to storing the data on the on-chain Blockchain; for instance, the users are not required to maintain both storage locations, thus reducing the computation and storage cost for off-chain storage. Also, users have complete control over their data in the on-chain storage \citep{Wst2017DoYN}.

\subsubsection{Off-Chain Storage}

In the Blockchain, off-line storage is also referred to as off-chain storage in which user transactions are stored on another system (or storage) other than the actual storage, in order to restrict users’ access. In fact, the off-chain system is utilised to restrict the read access to the Blockchain using different access control policies. Blockchain transactions are stored in a hash containing the actual information about the transactions stored inside them and no one can obtain the information from this hash \citep{ali2018blockchain}. Only the verifying party can access the hash information with the information about stored transactions on the Blockchain. There are many advantages in using the off-chain storage concept in Blockchain; for example, it can provide greater privacy for user transactions when user access is controlled and managed by access control policies. In this way, the user keeps personal information separate from the other Blockchain users in the network. However, using the off-chain method in the Blockchain, there are some disadvantages, such as a lack of user confidence and the need to distribute information across multiple storage locations through connecting hash references. Moreover, the cost of storing data in the off-chain system, along with the actual storage, is very high because users have to manage the record with extra computation power.

\subsection{Transaction Models}

One of the prominent features of Blockchain is distributed ledger in which the different users record the online generated transactions in an immutable way. The transactions in the Blockchain and their related applications are designed and maintained in a specific way. The basic idea behind the design of different transaction models is that they can resist a wide variety of attacks on Blockchain applications. The two most commonly used models in Blockchain applications are the Unspent Transaction Outputs (UTXO) model and the Account-based Transaction model.

\subsubsection{UTXO Model}

This model is the fundamental transaction model commonly used in Bitcoin and related cryptocurrencies applications to represent the currency transactions. In Bitcoin, the currency is in Bitcoins being recorded as a transaction in the user’s wallet. For the representation of Bitcoin in the wallet, a list consisting of unspent transactions is maintained by the user. This includes all details such as value, owner and time \citep{zahnentferner2018chimeric}. Anyone can see all the unspent transactions in the user’s wallet in the group, as the total balance of that user. The owners  sign Bitcoin transactions with their private keys. Anyone in the group can prove the transaction and the user’s ownership and authentication, with the owner’s public key. The UTXO model has become very successful in cryptocurrencies, especially in Bitcoin, for reasons of privacy and scalability of transactions \citep{delgado2018analysis}.

\subsubsection{Account-based Online Transaction Model}

This model represents the Blockchain transactions in a different format designed as an Account-based Online Transaction, a model in which the address (or account) of the sender is utilised to represent the transactions, as opposed to the unspent transactions output in the UTXO model \citep{shin2017first}. The account-based model is employed by the Ethereum application to generate and deploy the smart contracts on the Blockchain with the available accounts. The basic aim of the account-based model is to enhance the consensus algorithms’ efficiency and reliability by improving the verification time for blocks \citep{chalaemwongwan2018state}. As with the UTXO model, the balance in Ethereum applications are stored as the transaction and called ethers (or gas) in the Ethereum Blockchain with approved properties about the sender, such as signature, approval and balance of the sender. Unlike the UTXO model, the account-based online transaction model has unlimited space store the information of other users because it does not store any unnecessary details of ethers compared with the coins in the previous model. The account-based online transaction model is widely accepted in most Blockchain applications due to its many features, such as design simplicity, ease of understanding and security enhancement \citep{ma2017efficient}.

\section{Requirements for Blockchain-based Industry 4.0 Applications} \label{sec:4}

 \begin{figure*}[!ht]
    \centering
  \includegraphics[width=12cm,height=8cm]{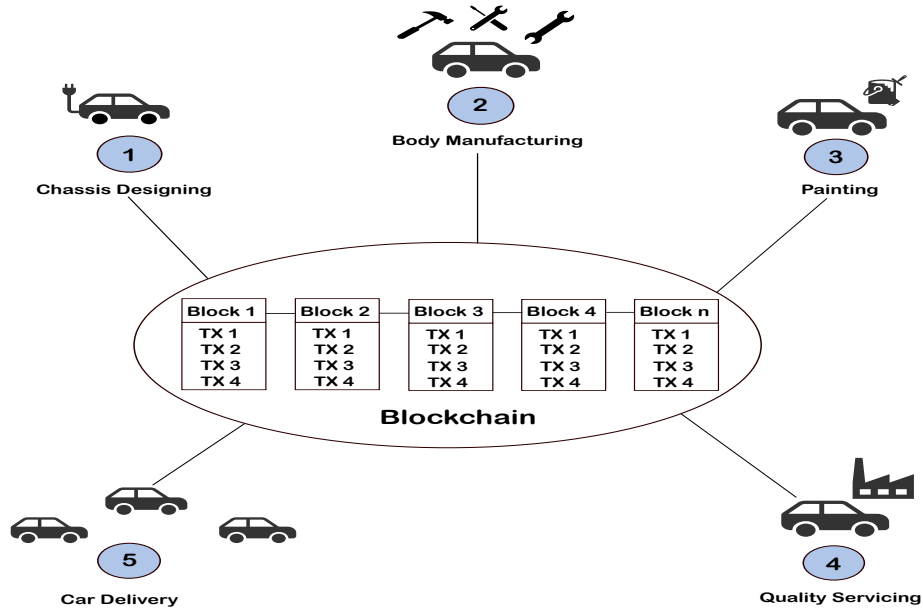}
  \caption{\colorbox{white}{Car Manufacturing Process in Smart Factory under Industry 4.0}}
    \label{fig:Industry_example}
\end{figure*}

We are in the industrial revolution 4.0, and this profoundly changes the way we live, work and communicate with each other. The term ``Industry 4.0'' is interchangeable with "fourth industrial revolution" and refers to a new stage in the coordination and planning of the industrial production process. With advanced digital technologies and the transition of Industry 4.0, the manufacturing process of Industry 4.0 has now changed completely, through a series of digital transformations, to achieve productivity and automation of the entire process. This transition of industrial processes enables traditional enterprises and factories to move towards the emergence of smart factories, referred to as Industry 4.0. The smart factories can be developed by integrating artefacts, operators, and the provision of background information into an industry framework through the internet \citep{lasi2014industry,sanchez2020autonomic}.

\textcolor{black}{The internet is the most important technology in Industry 4.0 since it is the foundation for most other technology drivers. By utilizing this technology, IoT and other associated innovations, including distributed network, fully automation, and competitive production networks, are primarily driving Industry 4.0. However, the scope of Industry 4.0 can be expanded by bringing leading technological players, propelling this advancement to the next level to benefit both industry and partners. IoT, Blockchain, big data, edge and cloud computing, robotics, artificial intelligence, and open-source software are the key technological players \citep{xie2021efficient}. The advantage of integrating these players with Industry 4.0 is the creation of an integrated and automated system, such as a cyber-physical system (CPS), that can transform the underlying industrial infrastructure and production processes into an autonomous and dynamic system in order to achieve process resiliency. The components in these highly integrated CPSs must interact and act intelligently to collaborate autonomously and accomplish a shared objective \citep{broo2021cyber}. Furthermore, real-time information sharing between different components can be achieved by the use of the web or a computing and data infrastructure that allows warehouse-scale computers to communicate with each another, which benefits organisations to improve their productivity and collaboration \citep{jiang2020design}.}

Ultimately, the goal of Industry 4.0 is to speed up manufacturing process, enhance operational effectiveness, and better serve customers while opening the door to new business models and possibilities that go beyond automation and to aid in real-time discovery. In Industry 4.0, the manufacturing process is integrated with smart devices, sensors, machines and humans, along with their behaviours, to periodically generate data used in different production processes, in order to achieve better quality and more sustainable environmental decision-making. In the manufacturing process, the modules collect data from the sensors embedded in the machine equipment, perform some computations and send them to the next module as an output transaction. Each module in the manufacturing application is linked to others, using some communication mediums. For most industrial setups, a series of processes carried out at various sites contribute to the system’s functionality as a whole \citep{oberer2018leadership, ghobakhloo2019adoption}.


\textcolor{black}{Industry 4.0 is not only concerned with the digital transformation of production or manufacturing, but also with the digital transformation of other industrial sectors and value creation processes. With the emergence of new technologies such as mobile data networking and network protocols, as well as the IoT stack components and security features used in Industry 4.0, it is possible to take a step forwards and leverage Industry 4.0 features, which can then be connected with other industrial sectors to form Industry 4.0-based applications. This transformation in Industry 4.0 has encouraged business and research communities to look beyond manufacturing processes and expand their horizons to include Industry 4.0 applications in other industry sectors such as healthcare, energy, financial, logistics and supply chain \citep{oztemel2020literature}.}


\textcolor{black}{With the increased interest in the development of Industry 4.0-based applications, different critical challenges have been raised. These issues are related to the design and performance of the application, security and privacy of users \citep{ahram2017blockchain}.
An example for design challenge that industries face is the use of centralised architecture, which can become a bottleneck and often results in problems with scalability and single point of failure in large industry setups. Additionally, since industrial applications process and store a vast amount of data, it is critical to consider storage-related issues such as data heterogeneity and data redundancy and security and privacy concerns regarding to data confidentiality and data integrity. As different industries continue to utilise unique features of Industry 4.0, they are seen as an appealing target for attackers. Therefore, security is considered as a critical factor in the successful implementation of Industry 4.0-based applications \citep{mohamed2019applying, fernandez2019review}.}

Nowadays, Blockchain technology is becoming more popular in multiple Industrial-based 4.0 applications due to its promising features, such as decentralisation, distributed immutable ledger, transparency, anonymity, autonomy, open source, verifiability and security. Such innovative Blockchain features can solve a number of problems which has risen dramatically in Industry 4.0 applications. Each application, which is explicitly based on the Blockchain structure, must comply with the requirements set out for the industrial system model, following the design phase, security and privacy of both users and data \citep{alladi2019blockchain, bodkhe2020survey}.

\textcolor{black}{Numerous blockchain architectures have been suggested for use in a variety of Industry 4.0-based applications. The summary of Blockchain architectures being used across different industry sectors is as follows, but these architectures are not limited to: 
\begin{itemize}
    \item \textbf{Codefi} \citep{codefi} is a blockchain-based financial architecture launched in September 2019. It consists of various product modules that work together to fuel the next generation of commerce and finance. Codefi's blockchain suite offers a significant solution for resolving the issues associated with traditional finance, scaling decentralised networks, and better access to web-based technology.
    \item \textbf{MedRec} \citep{MedRec} is a well-known implementation of the blockchain architecture for health care applications, which enables the secure and efficient storage of health-related data. In this architecture, each entity that participates in the case, such as the patient, the doctor, and the patient's insurance company, may update the patient's health record. \textbf{Medicalchain} \citep{Medchain} is another decentralised health care architecture focused on blockchain technology used in the United Kingdom to manage patient data. This architecture prioritises the users' needs while maintaining a single reliable version of the users' data on a distributed ledger.
    \item \textbf{PowerLedger} \citep{ledger2017power} architecture  is based on Blockchain technology that facilitates the buying and selling of energy resources based on an allocation market and prioritises this surplus energy within micro-grids or around the distribution. \textbf{Bankymoon} \citep{bankymoon} is another blockchain-based energy architecture that offers blockchain-enabled smart prepaid energy metres to schools and communities worldwide that lack access to affordable energy. 
    \item \textbf{Electronic Product Code Information Services (EPCIS)} \citep{lin2019food} is a blockchain-based architecture for food traceability that uses Electronic Product Code Information Services (EPCIS) services and demonstrates its advantages. In this architecture, the blockchain is used to keep track of data to a higher degree since less information is on-chain and more is off-chain using EPCIS. \textbf{OriginTrail} \citep{OriginTrail} is a blockchain-based architecture established in 2013 with the aim of bringing transparency to complex global supply chains. This platform is widely used in the food industry and informs users about their food products' location.
    \item \textbf{The IBM Watson IoT platform} \citep{ibmwatson} supports isolated blockchains for IoT data sharing, adding an extra layer of protection and integrity to IoT transaction flows. Another Blockchain architecture, \textbf{ADEPT (Autonomous Decentralized Peer-to-Peer Telemetry)} \citep{veena2015empowering}, uses Blockchain technology in IoT network and employs Ethereum, Telehash's functions and BitTorrent.
\end{itemize}
}

Developing secure Blockchain-based applications using Industry 4.0 guidelines, is a critical challenge that usually requires an appropriate relationship between the architectural components of the underlying domain and the Blockchain features in order to achieve optimal usability at various levels \citep{korpela2017digital}. Furthermore, security and privacy requirements are critical challenges for developers and researchers in order to ensure the proper compliance between users and industry partners \citep{gokalp2017development,qin2016categorical}.  For example, the Blockchain-based IoT application needs to accomplish domain requirements, as well as the security and privacy requirements for the Blockchain \citep{bahga2016blockchain}.

To meet the requirements of the Blockchain-based application in the Industry 4.0 domain, we show the example of the car manufacturing process in the smart factory and how it works under the Industry 4.0 definition and Blockchain in which each module involved in the manufacturing process is integrated with Blockchain technology, as shown in Fig \ref{fig:Industry_example}. With distributed ledger technology, the transactions of each module during the car manufacturing process are stored on a distributed, immutable ledger. A complete record of chassis designing, body manufacturing, painting, quality servicing and delivery of successful units is maintained throughout the process, followed by the distributed ledger’s data.

From this Industry 4.0 example, we draw two requirement perspectives for the design of secure Blockchain-based Industry 4.0 applications, including application and security. From an application perspective, we cover the design requirements of industrial applications related to their architecture, functional and non-functional, and performance that need to be met for decentralised and distributed Industry 4.0-based applications, together with security and privacy for both users and processes in Blockchain perspectives. The design requirements of industrial applications are discussed in detail in subsection \ref{section_design_requirement}.

We discuss users’ security and privacy requirements of Blockchain-based Industry 4.0 applications from a security point of view. Security and privacy are critical issues for Industry 4.0-based applications since there is a high chance of unauthorised data breaches or information leakage, resulting in critical data loss for Industry 4.0 based applications. For example, malicious attacks on sensing devices and between supply chain processes may disrupt the communication and services of the overall manufacturing processes and disclose personal information related to their identity and transaction. Therefore, the security of integrated modules and their generated data in the industrial production process is essential, requiring users’ security and privacy of their transactions in the field of application development Industry 4.0. The security and privacy requirements are discussed in detail in subsection \ref{section_security_requirements}.

\begin{figure*}[t]
    \centering
  \includegraphics[width=18cm,height=13cm]{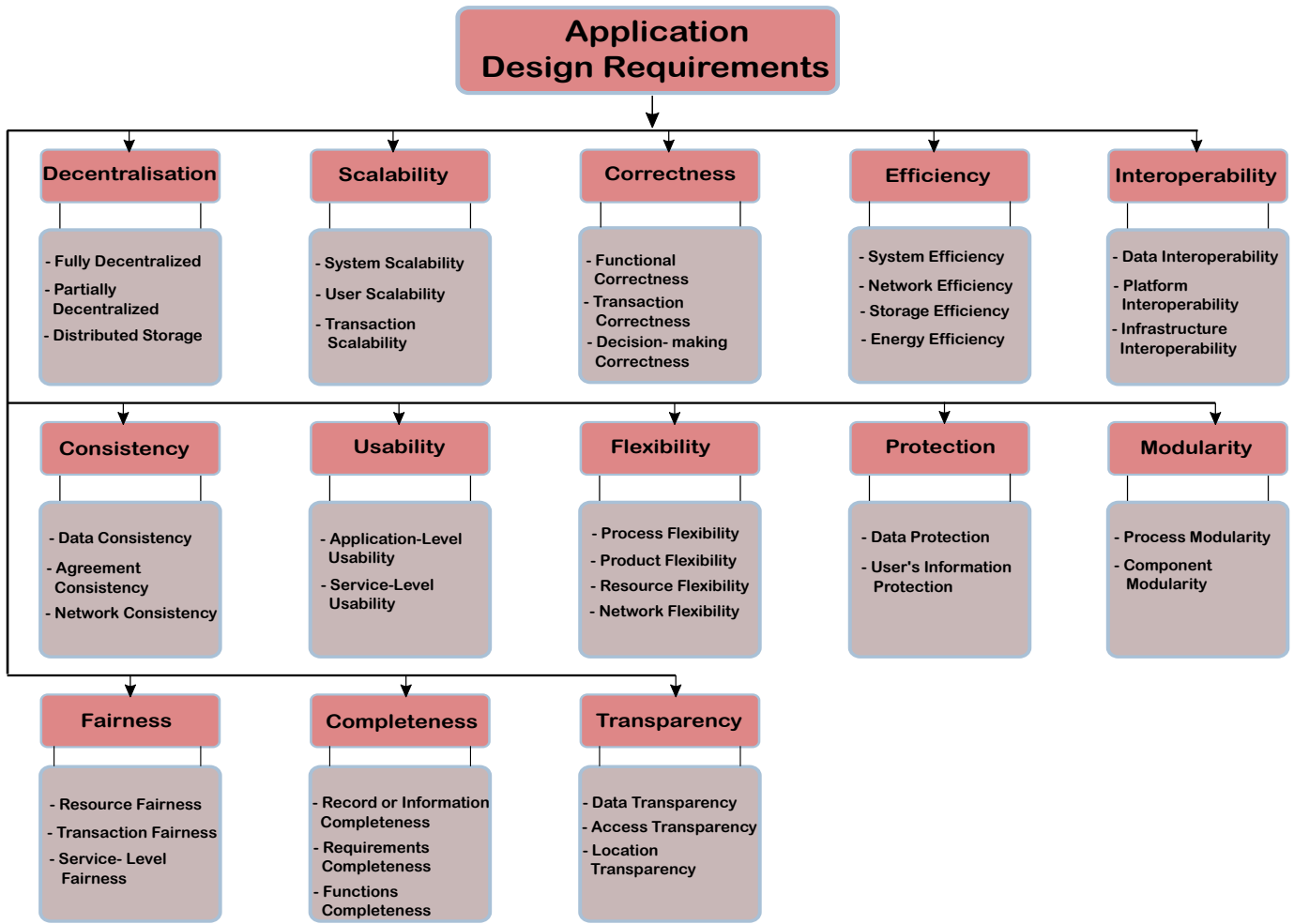}
  \caption{\colorbox{white}{A Taxonomy of Design Requirements for Blockchain-based Industry 4.0 Applications}}
    \label{fig:design_requirements}
\end{figure*}

\subsection{Application Design Requirements} \label{section_design_requirement}
This section goes into depth about the requirements and sub-requirements for designing Blockchain-based Industry 4.0 applications. We categorise each design requirement into its potential sub-requirements and describe them in accordance with industry perspectives. The requirements for the design of secure Blockchain-based Industry 4.0 applications include decentralisation, scalability, correctness, efficiency, interoperability, consistency, usability, flexibility, protection, modularity, fairness, completeness and transparency. We also investigated and outlined the potential sub-requirements for each design requirement and classified them rationally. The high-level taxonomy of requirements and sub-requirements for designing Blockchain-based Industry 4.0 applications is illustrated in Fig. \ref{fig:design_requirements}. 

Along with specifying the requirements and sub-requirements, we conducted an in-depth review and exploration of the various measuring criteria that state how to achieve these requirements rationally, as shown in table \ref{table:measuringcriteria}.

\subsubsection{Decentralisation}

Decentralisation is one of the key requirements for a Blockchain-based application to distribute the loads among the entities involved in the manufacturing process \citep{isaja2018distributed}. Traditionally, the manufacturing process of the industrial setup was managed by a single centralised party responsible for distributing and maintaining the overall computational loads on different modules involved in the process. This centralised server faced bottleneck and overhead storage issues if multiple requests were received from other processes at the same time \citep{kremenova2019decentralized}. Decentralised architecture removes a centralised entity’s requirement from the overall process, saving middlemen’s costs to ensure process verifiability and auditability. Decentralised architecture in Blockchain systems is designed according to the three types of Blockchain networks, public, private and consortium, managing access control for all network users \citep{asharaf2017decentralized}.

Aside from that, distributed storage is an essential prerequisite for any decentralised architecture that recognises the value of equality in a P2P network in which each node has equal rights to validate and verify transactions stored in the database \citep{aitzhan2016security, kopp2017design}. Since there is no central authority in the Blockchain network, each node has a dual responsibility to perform data computing and send updated copies of data to all other nodes in order to ensure consistency across the network. In this respect, each node manages its storage locally without relying on centrally controlled storage systems \citep{chatzopoulos2018privacy}.

The decentralised requirement can be further subdivided into sub-requirements such as:

\begin{itemize}
          \item \textbf{Fully Decentralised:} In a fully decentralised network, anyone can join the network as an individual node and send transactions to other nodes without the need for a trusted third party. A consensus process is used to create a trusting relationship between nodes. A public Blockchain is an example of a fully decentralised network.
          \item \textbf{Partially Decentralised:} In a partially decentralised network, some nodes are assigned to monitor and manage network operations, while other nodes may participate in the same way as in a completely decentralised network. A consensus mechanism, for example, is managed by a pre-selected set of nodes in this type of network. An example of a partially decentralised network is a consortium Blockchain.
          \item \textbf{Distributed Storage:} Distributed storage is a requirement for any decentralised architecture that emphasises equality between P2P nodes in order to validate and verify transactions stored in the database.
      \end{itemize}

\subsubsection{Scalability}

With the rapid pace of development of Blockchain applications and their popularity in Industry 4.0 over the last few years, scalability is emerging as a critical requirement \citep{ziegeldorf2018secure,zhang2018fhirchain,kang2018blockchain}. In real-time applications, data or transactions generated on a continuing process, each module will send multiple transactions to other modules. As a result, the entire network could be constrained under certain circumstances, such as traffic bottleneck and storage overflow \citep{Zheng2018}. For example, Bitcoin handles an average of five  transactions per second in one stream, extended to hundreds or thousands of transactions per second for financial applications \citep{wang2018designing}. Hence, scalability is an essential requirement for Blockchain-based applications to support the maximum number of users and transactions in the system.

The scalability requirement can be further subdivided into sub-requirements such as:

\begin{itemize}
          \item \textbf{System Scalability:} The system scalability addresses the requirements for scalable server capabilities to meet the expected future number of available interactions, as well as the response time per user request.
          \item \textbf{User Scalability:} The user scalability requirement is defined as the maximum number of concurrent users the system can accommodate without hindering its performance. 
          \item \textbf{Transaction Scalability:} Transaction scalability is often associated with some storage structures such as databases, in which scalability is described as the ability to react to an increasing number of user queries within a specific time frame in comparison to the increasing database's output performed using multi-processing systems. 
      \end{itemize}

\subsubsection{Correctness}

Correctness is a crucial requirement to analyse computational results and to measure the behavioural aspects of any real-time system \citep{ma2018blockchain, ziegeldorf2018secure}. There are many ways to determine the system’s correctness, such as experiments, simulations, mathematical analysis, logical evidence and formal modelling \citep{amoussou2018correctness}. In Blockchain applications, correctness can be measured by the number of transactions executed, block generation time, consensus time, transparency level and the integrity of the proposed system \citep{Lin2018BSeInAB}. Similar to the performance parameters, the correctness of security and privacy-preserving schemes for Blockchain-based applications is achieved by performing detailed security analysis and robustness against different types of privacy attacks.

The correctness requirement can be further subdivided into many sub-requirements, including the following:

\begin{itemize}
        \item \textbf{Functional Correctness:} A functional correctness requirement is concerned with properties that include deciding the relationship between inputs and outputs from industrial processes instead of other systems efficiency settings such as computation time, communication, and memory overhead.
        \item \textbf{Transaction Correctness:} The correctness of a transaction is concerned with the correct execution of the operations specified in that transaction in terms of their abstract meanings and data structure.
        \item \textbf{Decision-Making Correctness:} The requirement for correct decision-making is often based on a rational decision-making model, which ensures that decisions are made based on facts, systematic data collection, and analysis. Furthermore, this requirement ensures that the user is aware of the possible and reasonable values for the input data.
    \end{itemize}

\subsubsection{Efficiency}

Efficiency is the system’s competence to perform a variety of tasks with minimal effort but higher output. The basic parameters used to measure system efficiency are computational power, bandwidth communication and storage capacity \citep{fanning2016blockchain, guan2018privacy123}. The most prominent mode of measuring the efficiency of Blockchain applications is the miner’s ability to solve the challenges in a given time \citep{kang2018blockchain}. For example, in Bitcoin, PoW takes almost 10 minutes to solve a block, including the time needed to add it to the existing Blockchain. Although many consensus algorithms have been proven to be more efficient than others, their energy consumption is of concern \citep{li2018efficient}. For example, each consensus algorithm consumes a certain amount of energy (or power) from adherent hardware resources in order to solve a computational challenge such as a puzzle.

The efficiency requirement can be further divided into many sub-requirements, including the following:

\begin{itemize}
         \item \textbf{System Efficiency:} The term "system efficiency" refers to the highest performance level of any computation system, in which optimal results are obtained with the least number of inputs.
         \item \textbf{Network Efficiency:} Network efficiency is characterised as the efficient sharing or transmission of information to local and global networks such as the internet while maintaining the acceptable bandwidth.
         \item \textbf{Storage Efficiency:} The storage efficiency requirement focuses on storing and processing data in such a way that it requires the least amount of space while having little effect on output.
         \item \textbf{Energy Efficiency:} The degree of energy efficiency measures the extent to which particular results are attained with less energy expenditure.
     \end{itemize}

\subsubsection{Interoperability}

The interoperability design requirement enforces process integration between the different components of decentralised applications in order to facilitate efficient interaction and communication \citep{kuo2018modelchain, zhang2018fhirchain, zhang2018towards}. Since interoperability derives from the idea of secure communication between different network components, it enables decentralised applications to exchange information over the network using a secure Blockchain \citep{dagher2018ancile}. Thus, the Blockchain platform needs to propose a seamless exchange of data between different Blockchains. Furthermore, the inter-operability feature is more challenging when considering that Blockchain applications must not cross the boundaries defined for the inter-operability of the system employing fair access mechanisms.

The interoperability requirement can be further subdivided into sub-requirements such as:

\begin{itemize}
    \item \textbf{Data Interoperability:} Data interoperability enables various common frameworks to construct, exchange, and handle data in order to share definitions, comprehend context, and accept collective responsibility.
    \item \textbf{Platform Interoperability:} Interoperability across platforms allows individuals and applications to explore, access, integrate, and analyse data on a single platform. Furthermore, it promotes system flexibility through the use of standardisation software bundles, metadata, and identifiers. 
    \item \textbf{Infrastructure Interoperability:} Infrastructure interoperability is concerned with enhancing and expanding information technology infrastructure to a broader scale, such as clouds, to ensure that all applications and related technologies work seamlessly. Furthermore, infrastructure interoperability is based on securely connecting new and existing systems to guarantee data consistency and data security.

\end{itemize}

\subsubsection{Consistency}

Blockchain is recognised as a leading technology for maintaining the consistency of transactions stored in the distributed databases \citep{zhong2019secure, zhang2018fhirchain}. As transactions in Blockchain applications are generated periodically, each node is responsible for sending updated copies of these transactions to other nodes over the network in order to maintain transaction consistency \citep{rottondi2017privacy, gao2018blockchain}. Therefore, maintaining and ensuring the Blockchain system’s degree of consistency is a challenging requirement for real-time industrial applications.

The consistency requirement can be further subdivided into many sub-requirements, including the following:

\begin{itemize}
          \item \textbf{Data Consistency:} In an industrial environment, data consistency ensures that the original logics can be correctly simulated. Furthermore, it implies that all operating processes have access to the same data in order to satisfy data integrity constraints.
          \item \textbf{Agreement Consistency:} Agreement consistency is characterised as the realisation of rules and regulations between different processes implemented by some regulatory authorities, such as smart contracts in Blockchain.
          \item \textbf{Network Consistency:} Network consistency is an essential requirement that focuses on evaluating system behaviour and functionality using index measures. These measurements are based on three factors: network hardware, software, and technology configurations.
      \end{itemize}

\subsubsection{Usability}

With the decentralisation aspect of Blockchain technology, many users choose to solve the scalability and efficiency issues in traditional IoT systems \citep{dorri2017blockchain}. However, most Blockchain systems achieve the essential functionality needed to process and validate transactions over the network without an in-depth look at usability issues that may prevent users from using such systems in their domains. Therefore, usability is one of the key requirements, specified to meet customers’ essential needs at first sight, so that users can feel more secure in communicating with different Blockchain systems \citep{kopp2017design}. As a result, the current requirement is to provide users with an easy-to-use interface to enhance customer experience and satisfaction.

The usability requirement is further subdivided into various sub-requirements such as:

\begin{itemize}
           \item \textbf{Application-Level Usability:} Application-level usability is concerned with an analysis of user experiences with the application. The resulting data is used to enhance the system's capabilities and suggest further changes to make the interface more interactive and accessible.
           \item \textbf{Service-Level Usability:} In comparison to application-level usability, service level usability is characterised as any system's ability to react to user requests in order to measure the user's expectation on some scale. At this level of usability, each user request is weighed against the services offered by the system. 
       \end{itemize}

\subsubsection{Flexibility}

Blockchain technology has proven to be a potential solution for the requirements of various business and Industry 4.0 that have arisen in their existing systems, such as the efficiency of secure and reliable transactions without a central entity. A flexible Blockchain system can provide a basic platform for other technologies to integrate effectively, deploy different modules and deliver effective results \citep{ma2018blockchain,dagher2018ancile, Lin2018BSeInAB}. Besides, there is also a need to optimise performance for integrated Blockchain systems; therefore, addressing the flexibility of different applications is a challenging requirement. The system should have inherent features provided by the different Blockchain technologies \citep{ellis2018flexibility}.

The flexibility requirement can be further subdivided into the following sub-requirements:

\begin{itemize}
          \item \textbf{Process Flexibility:} In the industrial context, process flexibility is an essential requirement that uses the principle of process management to effectively respond to critical system operations concerning outside factors such as increases or decreases in supply or demand. Using process flexibility can increase system outputs in terms of goods while also lowering the cost of external factors such as time.
          \item \textbf{Product Flexibility:} Product flexibility requirement can be measured in terms of adaptability for any potential changes in the product, including new designs and variations. Flexible product design reduces the costs of redesign and allows for quick customer response through better efficiency.
          \item \textbf{Resource Flexibility:} Being resource-flexible is frequently evidenced by the capabilities of the resources to handle a wide variety of manufacturing activities in an efficient way.
          \item \textbf{Network Flexibility:} In the industrial context, network flexibility is defined as the underlying system's ability to effectively handle processes that must be executed and migrated between different modules. 
      \end{itemize}

\subsubsection{Protection}

Blockchain technology is designed as protected since it uses the immutable feature to store data in an append-only fashion. It is not, therefore, practical for everyone to modify the data stored in the Blockchain databases \citep{ma2018blockchain, wang2018designing, hussein2018medical, zhong2019secure, zhang2018fhirchain}. Besides, Blockchain technology has demonstrated its potential to achieve data integrity by using the Merkle hash tree concept, in  which block hashes are interlinked so that all network nodes can easily detect any change in hash values. Given this, the key requirement here is to develop a security model that protects the user from unauthorised attempts to obtain personal information.

The protection requirement can be further categorized into two sub-requirements such as:

\begin{itemize}
          \item \textbf{Data Protection:} The data protection requirement focuses on preventing sensitive information from being altered, corrupted, or lost. Nowadays, data protection is becoming a significant issue in industry as the volume of data reaches an unprecedented scale.
          \item \textbf{User's Information Protection:} User information protection aims to safeguard the personal information of individuals involved in the overall processing of industrial setups.
      \end{itemize}

\subsubsection{Modularity}

To maximise flexibility and decouple the hierarchy of modules, modularity allows different interrelated organisations to join and use network resources in order to provide comprehensive services with the power of reusability \citep{ziegeldorf2018secure, zhong2019secure, mohamed2019leveraging}. Over the years, Blockchain technology has become more popular among the public to solve various problems; developers can build and develop decentralised applications using different languages that run on heterogeneous platforms \citep{zhang2018fhirchain}. Efforts are being made by industry and research communities to address the issue of modularity. One example is the \colorbox{white}{the Komodo \citep{komodo}}, an open source Blockchain modular framework designed to facilitate the process of integrating different end-to-end communication modules between users, in order to address the issues of scalability, protection and inter-operability in the Blockchain network.

The modular requirement is further subdivided into two sub-requirements, which include the following:

\begin{itemize}
          \item \textbf{Process Modularity:} Process modularity is a requirement that focuses on improving system efficiency by breaking down a single extensive process into multiple sub-processes that can operate in parallel on multiple machines.
          \item \textbf{Component Modularity:} In terms of component-level modularity, a modular design is often interpreted as partitioning functions into multiple discrete, compact, and scalable modules, in which extensive use of well-defined standardised interfaces is needed.
      \end{itemize}

\subsubsection{Fairness}

Fairness is one of the key requirements for Blockchain applications to achieve trust between different industries and the proposed security models \citep{kosba2016hawk, Zyskind2015EnigmaDC,zhong2019secure}. In other words, fairness is achieved by providing middleman agreements, that is smart contracts, which comply with the rules and conditions in order to facilitate the communication process between sender and receiver \citep{hardwick2018voting}. These contracts define the logic and conditions used to allow the parties to interact with each other without trusted third parties. As a result, there is a critical need to design security schemes for Blockchain applications that provide fairness to all users, keeping them actively involved and remaining part of the Blockchain.

The fairness requirement can be further subdivided into many sub-requirements, including the following:

\begin{itemize}
          \item \textbf{Resource Fairness:} Resource fairness is an essential requirement in industrial setup since it enables a system to allocate resources equally among the various processes running in the system. A single user may completely overload the system with its transactions, leading to poor performance for other users.
          \item \textbf{Transaction Fairness:} In the Blockchain, transaction fairness is often associated with some payment systems, in which fair transactions are needed for promoting fair payment to those who participate in and join the network for mining. 
          \item \textbf{Service-level Fairness:} Service-level fairness encourages the equality of resources and software running on the system to all network users.  
      \end{itemize}

\subsubsection{Completeness}

Completeness, as a design requirement, aims to ensure users’ specific needs and requirements in order to complete any application. In Blockchain-based applications, the security and privacy models are deemed complete if they prove the satisfactory computational requirement and comprehensive security analysis, using multiple proofs and logic \citep{kosba2016hawk,Zyskind2015EnigmaDC}.

The completeness requirement can be further subdivided into many sub-requirements, including the following:

\begin{itemize}
          \item \textbf{Record or Information Completeness:} Data completeness is expressed as an expected degree of completion of data, in which optional data is often discarded at some level. As a result, as long as the data follows the standard and specifications, it is considered complete. 
 
          \item \textbf{Requirements Completeness:} An individual requirement is considered complete if it contains all required information to communicate the message to prevent uncertainty and requires no amplification to maintain adequate implementation and verification.
 
          \item \textbf{Functions Completeness:} A function completeness must ensure.
      \end{itemize}

\subsubsection{Transparency}

Transparency is one of the most demanding requirements of Blockchain applications, specifically for public Blockchain users \citep{wang2018designing}. Although completely different from the implementation point of view \citep{zhong2019secure}, many users are confused with the concepts of privacy and transparency, For example, in Blockchain applications, cryptography primitives are considered the most powerful means of achieving transparency of transactions linked to the respective users. Therefore, regardless of the Blockchain’s open nature, the users’ transactions must be transparent and invisible to other users on the network.

The transparency requirement can be further subdivided into many sub-requirements, including the following:

\begin{itemize}
          \item \textbf{Data Transparency:} Data transparency is an essential requirement that many industries today integrate into their everyday routine processes in order to facilitate open communication, so everyone has access to the same information.
          \item \textbf{Access Transparency:} Access transparency enables objects to be accessed using the same access functions regardless of whether they are static or mobile. Furthermore, the interface required to access an object should be identical to the object's position in the system.
          \item \textbf{Location Transparency:} The ability to access entities and system resources without knowing their specific location is referred to as location transparency.
      \end{itemize}

\begin{sidewaystable*}
    \centering
\caption{Designing Blockchain-based Industry 4.0 Applications: Requirements, Sub-Requirements and Measuring Criteria}
\label{table:measuringcriteria}

   \footnotesize	    
    
    \begin{tabular}{p{0.09\textwidth}|p{0.16\textwidth}|p{0.7\textwidth}}
    
      \toprule 
     \textbf{Requirements} & \multicolumn{1}{c|}{\textbf{Sub-Requirements}}& \multicolumn{1}{c}{\textbf{Measuring Criteria}} \\\midrule
      
      \textbf{Decentralisation} &  
      
      \begin{itemize}
          \item Fully Decentralised
          \item Partially Decentralised
          \item Distributed Storage
      \end{itemize}
      
      &

      \begin{itemize}
          \item \textbf{Network:} Fully decentralised or partially decentralised
          \item \textbf{Storage:} Distributed ledger technology (DLT) \citep{viriyasitavat2019blockchain}
          \item \textbf{Control:} Fully centralized (public Blockchain), centralized (private Blockchain), partially centralized (consortium Blockchain) \citep{maesa2019blockchain}
          \item \textbf{Communication Topology:} P2P \citep{hao2020towards}
          
      \end{itemize}

       \\\midrule
      
      \textbf{Scalability} &
      
      \begin{itemize}
          \item System Scalability
          \item User Scalability
          \item Transaction Scalability
      \end{itemize}
      
      & 
      \begin{itemize}
          \item \textbf{Transaction Per Second (TPS):} Used to compute the number of transactions processed and recorded on the Blockchain per second. 
          \item \textbf{Block Creation Time:} The amount of time taken to build a new block \citep{bodkhe2020survey}. 
          \item \textbf{Block Size:} Block size (in bytes) for storing a number of transactions.
          \item \textbf{Throughput:} The transaction throughput is characterised as the rate at which valid transactions are accepted and stored by the Blockchain within a given time frame.
          \item \textbf{Response Time Per User Request:} The system's initial response to user requests submitted to the system.
          \item \textbf{No. of Open Connections:} Determine the number of connections per system needed to accommodate a large number of users.
          
      \end{itemize}

      \\\midrule

    \textbf{Correctness} & 
    
    \begin{itemize}
        \item Functional Correctness
        \item Transaction Correctness
        \item Decision-making Correctness
    \end{itemize}

    & 
    
    \begin{itemize}
        \item \textbf{Model Driven Engineering (MDE):} A technique for addressing software development complexity by using models at different levels of abstraction \citep{lu2020integrated, lu2019blockchain, xu2021decision}.
        
        \item \textbf{Experiments and Simulations:} A collection of performance parameters used to determine the system's performance \citep{kalra2018zeus, zhang2021trustworthy, madine2021application}.
        
        \item \textbf{Formal Verification:} A formal specification language used to construct complex software systems by applying mathematical methods or techniques \citep{mavridou2018designing}.
        
        \item \textbf{Mathematical Modelling:} The use of mathematics to predict and make decisions in the real world \citep{cheng2020design}.
        
        \item \textbf{Logical Evidences:} The logical proof is used to validate or invalidate an idea using certain logics, in which deductive reasoning can be used to reach a conclusion to provide logical evidence \citep{governatori2018legal}.
    \end{itemize}
     \\ \midrule
            
     \textbf{Efficiency} &  
     
     \begin{itemize}
         \item System Efficiency
         \item Network Efficiency
         \item Storage Efficiency
         \item Energy Efficiency
     \end{itemize}

     & 
     
     \begin{itemize}
         \item \textbf{Throughput:} The amount of time it takes to append correct records to blocks. In general, it is measured as the total number of committed and saved records (after being validated) divided by the total time to validate and save records.
         \item \textbf{Latency:}  The time interval between when a transaction is submitted and when it has been written to the ledger \citep{xu2021latency}. 
         \item \textbf{Bootstrap Time:} The time it takes to load all of the information and data needed to create a block \citep{stoykov2017vibes}.
         \item \textbf{Bandwidth Overhead:} The propagation of all information associated with each block in the Blockchain network \citep{jin2019reducing}.
         \item \textbf{Transaction Size:} The size of a single transaction (in bytes) stored in each block, which results in the development of the Blockchain \citep{min2016permissioned}.
         \item \textbf{Block Size:} The total number of transactions saved on each block \citep{singh2021multi}.
         \item \textbf{Data Provision:} Ensure that the data is open to all users with sufficient permissions and in a protected manner \citep{paik2019analysis}.
         \item \textbf{Computational Complexity:} Current hash rate of a consensus mechanism, such as PoW, in a public Blockchain \citep{sedlmeir2020energy}.
         
     \end{itemize}

      \\

      \bottomrule
    \end{tabular}

  \end{sidewaystable*}

  \begin{sidewaystable*} 
    \centering

   \footnotesize

 \begin{tabular}{p{0.09\textwidth}|p{0.15\textwidth}|p{0.7\textwidth}}
       
   \toprule 
     \textbf{Requirements} & \multicolumn{1}{c|}{\textbf{Sub-Requirements}}& \multicolumn{1}{c}{\textbf{Measuring Criteria}} \\\midrule

      \textbf{Interoperability} & \begin{itemize}
          \item Data Interoperability
          \item Platform Interoperability
          \item Infrastructure Interoperability
      \end{itemize}

      &

      \begin{itemize}
          \item  \textbf{The usage of APIs:} The interaction of two Blockchain networks through a specially built application programming interface such as smart contracts \citep{belchior2020survey}.    
          \item \textbf{Multi-Chain:} Using crypto-dependant tools such as side chains to insure interoperability \citep{chang2021synergychain}. 
          \item \textbf{Separate Blockchain:} An intermediary Blockchain sitting between the two existing Blockchains \citep{hardjono2019toward}.
          \item \textbf{Off-chain:} Using middleware such as state channels, or atomically swap boxes \citep{gordon2018blockchain}. 
          
      \end{itemize}

  \\\midrule

      \textbf{Consistency} &
      
      \begin{itemize}
          \item Data Consistency
          \item Agreement Consistency
          \item Network Consistency
      \end{itemize}

      & 
      
      \begin{itemize}
          \item \textbf{Consensus Mechanism:} The use of a consensus mechanism by different parties to reach an agreement on the same data stored in a Blockchain \citep{da2021embedding,perera2020blockchain}.  
          
          \item \textbf{Synchronisation Algorithms:} The synchronisation algorithms can be used to determine the consistency of different components, as well as the data transmitted over the network \citep{bonneau2015sok}.
          
          \item \textbf{Block Structure Elements:} Merkle Hash Tree, timestamp, Byzantine \citep{kolb2020core}.   
          
          \item \textbf{Smart Contracts:} Smart contracts also maintain agreement consistency across Blockchain communication \citep{nguyen2019blockchain, kosba2016hawk} .
      \end{itemize}

 \\\midrule

       \textbf{Usability} & 
       
       \begin{itemize}
           \item Application-level Usability
           \item Service-level Usability
       \end{itemize} &

      \begin{itemize}
          \item \textbf{Usability Test Approaches:} \citep{sundarakani2021big, yli2016current}
          
          \item \textbf{Combination of Different Metrics:}  (Effectiveness, Efficiency, Fault tolerance, etc.) \citep{li2021industrial}
      \end{itemize}

      \\\midrule

      \textbf{Flexibility} &

      \begin{itemize}
          \item Process Flexibility
          \item Product Flexibility
          \item Resource Flexibility
          \item Network Flexibility
      \end{itemize}

      & \begin{itemize}
          \item \textbf{Asynchronous Operations:} Asynchronous operations are used to determine the flexiblity of system modules \citep{leng2019manuchain}.
          \item \textbf{Decision-Modelling Approach:} Define the hypothesis and then evaluate it using qualitative methods \citep{werner2020blockchain}.
          \item \textbf{Other Proposed Frameworks:}  \citep{das1996measurement, d2018decentralized, fragapane2020increasing}
          
      \end{itemize}  \\\midrule

     \textbf{Protection} &
     
     \begin{itemize}
          \item Data Protection
          \item User's Information Protection
      \end{itemize}

     & \begin{itemize}
          \item \textbf{Privacy-Preserving Approaches:} Various privacy-preserving approaches are used to secure data and user information transmitted over a network \citep{peng2020privacy}.

      \end{itemize} \\\midrule
      
      \textbf{Modularity} &

      \begin{itemize}
          \item Process Modularity
          \item Component Modularity
      \end{itemize}

      & \begin{itemize}
          \item \textbf{Business Process and Modelling approaches:} \citep{mohamed2019applying}
          
          \item \textbf{Quantitative Modelling Approaches:} Using quantitative modelling methods to assess modularity \citep{fahimnia2015quantitative}
      \end{itemize} \\

      \bottomrule
    \end{tabular}

  \end{sidewaystable*}

  \begin{sidewaystable*} 
    \centering

   \footnotesize

 \begin{tabular}{p{0.08\textwidth}|p{0.15\textwidth}|p{0.6\textwidth}}
       
    \toprule 
     \textbf{Requirements} & \multicolumn{1}{c|}{\textbf{Sub-Requirements}}& \multicolumn{1}{c}{\textbf{Measuring Criteria}} \\\midrule

      \textbf{Fairness} &
      \begin{itemize}
          \item Resource Fairness
          \item Transaction Fairness
          \item Service-level Fairness
      \end{itemize}
        & 
        
        \begin{itemize}
            \item \textbf{Incentive Mechanism:} Provide equal incentives to users who participated in the creation of the block \citep{goel2018resource, danzi2017distributed}.
            \item \textbf{Time Commitment:} Before the timer runs out, the committer party must reveal a secret \citep{li2020blockchain}.
            \item \textbf{Price Calculation:} Maintain the rates paid for services \citep{pathak2021aerialblocks}.
            
            \item \textbf{Usage Intention:} Ensure that the specifics of the usage intention are accurate \citep{sun2021using}.

        \end{itemize}

             \\\midrule

      \textbf{Completeness} & 
      
      \begin{itemize}
          \item Record or Information Completeness
          \item Requirements Completeness
          \item Functions Completeness
      \end{itemize}

       & 
       
       \begin{itemize}
           \item  \textbf{Information Centred Approach:} The relevant information is available to the relevant authorities at the right moment \citep{engelhardt2017hitching, leng2020blockchain}.

           \item \textbf{Software Validation Strategies:} Software validation techniques can be used to insure that functional and non-functional requirements are met \citep{lu2018blockchain, porru2017blockchain}.
       \end{itemize}

        \\\midrule

      \textbf{Transparency} & \begin{itemize}
          \item Data Transparency
          \item Access Transparency
          \item Location Transparency
      \end{itemize}
      
      &
      \begin{itemize}
          \item \textbf{Smart Contracts:} The use of smart contracts is one method for achieving transparency in industrial Blockchain-based systems  \citep{baralla2021ensuring}.
          
          \item \textbf{Encryption Mechanisms:} One of the tools for achieving transparency in Blockchain-based systems is encryption \citep{mitani2020traceability}.
          \item \textbf{Access Control Policies}: Access control mechanisms are also used to achieve transparency in Blockchain-based systems \citep{bertino2019data}.
      \end{itemize}

        \\

      \bottomrule
    \end{tabular}

  \end{sidewaystable*}



\begin{figure*}[ht]
\centering
  \includegraphics[width=18cm,height=10cm]{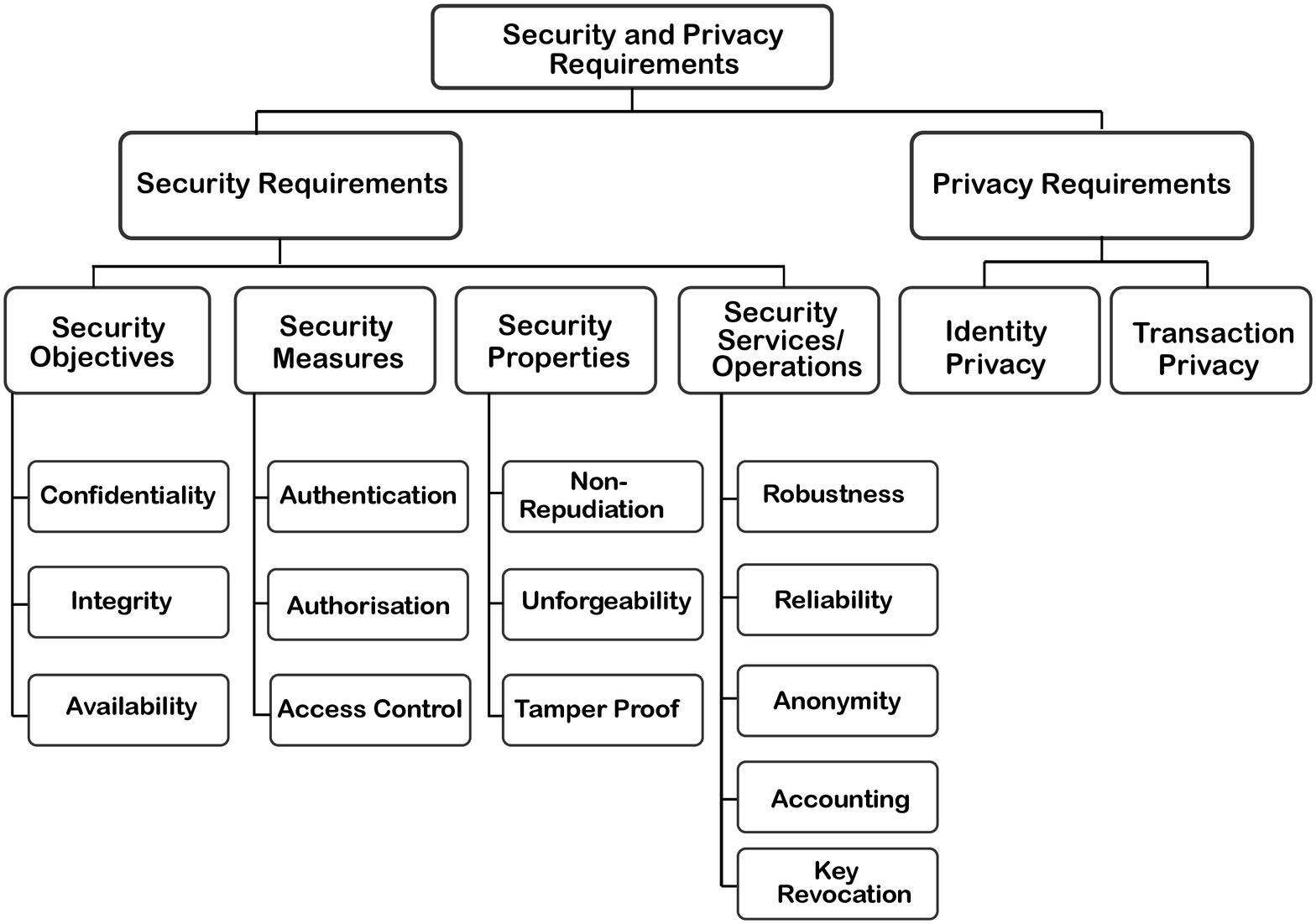}
  \caption{\colorbox{white}{A Taxonomy of Security and Privacy Requirements for Blockchain-based Industry 4.0 Applications}}
  \label{fig:SubTaxonomy2}
\end{figure*}

This subsection explains various security requirements for Blockchain-based applications in detail. These requirements are subdivided into different security categories, such as objectives, measures, properties, services and operations, as shown in Fig. \ref{fig:SubTaxonomy2}. In privacy requirements, we discussed two key types of privacy: identity privacy and transaction privacy.


\subsection{\textcolor{black}{Security and Privacy Requirements}} \label{section_security_requirements}

This subsection explains various security requirements for Blockchain-based applications in detail. These requirements are subdivided into different security categories, such as objectives, measures, properties, services and operations, in order to ensure secure communication between other network nodes.

\subsubsection{Security Objectives}
Confidentiality, integrity and availability are the primary security objectives of any system to provide a secure environment for communication and ensure the safe utilisation of the system and network resources efficiently. Here, we discuss each security objective set to achieve security measures in Blockchain-based applications.

\begin{itemize}
    \item \textbf{Confidentiality:} This is the critical security objective of any system defined to protect data from unauthorised access. Confidentiality allows individuals or groups of people to communicate with each other securely and prevents any unauthorised person from accessing their data \citep{Chang2017BlockchainDD}. Since the data contains users’ personal and classified information, it is necessary to protect personal information from others. In the Blockchain, a public network is open for everyone and it allows the users to send their transactions publicly to others in the network. Therefore, the key challenge of the public Blockchain network is to keep data secure. No unauthorised person is allowed to read and access data from the network \citep{English2016DisintermediationOI}.

    \item \textbf{Integrity:} Integrity is another essential security objective that ensures the validity and reliability of data throughout its life cycle \citep{Hameed2016AZW}. An adversary can change or modify the contents of data stored in some databases or communicate through a network channel. Blockchain technology combines many elastic features to provide security for Blockchain data; the immutable distributed ledger provides sophisticated services resistant to data modifications. Once transactions are recorded on an immutable ledger, it is impossible to change or delete them later \citep{Gaetani2017BlockchainBasedDT}.

    \item \textbf{Availability:} This refers to the state of any resource or system’ available to perform certain functions within a given timeframe \citep{7973733}. The availability of Blockchain can be defined as the interaction of validated Blockchain transactions. Interaction mainly depends on the availability of two Blockchain functions, such as reading and writing. The read-availability response is always higher than the written-availability response in the Blockchain system. Therefore, variation in the measured time can lead to a Blockchain system failure in terms of availability \citep{Weber2017OnAF}.

\end{itemize}

\subsubsection{Security Measures}
Security measures define the criteria for limiting unauthorised users’ access to personal information from some storage locations, thereby protecting system resources from malicious attempts. These measures enable users to perform legitimate actions. Defining security measures for any system, authentication, authorisation and access control are the primary methods to define essential access criteria that help protect the system from malicious activities. 

\begin{itemize}
    \item \textbf{Authentication:} Authentication is defined as an essential security measure for any security system that integrates specific human details, such as name, age and password. To access a complete system or some resources, it is important to describe here that authentication is an essential security measure for all Blockchain systems to grant permission to authorised users \citep{googlepatent}. Digital Identities (IDs) are used in Blockchain systems with several other features, such as public and private keys, to access system resources \citep{Mann2016TwofactorAF}.

    \item \textbf{Authorisation:} This security measure is intimately connected to the authentication process to gain access to system resources. In most cases, authentication is the first critical step to determine whether a particular user is permitted to access system resources \citep{Okada2017ProposedCO}. For authorisation in Blockchain systems, each system defines its own rules and policies that apply to users who complete the authentication procedure \citep{Cachin2016ArchitectureOT}.

    \item \textbf{Access Control:} Like authorisation, access control is another critical security measure that defines an authorised user’s conditions for accessing system resources. Access control policies usually provide users with information and control of the resources of critical systems \citep{Maesa2017BlockchainBA}. For example, Blockchain network types, including public, private and consortium Blockchains, are based on different access control policies used to restrict access by a person or a group of users \citep{Pilkington2015BlockchainTP}.

\end{itemize}

\subsubsection{Security Properties}

Security properties are defined as a set of features that enforce security on the computer system. They are the challenging design part of most real-world applications. For example, from the Blockchain perspective, being tamper-proof is the desirable feature that encourages the users to save a record on distributed databases in an immutable way. Similarly, non-repudiation and unforgeability are also fundamental security properties for any Blockchain system. Details on security properties, along with their related characteristics, are given below.
\begin{itemize}
    \item \textbf{Non-Repudiation:} In modern conversation systems, non-repudiation is an agreement to secure transforming assets between two parties \citep{cucurull2016distributed}. In non-repudiation, both the sender and receiver cannot deny when they are committed to send and receive the messages from each other \citep{smith2016ibm}. For example, non-repudiation assures others about the creator and origin of data when each user is responsible for creating their data. Besides this, non-repudiation also confirms the data integrity of the users in the network \citep{mendes2018anonymized}.

    \item \textbf{Unforgeability:} Unforgeability is a security property that validates the authenticity of both the user and the data, and confirms that it is sent by legitimate users in the network \citep{muzammal2019renovating}. In the Blockchain, the user generates and broadcasts the transactions to all other users in the network for verification purposes \citep{tasca2017taxonomy}. The transactions are verified and added to the Blockchain using some mechanisms called consensus algorithms.

    \item \textbf{Tamper-Proof:} Blockchain utilises the distributed ledger concept for maintaining and storing the transactions data in a permanent and tamper-proof way. In general, the transactions stored in containers consist of a header unit and a data unit called blocks. Each subsequent block is connected to previous blocks using hash functions in order to form a complete chain of blocks. Once the data is added into the Blockchain, it is computationally impossible for the nodes to change the data unless the majority of the network nodes agree to do so. This requires much computational power and many energy sources. Therefore, the tamper-proof security requirement is desirable for the Blockchain. Most of the applications based on this technology prevent fake access from unauthorised users which would harm the stored data.

\end{itemize}
\subsubsection{Security Services / Operations}
Security services are defined as basic operations and protocols, and they require any system or application to provide an adequate, secure environment for data communication over the network. These services can be derived from the available third parties or the system can inherently implement the required services for each operation or task. In this sub-section, we place robustness, reliability, accounting and key-revocation as the underlying security services for Blockchain-based applications.

\begin{itemize}
    \item \textbf{Robustness:} The robustness feature of Blockchain technology provides guarantees to individuals or groups of people about their data stored on distributed ledger in which each node maintains and updates the distributed ledger \citep{li2014robust}. Blockchain uses the decentralised and distributed network feature to store identical information across the whole network in a P2P fashion \citep{kalra2018blockchain}.

    \item \textbf{Reliability:} Reliability is the primary feature of Blockchain technology and it is supposed to be a key element of each Blockchain application. In general, Blockchain’s reliability depends upon many constructive factors such as decentralisation, distributed ledger, immutability and security \citep{fujimura2015bright}. Collectively, these factors play an important role in achieving the reliability of Blockchain and they store data in a better way \citep{zheng2017overview}.

     \item \textbf{Anonymity:} This refers to hiding the personal information or details of users from the outside world. In other words, anonymity means remaining part of the system without showing any real identity to others \citep{Mser2013AnonymityOB}. In a public Blockchain, all transactions are stored at the immutable ledger and are publicly available to all other network users. Taking Bitcoin as an example, the pseudo-anonymous technique is used to link an actual person’s transactions with their given addresses. However, this is not a completely anonymous method for hiding the details of transactions \citep{Martins2011IntroductionTB, maxwell2013coinjoin}.

    \item \textbf{Accounting:} This service refers to keeping the record of all system resources used by authorised users on the network. In the Blockchain system, accounting refers to maintaining the following system activities, such as node behaviour, session maintenance, network resources and wallet information \citep{dai2017toward, herlihy2016blockchains}.

    \item \textbf{Key-Revocation:} Key Revocation is a cryptography mechanism that requires users to update secret keys periodically to make it impossible for adversaries to know and obtain keys from valid users. In a Blockchain, the key revocation approach requires key certification authorities to update each certificate before sending it to valid users \citep{axon2015privacy}.

Based on the above-mentioned security and privacy challenges in Blockchain-based applications, we provide a detailed description and analysis for each of them and conclude that confidentiality, integrity, availability, authentication, authorisation, access control, non-repudiation and privacy are the main security challenges mostly encountered by Blockchain-based applications. However, there are some additional security challenges such as unforgeability, robustness, reliability and key-revocation that require researchers to address them in Blockchain-based security and privacy solutions.


\end{itemize}

\subsubsection{Privacy Requirements}
In this subsection, we study two broader categories of the privacy requirements of Blockchain-based applications: identity privacy and transactional privacy.


\begin{itemize}
    \item \textbf{Identity Privacy:} In the network world, each object is identified by means of unique information referred to as the identity of a specific object. The object can be a user or a device in the network. In Blockchain systems, identity leakage threat is the main concern because an attacker can use different methods such as behaviour analysis and pattern matching to obtain the user’s personal information from the transactions \citep{zyskind2015decentralizing}. Personal information refers to the unique credentials related to the users such as ID, password, address, private key and balance, in some of the cases using the Bitcoin and Ethereum applications. Therefore, this issue can reveal the identity of Blockchain users and impose a serious threat to identity privacy \citep{kravitz2017securing}.

\item \textbf{Transaction Privacy:} In a Blockchain, the distributed immutable ledger consists of several verified transactions updated and maintained by the network users. Each user is responsible for sending the updated transaction to all other network users and updating the distributed ledger. However, during the transmission or interchange of transactions among Blockchain users, the transaction’s information can easily be obtained by an adversary in the network \citep{kravitz2017securing}. The adversary follows different approaches to steal personal information from a transaction related to a specific user. The transaction graph pattern method is one of the common methods that link and retrieve users’ personal information from a transaction \citep{ron2013quantitative}. Considering the given scenario, the privacy of transactions is the first and foremost requirement before sending an update to the Blockchain’s distributed ledger.

\end{itemize}

\section{Discussion on Security and Privacy Requirements for Blockchain-based Industry 4.0 Applications} \label{sec:5}



\textcolor{black}{The interest in Blockchain technology and its implementation in Industry 4.0 has evolved to capture the new opportunities. Many Blockchain-based applications have been developed and deployed across industries such as energy, finance and banking, healthcare and supply chain and logistics. Also, the adaptation of Blockchain technology in IoT, big data and cloud, crowdsensing and eCommerce technologies has been explored. With the use of Blockchain technology across industries, considerations about security and privacy requirements are crucial.} 

This section presents the security and privacy requirements of various Blockchain-based Industry 4.0 applications. For each application, we provide an overview, present integration challenges with Blockchain technology and discuss further security and privacy requirements to meet the needs of a secure environment. We also discuss how these requirements could be met effectively using security enhancement techniques.

\subsection{\colorbox{white}{Financial Industry}}

\begin{figure*}[t]
    \centering
\includegraphics[width=16cm,height=10cm]{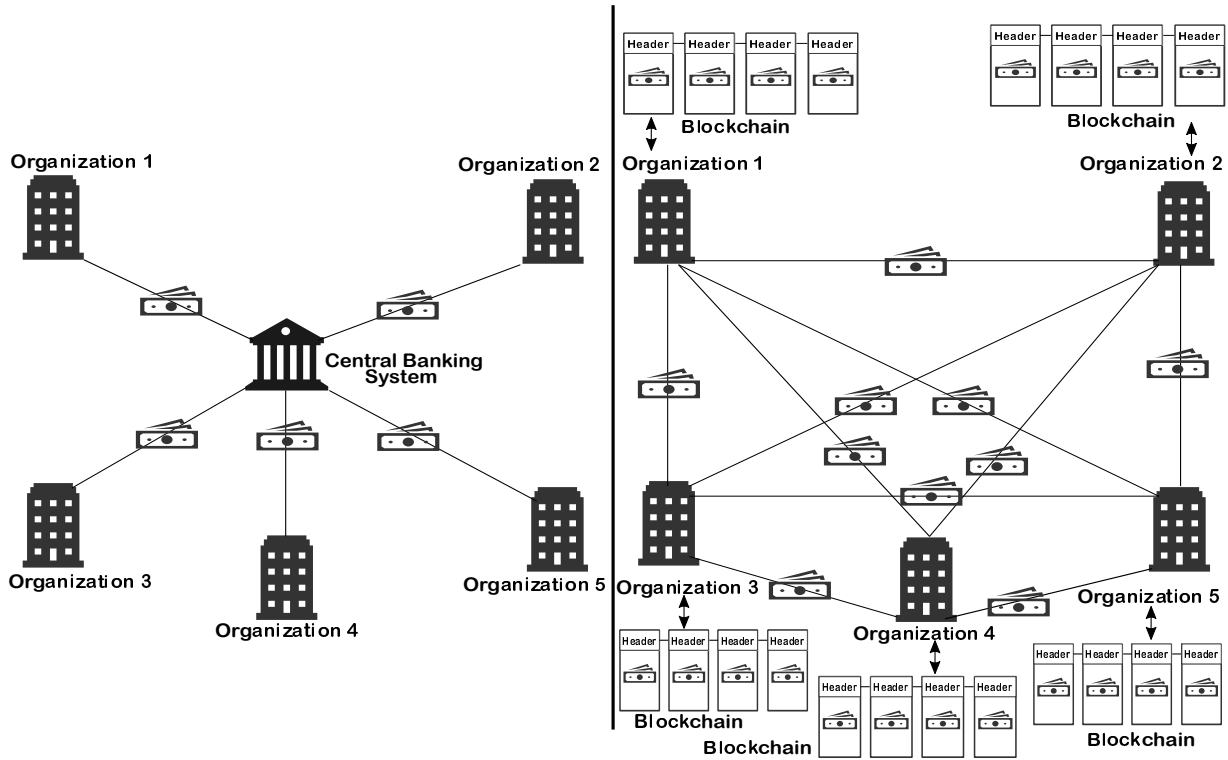}
  \caption{\colorbox{white}{Difference between Current and Blockchain-based Financial Systems}}
   \label{fig:Financial}
\end{figure*}


The industrial age influence is particularly recognised in financial services, including online payments, digital loans, currency trading and so on. The financial sector has greatly benefited from industrial automation 4.0 \citep{bilan2019influence}. In the financial sector under Industry 4.0, different banks, insurance companies, brokerage firms, investment firms and other financial institutions benefit from the broad positive growth of digital innovation \citep{feshina2019industry}. The changes in Industry 4.0 have affected the financial services sector in several respects. For example, it offers the quickest way to carry out global financial services with less human effort and lower costs \citep{mhlanga2020industry}. Additionally, the emergence of Blockchain-based financial services has an enormous effect on financial institutions. For example, with Blockchain technology and smart contracts, intermediaries, centralised administration, fraud and burglary can also be minimised. Blockchain technology also makes it easier for the transparency mechanism to reduce the liability of financial services \citep{chang2020blockchain, knezevic2018impact}.

One of the key principles of Blockchain technology is to build a trustworthy and transparent relationship across multiple platforms from which users can obtain and share their data anonymously, storing transactions on a tamper-proof distributed ledger. Furthermore, the distributed ledger also plays an important role in many cross-organisations in which people can trust each other and contribute to making the financing process easier and more efficient. Banking organisations are now using Blockchain as a key objective of transferring their digital assets to other banks located nationally or internationally through distributed ledgers. \textcolor{black}{In the context of financial industry, the role of decentralised Blockchain network is to eliminate financial intermediaries by allowing each financial party to contribute to the network with the implementation of a P2P network in order to facilitate direct money transfers. Additionally, each financial group maintains a distributed ledger in which each block comprises a finite volume of financial transactions, such as in banking. Fig. \ref{fig:Financial} shows the difference between traditional financial systems based on centralised architecture and current Blockchain-based financial systems. This figure clearly shows that the intermediary bank was solely responsible for managing all transactions between various organisations in conventional banking systems. However, with the introduction of Blockchain to the financial sector, the role of the middleman is removed, and the majority of nodes in the network must verify the transaction's validity.}

With the integration of public Blockchain and distributed ledgers, the stored transactions may reveal \textit{``sensitive and trading information''} to others. Another critical challenge the distributed ledgers face is to audit or verify the stored transactions since there is no central party within the whole process that can verify the transactions later. To address the integration problems of distributed ledgers with Blockchain technology, Wang and Kogan \citep{wang2018designing} designed a Blockchain-based privacy preserving scheme to \textit{``protect the financial information''} of banking users. The proposed system utilises zero-knowledge proof and homomorphic encryption to secure financial transactions in a private Blockchain network.

Most of the Blockchain systems specially designed for financial transactions face the \textit{``issues of transaction cost, propagation delay and high latency''} in the network. To address these critical issues in the Blockchain network, Zhong et al. \citep{zhong2019secure} suggested a secure and lightweight payment scheme to make efficient use of Blockchain-based financial system services. The proposed scheme takes advantage of both digital signatures and a one-way hash function to provide a guarantee to achieve security properties and the robustness of the system. In the proposed system, off-chain storage is utilised to store and access effectively the data from different remote locations.

A further improvement in the proposal for Blockchain technology is the concept of smart contracts that allow users to share information and perform automatic tasks over a decentralised network without any trusted third party. However, users in a decentralised network face a vital challenge as \textit{``distributed privacy''} requires significant effort from the researchers to overcome this challenge. With this intention, Kosba et al. \citep{kosba2016hawk} presented the Blockchain-based cryptography model to preserve the privacy of the user’s financial transactions. The proposed model has become very popular and was given the name “Hawk”. In Hawk, the zero-knowledge proofs used to store the transactions on the Ethereum Blockchain in unclear format achieve the security and privacy of users’ transactions. The Hawk project’s distinguishing feature is its exclusion of cryptography operations from the actual program, as the cryptography compiler generates the operations for the specific program on run time to protect the data inside the transaction. Another work \citep{kopp2017design} is proposed by Kopp et al. to overcome the problem of \textit{``transactional privacy''} in the payment system. The system combines the features of the distributed system, that is, cloud, with privacy enhancement mechanisms such as ring signatures and one-time addressing to protect users’ privacy and their financial transactions over a public Blockchain network.

In Blockchain-based financial systems, two types of clients, (i) a full client and (ii) a light client, are involved in the payment process to efficiently perform the financial transactions. The full client’s responsibility is to authenticate those light clients who have a valid address with which to do so. Since the full clients send the payment to the light clients, the attacker can easily track and obtain the full clients’ details by performing attacks on the light clients. To illustrate this concept, Kanemura et al. \citep{kanemura2017design} put forward the idea of \textit{``deniability''} to preserve the privacy of the light client in the Blockchain payment system. In addition to the privacy metric, the bloom filters also prove deniability using identical patterns of addresses matching metric parameters. Another major challenge found in the Blockchain-based finance system (primarily based on the principle of Know-your-Customer) is the \textit{``validation''} of the customers’ transactions with which respective bank authorities can disclose the personal and confidential information of the customers involved in the verification process. To meet this challenge, Bhaskaran et al. \citep{bhaskaran2018double} proposed the public Blockchain-based data validation scheme which utilises smart contracts in a combination of double-blind sharing scheme to protect customer data. Another similar work was presented by Biryukov et al. \citep{biryukov2018privacy} to protect the \textit{``identity of customers''} in a decentralised public network. The proposed system utilised the Ethereum smart contracts to solve the challenges faced by a centralised system, such as identity leakage and disclosure of other personal information.

With regards to privacy issues, most of the research aims to solve transactional privacy in finance and payment related systems with high accuracy. However, a few  privacy-preserving schemes face an issue of trade-off; the more the use of expensive protocols, the less the speed of the transaction. An efficient Blockchain-based approach called FPPP (Fast and Privacy Preserving Blockchain) is modelled by Li et al. \citep{li2018fppb} to achieve the performance of a system with guaranteed privacy of transactions. In this approach, an Ethereum Blockchain is enforced with an off-chain storage system to record many transactions which can significantly enhance the computation time and speed of the transaction process. Moving further to another approach, Ziegeldorf et al. \citep{ziegeldorf2018secure} proposed the CoinParty scheme which efficiently utilised the distributed mixing method to combine the different identities for the protection of \textit{``financial privacy''}. The CoinParty method used both mix nets and threshold signatures to preserve users’ anonymity and scalability in a decentralised public environment. Another coin mixing scheme for Bitcoin and their related cryptocurrencies was proposed by Liu et al. \citep{liu2018unlinkable}, enabling the users to unlink their \textit{``identities''} from the coins without the need of any central party. Furthermore, the scheme utilises the different cryptography primitives such as ring signatures and elliptic curve digital signature schemes to preserve the privacy of the transactions.

\begin{figure*}[t]
    \centering
\includegraphics[width=14cm,height=10cm]{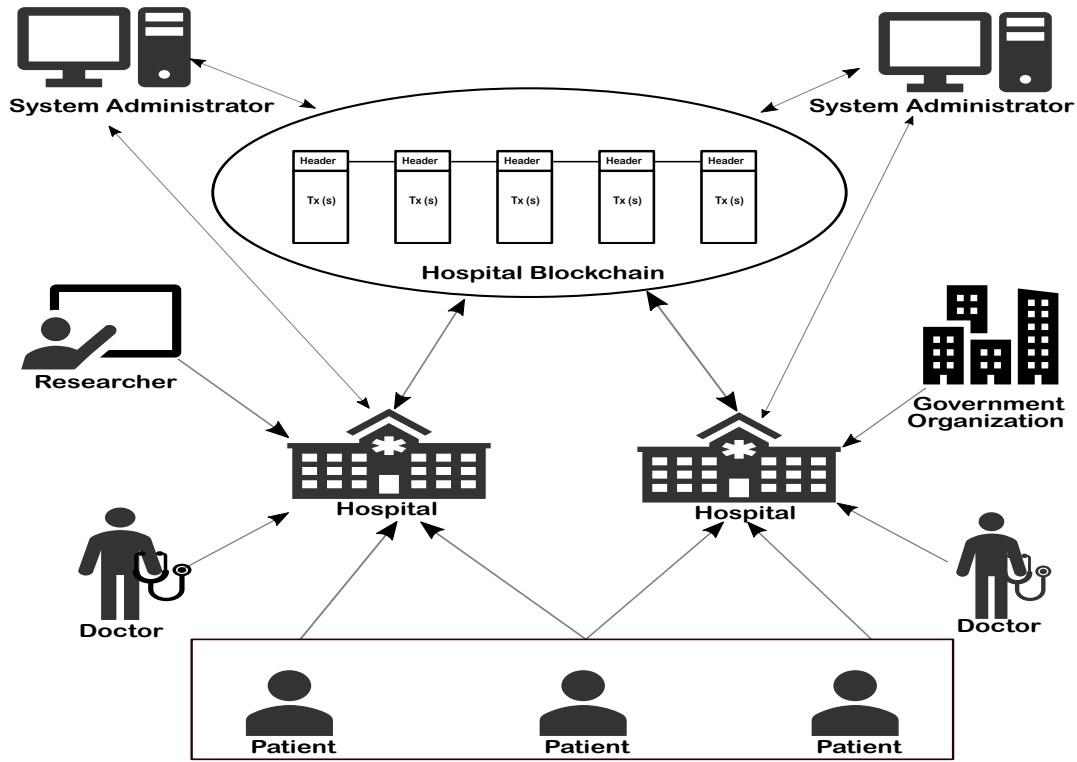}
  \caption{\colorbox{white}{Blockchain-based Healthcare Industry Model}}
   \label{fig:Medical}
\end{figure*}

\subsection{\colorbox{white}{Healthcare Industry}}



The existing industry healthcare models, particularly those using centralised architectures \citep{thota2018centralized}, have identified many vulnerabilities, including single-point failure and unauthorised modifications, as well as further security and privacy issues \citep{hathaliya2020exhaustive}. Since the conventional models are no longer reliable and stable, patient data is no longer maintained in this situation. Privacy of patient data is also key to effective health care management \citep{dwivedi2019decentralized}. These problems and challenges can be addressed by applying Blockchain technology sponsored by Industry 4.0 to achieve data security, including integrity and privacy, and eliminating a single point of failure issue \citep{hussien2021blockchain}.

Industry 4.0 is also transforming the healthcare sector, in a similar way to other industries, in order to embrace the adoption of innovative technologies. In the healthcare industry, patient data is considered the most valuable source of information to regularly monitor patient health and make critical decisions that assist doctors and researchers to improve the diagnostic learning process effectively, providing an efficient way to address health-related issues. The healthcare industry makes use of the IoT, along with cloud computing and big data services, to collect and store a massive amount of patient data that enables doctors, researchers and health workers around the world to build a digital global healthcare ecosystem called Healthcare 4.0 \citep{aceto2020industry, fayoumi2021integrated}.

Blockchain technology has proved to be a promising technology due to its characteristics of transparency, immutability and security, as well as its ability to connect multiple organisations through the decentralised and distributed aspects of the network. The rise of Blockchain technology in the healthcare industry empowers electronic health records (EHRs) and telecare medicine by keeping patient data secure and anonymous while opening the door to medical researchers to perform reliable analysis. Moreover, Blockchain has made healthcare transactions more transparent and accessible, enabling patients to know more about their treatment options and providers \citep{hussien2021blockchain, mcghin2019blockchain}. 

\textcolor{black}{The role of Blockchain in the healthcare industry is multifaceted. For example, Blockchain is reliable in terms of network structure since there is no centralised structure for a malicious user to target the patient data stored in a single location. Further, Blockchain technology enables patients to have full access to their medical records and history in a secure way. Accessing medical records in conventional systems can be difficult since they are usually spread across multiple healthcare facilities; however, Blockchain technology and primarily distributed ledger technology can be applied to safely access and exchange patient medical records \citep{fan2018medblock}. The distributed ledger technology enables the secure transfer of patient health records between doctors, researchers, and government agencies \citep{yue2016healthcare}. This ledger also helps in the effective and safe management of medication supplies and assists healthcare researchers in unlocking the human genome. Additionally, Blockchain technology can enhance the security and quality of mobile apps and remote monitoring machines used in the healthcare industry. Fig. \ref{fig:Medical} reflects the Blockchain-based EHR model to provide a better understanding of the Blockchain role in the healthcare industry, and the involvement of individuals in the system, such as patients, doctors, medical researchers, and government organisations who can interact and communicate securely with each other using Blockchain technology.}

By following the idea of implementing Blockchain technology for the healthcare industry, Zhang et al. \citep{zhang2018fhirchain} designed the FHIRChain, a Blockchain-based scheme to meet the specific needs and requirements of national health infrastructure organisations who \textit{``control and manage''} further health-related organisations and sectors. The proposed scheme carefully determines the specifications of health organisations in order to bridge the gap between patients and service providers, such as hospitals. In EHR systems, \textit{``access control''} is a crucial task that determines the access of personal data given to the right person for the right purposes. Therefore, to solve the access control challenge specifically for EHR, Hussein, et al. \citep{hussein2018medical} proposed a private Blockchain-based data-sharing scheme that allows doctors to access the sensitive data of patients who have granted access rights. This scheme uses the discrete wavelet transformation  technique to ensure the privacy and anonymisation of clinical data. The proposed scheme also employs the query service interface at which the genetic algorithm technique can access and optimise the Blockchain data. The experimental result determines that the scheme is scalable and robust against different types of security attacks. Similar to this scheme \citep{hussein2018medical}, Dagher et al. \citep{dagher2018ancile} implemented the Ancile, a Blockchain-based \textit{``access control''} scheme to utilise the private and sensitive information of patients without disclosing this information to others. In addition to access control, the scheme also confirms patients’ data privacy, which would not be disclosed to others while accessing the data via multiple platforms.

Another significant contribution is put forward by Yue et al. \citep{yue2016healthcare} to solve the problems between patients and doctors about the \textit{``controlling of sensitive information''}. The proposed scheme empowers the patients by giving them the appropriate control over their data and allows them to send and securely receive data. Another “secure data sharing” scheme (MeDShare) was proposed by Xia et al. \citep{xia2017medshare} to solve the disputing correspondence among different public parties using Blockchain technology. \textit{``Data reliability''} is another major challenge in electronic health data that requires substantial contributions from the researchers to propose secure solutions for the reporting of patients’ data at different levels. Kuo and Ohno-Machado proposed a solution to a particular challenge \citep{kuo2018modelchain} that implements the Blockchain-based privacy preserving public model called ``ModelChain'', in order to preserve the privacy of patient data by using machine learning techniques.

Sun et al. \citep{sun2018decentralizing} proposed the Blockchain-based privacy preserving scheme that utilised the patient’s attributes in attribute-based signatures so as to protect personal information. In the proposed model, both on-chain and off-chain storage systems are employed with the private Blockchain system so that on-chain is used to store the original data, whereas off-chain is used to store indexes of data. Another privacy preserving scheme called BSPP (Blockchain-based secure and privacy preserving) was proposed by Zhang and Lin \citep{zhang2018towards} to \textit{``protect the personal data''} from different health-related organisations involved in the whole process. In the proposed scheme, both private and consortium Blockchains construct the data structure, which can securely store the patient’s data.


Similar to the scheme presented in \citep{zhang2018towards}, Guo et al. \citep{guo2018secure} utilised the attribute-based signature scheme to provide the  \textit{``validation of health record''} stored on the public Blockchain. In the proposed scheme, many authorities can sign and send data without disclosing patients’ personal information unless it requires evidence. To solve the \textit{``authentication and accountability challenges''} of the patient health record, Azaria et al. \citep{azaria2016medrec} proposed the MedRec, a novel Blockchain-based methodology to secure the medical data of patients in a decentralised public environment. In the proposed architecture, patients can easily control and access their own data from remote places. Ji et al. \citep{ji2018bmpls} made a significant contribution to the Blockchain-based EHR application, investigating a \textit{``location sharing problem''} regarding patients in a telecare medical information system. To identify location issues of health data, a Blockchain-based location preserving scheme called BMPLS was proposed to achieve multi-level location sharing protection by employing a Merkle hash tree which can store the patient data in an hierarchical form. Apart from the different security challenges in the EHR, there have been many issues in distributed applications such as \textit{``retrieving, indexing, and aggregating''} of data collected from multiple domains. To solve the given challenges, Zhou et al. \citep{zhou2018distributed} proposed the distributed \textit{``data vending''} framework that utilises indexing and embedding methods to save similar data into different locations with the calculated indexes.

\subsection{\colorbox{white}{Transport and Logistics}}

\begin{figure*}[t]
    \centering
\includegraphics[width=14cm,height=10cm]{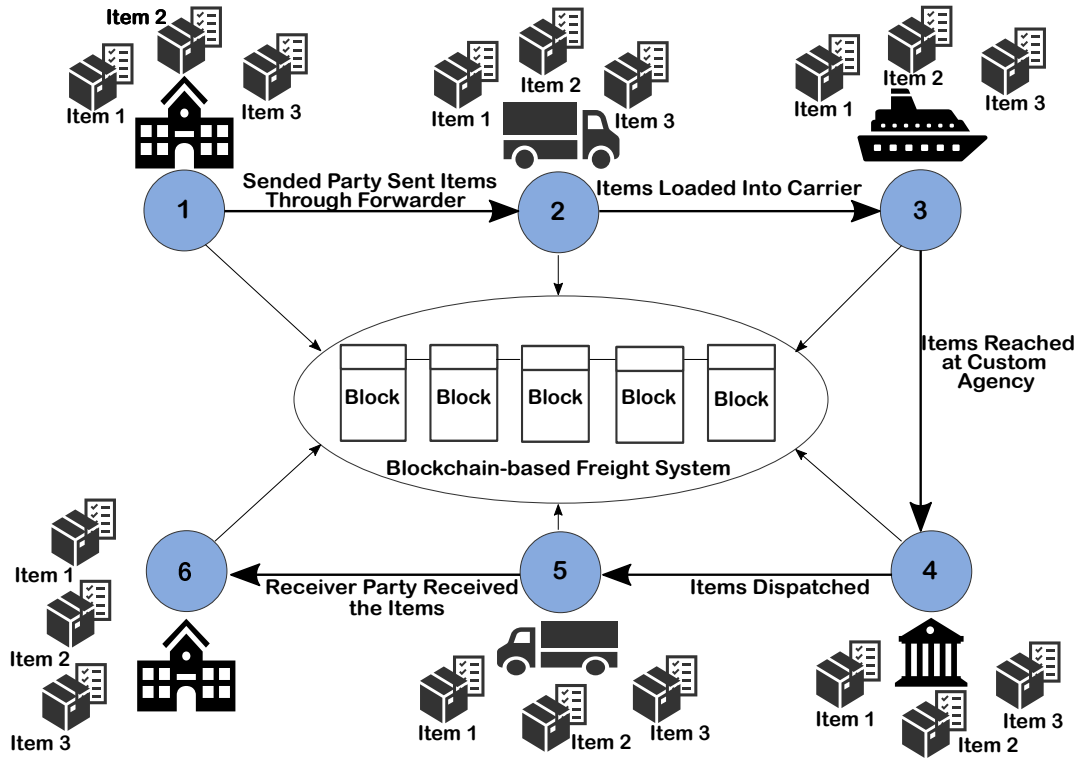}
  \caption{\colorbox{white}{Blockchain-based Transport and Logistics – A Complete Process}}
   \label{fig:Freight}
\end{figure*}

Industry 4.0 is driving substantial and pervasive change within the transport and logistics sector because of a rise in supply chain demand and usage of modern and emerging technologies, as well as the potential to build the industry process’s digital supply chain \citep{viriyasitavat2021augmenting,rossit2019data, molka2018large}. Transport and logistics sectors are following a trend towards increased automation levels so that these sectors integrate their business models to access different market segments. The increased availability of information through the physical internet and open standards encourages manufacturers to reposition their supply chains to support social and data-driven market dynamics and innovation in traceability and acquisitiveness \citep{barreto2017industry, vieira2020supply}.


In supply chains, formal concepts being applied to real-world shipping processes, with flexible, sustainable and online shipping varying from a large container to a small box, are becoming industry standards worldwide. These containers are constantly tracked and monitored and guided through the IoT \citep{golpira2021review}. The shipping industry is an important player in virtually every industry and operation that manages the shipment of vast containers from one place to another. These containers consist of various items sent to a specific destination, assuming that it does not include illegal or mislabelled items. However, the central agencies cannot check and perform audit procedures on each container’s items due to bulk quantities and time-restrictions \citep{tang2019strategic}.

Moreover, the audit procedure includes tracking and selecting random items from the container with detailed information, such as item id, company name and addresses of both sending and receiving parties. Mostly, the auditing agencies stored the data about all items at a single location or server, which is easily accessible for every auditing party to audit. On the other hand, centralised systems are more at risk when authorised access or a single point of failure attack can breach the privacy of shipping information. However, with the advent of Blockchain technology, decentralisation and immutability features allow the freight system to transfer freight items from one place to another in a secure manner. 

\textcolor{black}{The role of Blockchain technology in transportation and logistics is to ensure data integrity and security in the ecosystem, as the entire network contributes to data validation. Moreover, Blockchain-based logistics systems facilitate document sharing through a shared distributed ledger, obviating the need for manual paper-based processes. The use of smart contracts speeds up and automates the customs clearance and approval process, resulting in reduced processing times for items at customs checkpoints.
Fig. \ref{fig:Freight} illustrates the Blockchain-based transport and logistics framework, which shows the complete process of transferring the items from one destination to others, through the shipping agencies.}

To protect the \textit{``data privacy''} of shipping items, Vos et al. \citep{vos2018defend} proposed DEFEND, a secure decentralised Blockchain-based platform that protects the privacy of containers and the store items. In this system, the sending agency carries out the following tasks: make a claim of sending items, encrypt the claim and send it to other destination agencies. At the destination side, the claim can only be accessed and decrypted by the destination custom agency. The proposed scheme’s main contribution is the partitioning of the data among the different parties involved in the Blockchain system. Moreover, experimental results and performance analysis claim that the proposed scheme is efficient for both customs agencies and economic operators so as to perform risk analysis on items without causing the delay.

\subsection{\colorbox{white}{Energy Industry}}
\begin{figure*}[t]
    \centering
\includegraphics[width=14cm,height=10cm]{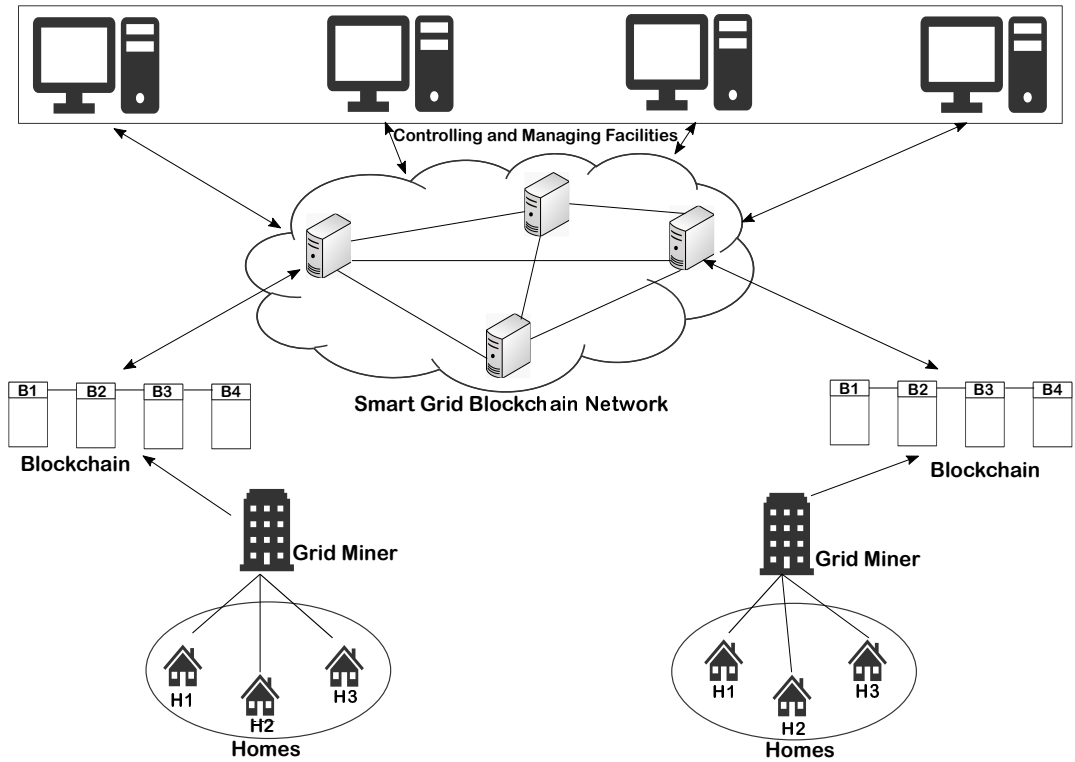}
  \caption{\colorbox{white}{Blockchain-based Smart Grid System Model}}
   \label{fig:Smart_Grid}
\end{figure*}

Industry 4.0 has provided the path for continuous and widespread grid-based use of alternative energy sources, by implementing a flexible framework called the SG in order to manage increasing demands in overall energy consumption. More precisely, in the area of Industry 4.0, the SG is developed by the convergence of electricity networks and state-of-the-art information and communication technologies (ICTs), including smart meters, smart information processing units and advanced communication protocols to address numerous barriers and vulnerabilities such as increased energy consumption, faulty transmission, increased construction costs and less efficient delivery in conventional metering systems. The purpose of designing the SG is to control and track the energy resources of users and suppliers using smart digital metres in a more effective, accurate, functional, scalable, safe and cost-effective manner \citep{faheem2018smart, khan2017design}.

In the SG scenario, one unit called the grid collects the real-time data (also referred to as meter reading in a conventional meter) from smart meters deployed at different locations such as homes and industries. Most of the data collected are in low-level operations and this allows data analysts to discover significant data outcomes and help coordinate subsequent usage, followed by even more complex analytics and planning \citep{dragicevic2020conceptual}. This technology has become a significant part of any country’s success, encouraging its power consumers to use smart meters to manage and efficiently control power consumption. However, one crucial challenge found in the SG is the \textit{``leakage of personal information''} through the streaming data. Personal information can include the details of household users, such as billing information, metering units and address, and these can cause severe security threats and loss of privacy for both consumers and providers \citep{al2019iot}.

Blockchain technology is an emerging technology that can be utilised to provide a solution that enables secure, transparent and efficient energy transactions. Indeed, it provides a decentralised management and P2P energy trading. \textcolor{black}{In SG architectures, the primary role of the Blockchain technology is to handle network transactions. Each entity in the SG, including producers, customers, distributors, and managing authorities, communicates with others operating as Blockchain nodes, and their interactions are recorded on the Blockchain as a transaction. There are usually two types of entities in the Blockchain-based SG architecture: light nodes and full nodes. Light nodes are typically customers that use electricity and pay their bills, while full nodes are those nodes that handle electricity and participate in the Blockchain mining process. Further, smart contracts are typically used to enforce transactions in Blockchain-based SG. These transactions include the payer's and payee's remaining balances, balance deductions, benefit or loss on the grid side, and so on \citep{agung2020blockchain}. Fig.12 indicates the Blockchain-based SG model in which each grid is acting as a miner to manage and control the group of homes using the Blockchain architecture.}

As we mentioned earlier, streaming data can disclose the household users’ private information, which causes further security and privacy problems in the SG environment. An attacker can also obtain power consumption history by tracking and analysing the different behaviour patterns of users, such as on and off timing for a job, switching off the house lights and recharging smart meters. To tackle these issues, Guan et al. \citep{guan2018privacy123} proposed the Blockchain-based \textit{``privacy''} preserving and data aggregation method for the SG in which the users are divided into separate subgroups and each subgroup head can record the data of their sub-users. To protect the users’ information, each head employs the pseudonym technique to hide the streaming data during data transmission to other neighbouring heads.

Similar to the previous scheme \citep{guan2018privacy123}, Aitzhan and Svetinovic \citep{aitzhan2016security} designed a private Blockchain-based decentralised system that provides secure \textit{``end-to-end communication''} between the smart meter and the SG without the need for any trusted third party. A multi-signature scheme and anonymous encrypted messages are used to transform the trading transactions anonymously between the different end-users to preserve transaction privacy. Rottondi and Verticale \citep{rottondi2017privacy} made another contribution towards the \textit{``privacy preservation''} proposal in a SG network when they implemented the Blockchain-based smart metering architecture in which public users can transform their data to the SG in a secure way. In the proposed system, a secure multi-party protocol which is utilised for encryption and authentication purposes, guarantees users the correctness and authenticity of their data without them being disclosed by the SG.

\subsection{Technology Industry}
The technology industry includes internet-of-things, big data and cloud computing, crowdsensing, and eCommerce as Blockchain-based Industry 4.0 applications.
\subsubsection{Internet of Things}
\begin{figure*}[t]
    \centering
\includegraphics[width=14cm,height=10cm]{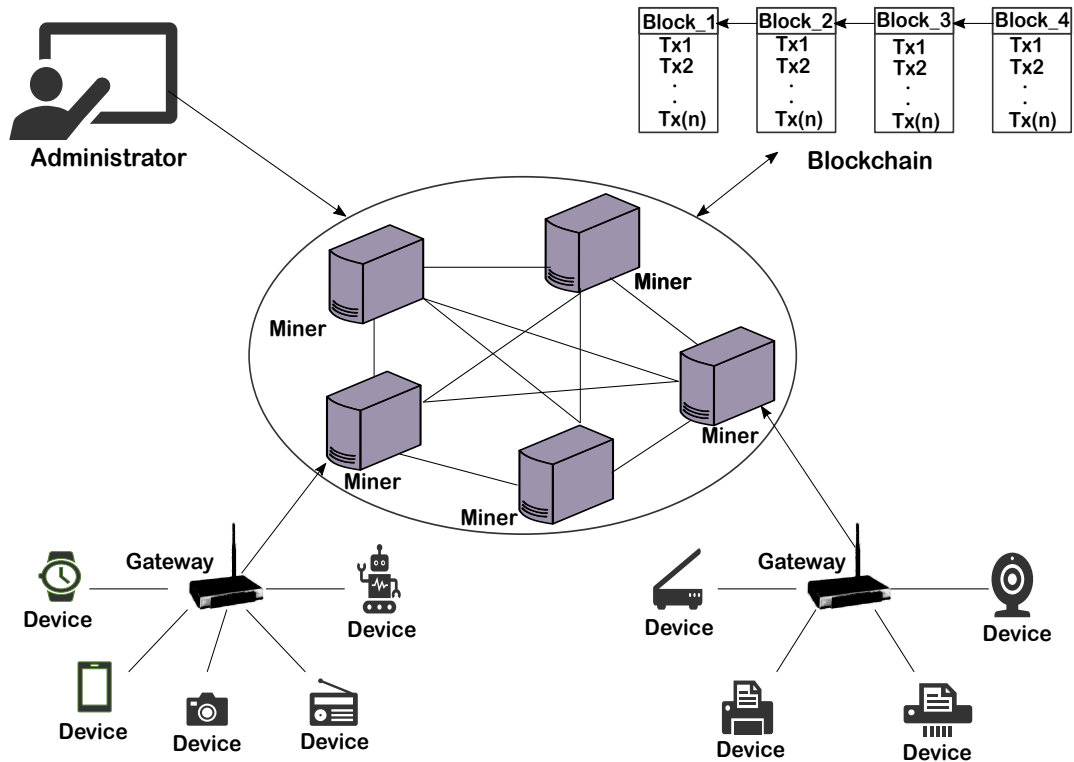}
  \caption{\colorbox{white}{Blockchain-based Internet of Things (IoT) Network}}
   \label{fig:IoT}
\end{figure*}

Industry 4.0 has used the IoT and its associated technologies and protocols to conduct digital manufacturing in which almost all embedded devices, such as robots, machines and tools, have sensors to gather data from the environment and function accordingly. IoT has changed the lifestyle of humans by implicating ubiquitous applications at every step, which help to perform daily tasks automatically and more efficiently. In conjunction with Fog computing, IoT plays a significant role in offering time-sensitive services such as disaster related services, smart transportation and smart health services \citep{ujjwal2019cloud, battula2019micro}. However, the improvement of the IoT’s current paradigms can only occur with continuous research and development in the underlying technology. Although this technology simplifies and facilitates human life, it also introduces numerous performance, security, and privacy concerns, as well as other risks to human life. Thanks to Blockchain technology that filled the missing gap between IoT services and security challenges due to its tremendous features such as decentralisation, immutable ledger, transparency and data auditability.

\textcolor{black}{The role of Blockchain technology in IoT has the ability to alleviate security and performance concerns. The use of Blockchain technology in IoT-based applications adds another security level, which is almost impossible for the attackers to get access to the network. The overall contribution of Blockchain to IoT systems, either full or partial, can be summarised as follows: the distributed ledger in a Blockchain system is immutable, eliminating the need for trust between the parties involved. No one organisation has complete control over the massive amounts of data produced by IoT devices. Blockchain technology offers a significantly higher degree of security, making it almost impossible to bypass existing data records. The transparency features of Blockchain allow anyone with authorisation to access the network to monitor previous transactions. By facilitating trust between stakeholders, Blockchain enables IoT companies to cut costs by removing the processing overhead associated with IoT gateways \citep{makhdoom2018blockchain}. Fig. \ref{fig:IoT} illustrates one example of integration of a decentralised IoT network with Blockchain technology. In this illustration, the IoT devices act as the light nodes that are resource-constrained in nature and whose security such as authentication and authorisation is desirable and the scalability challenge.  The miners' nodes are the full nodes and work as gateway nodes responsible for providing IoT devices' secure interaction to the Blockchain. Miners nodes are also responsible for mining the transactions for validation and adding purposes into the Blockchain. Administrator authority is responsible for deploying the entire system and managing the behaviour of nodes in the network.}

To solve the \textit{``user's privacy''} challenge in IoT, Cha et al. \citep{cha2018blockchain} proposed the Blockchain-based scheme for IoT users that provides a secure gateway to achieve the privacy of users in the IoT network. The secure gateway restricts unauthorised access to users’ personal data from the Blockchain in the proposed scheme. To achieve the \textit{''authentication of users''} in IoT, a digital signature scheme is utilised to manage users’ access to resources. Another Blockchain-based IoT scheme is proposed by Wan et al. \citep{wan2019blockchain} to achieve security and privacy among different industrial processes. In the proposed scheme, existing industrial models are studied and then the weaknesses analysed in order to overcome any potential challenges of industrial applications. The devices in IoT are subject to different types of constraints, such as computation and limited memory. Therefore, there is a need for third parties or cloud service providers to process and store a massive amount of data on them. Although cloud computing is considered the most potent resource for computation and storage of data, this technology has its security and privacy concerns that need to be overcome. To address the \textit{``access control challenges''} for IoT devices, Le and Mutka \citep{le2018capchain} proposed the CapChain, a novel access control scheme that allows the IoT devices to store and manage their data on a public cloud without disclosing any personal information. An anonymisation technique is utilised to protect sensitive information, such as identities, from adversaries in order to preserve user privacy. For the \textit{``protection of sensor data''}, Chanson et al. \citep{chanson2019privacy} proposed the certification-based Blockchain scheme to achieve data integrity in the IoT network. The certification authority allows the users to perform authentication steps in three different stages to prevent malicious activities in the system.

Nowadays, IoT-based smart home applications are becoming very popular and many home appliances are connected to the internet to control and manage the home environment remotely. The increasing demand for smart home devices raises different problems in terms of \textit{``security and privacy''}, efficiency and scalability. To cover such needs for designing a smart home system, Singh et al. \citep{singh2019sh} proposed the Blockchain-based smart home network to achieve \textit{``secure communication''} between IoT devices. In the proposed system, the multi-correlation technique analyses the network traffic that contains malicious data and information. The security analysis of the proposed system claims the high efficiency and throughput of the system. Another potential design of IoT-based smart home was proposed by Dorri et al. \citep{dorri2017blockchain2} to guarantee the \textit{``security and privacy''} of home users. In their proposed Blockchain design, three major components are used in a complete smart home setup: cloud, overlay network and home appliances. On-chain storage is used to store the local processing data, whereas off-chain storage, such as cloud, is utilised to store future data which can only be retrieved through separate transactions.

\subsubsection{Big Data and Cloud Computing}

Big data analytics refers to the use of advanced computing techniques implemented by many enterprises and industries to discover significant patterns in broad datasets, in order to help businesses detect trends and the impact of consumer preferences. Within Industry 4.0, data analytics plays a critical role in smart factories, where equipment captures relevant machine data to predict when maintenance and operations are required. Manufacturing companies use big data analytics in the same way that manufacturers use it, with the exception of emphasis. In manufacturing, numerous distributed sensors, which are deployed through cloud computing and IoT technologies, help the manufacturer uncover patterns which enable them to manage the supply chain more efficiently \citep{elmamy2020survey}.

Cloud computing is a stack of on-demand different resources and services provided to users to deliver different operations appropriately. The resources and services provided to Cloud users are controlled and managed by Cloud Services Providers (CSPs), which monitor and determine the applicability of on-demand access with the available resources. With the rapid pace of the development of cloud computing applications, numerous industrial organisations utilise cloud computing services for their data’s extensive computations and storage. 

The partnership between the Cloud and Industry 4.0 is a winning one, as both technologies required time to develop and gain acceptance within the broader industry communities. This incorporation enables businesses to fundamentally and profoundly reconsider their entire spectrum of digitisation processes and modify their current architectures to accommodate a broader industry market. Additionally, all of this occurred with increased versatility across the globe, from consumer response to cost control and proper management. With the convergence of big data and cloud computing services into Industry 4.0, millions of people today can use devices and applications daily that contain highly complicated data in an ever-changing technological environment. The growing trend of digital innovation, such as cloud computing, big data, and the IoT, creates new connectivity and knowledge-sharing opportunities. However, with such a large volume of sensitive data, it must be handled and secured effectively and continuously. The convergence of Blockchain technology and cloud computing brings the industrial community into a new era of data protection and service availability. The majority of cloud research problems can be resolved by leveraging Blockchain characteristics such as decentralisation, immutability, and transparency.

\textcolor{black}{The role of Blockchain technology into cloud computing and big data is many-fold. For instance, there is no single authority responsible for data management, eliminating the risk of a single point of failure. Depending on the level of data protection needed, Blockchain technology offers better deals than traditional providers in terms of security using an immutable ledger. In terms of storage, a cloud storage network could be managed by the nodes that assist with transaction facilitation, with the nodes granting the user access to storage on their devices. Additionally, advanced cryptographic technologies used in Blockchain can partition user data stored in the cloud into small bits, encrypt them for an extra layer of data security, and distribute them across the network.}

From the security point of view, CSPs  are solely responsible for giving proper access to cloud users using the different authentication and authorisation services available. However, the prevalent issue revealed in cloud computing services is a single point of failure, which mostly leads to disclosing the personal information of cloud users. For the alleviation of the \textit{"authentication issue"} occuring in cloud computing, Lu. et al. \citep{Lu2018AnPC} proposed the Blockchain-based decentralised authentication system for storing a complete access control list of users on the Blockchain. The proposed system provides authorisation and accounting features which utilise virtual coin features commonly available for the digital currencies system. For security enhancement, a simple one-way hash function is employed to secure the links between users’ transactions; this confirms transparency at a higher level. Another Blockchain-based scheme for trusted \textit{``data sharin''} with the cloud service provider was designed by Zheng et al. \citep{Zheng2018}, in which unauthorised users were restricted from performing the malicious modifications on data. In the proposed scheme, Paillier cryptography was utilised to achieve the confidentiality of data stored at distributed databases.

For the \textit{``protection of data''}  from unauthorised access, Fan et al. \citep{8320551} proposed the Blockchain-based privacy preserving scheme to protect users’ information being communicated over a content-centric 5G mobile network. The proposed scheme successfully established a secure connection between different service providers and users for transferring data. In addition to data protection, the access control mechanism is also used to ensure access to cloud resources. 
\begin{figure*}[t]
    \centering
\includegraphics[width=14cm,height=7cm]{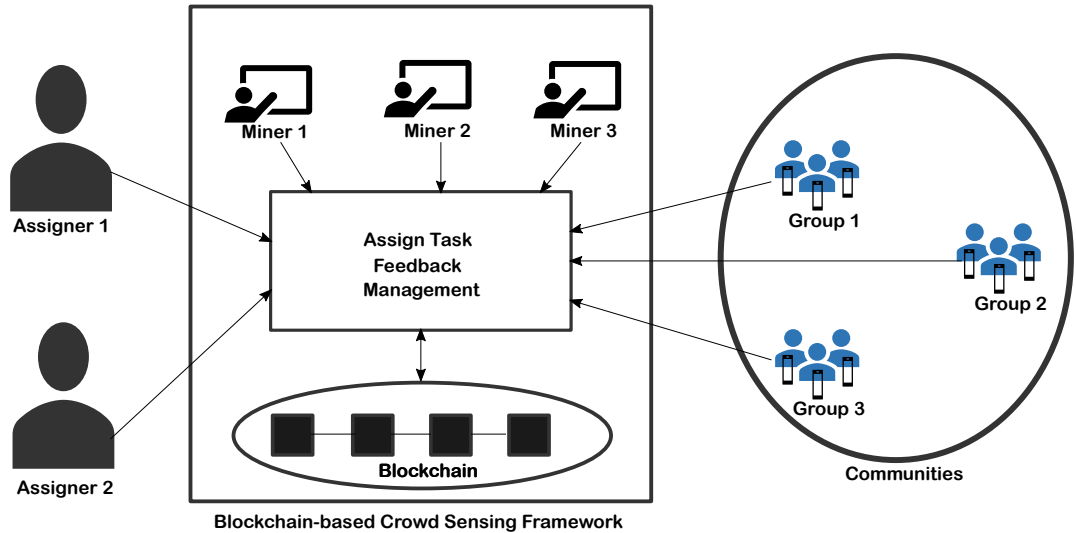}
  \caption{\colorbox{white}{Blockchain-based Crowdsensing Framework}}
   \label{fig:Crowd}
\end{figure*}

\subsubsection{Crowdsensing}

Giving freedom of knowledge to people in order to obtain information through some valuable resources has become a popular concept, used to discover innovations in information and communication technologies. One of the most popular ways of providing leverage to the crowd about real knowledge discovery is crowdsensing \citep{shu2017mobile}. In Industry 4.0, crowdsensing has been proven to solve diverse problems effectively, while lowering costs and extending the reach of ideas. The use of individuals’ data to implement processes and projects differs greatly in every sector, from production to delivery, depending on intent. Computational frameworks with high throughput are important in streamlining organisational processes and also assist in connecting decisions to crowd-sourced information and expertise through decision making projects. Crowdsensing aids in monitoring ecosystems and mapping and exchanging information and knowledge get by the peoples. For instance, in energy sector, crowdsensing can help minimising the building's energy consumption by monitoring users behaviour and thermal comfort. In industry, crowdsensing can minimise maintenance cycles and help fix machinery issues by tracking environmental conditions and failure. Crowdsensing also assists in monitoring ecosystems and mapping, and exchanging information and knowledge gained by the people \citep{vianna2020role}.

Human beings are equipped with sensing devices in the crowdsensing approach to sense the data from surrounding environments and to take useful actions over data in the industrial process. The quality of data captured by sensing devices depends on the number of people and their level of competency, such as primary, average and high skilled users. However, the limitation found in the crowdsensing process is that the data of users who participated in the data capturing process can be leaked during the sensing, which prevents further users from joining the crowdsensing network. Furthermore, the crowdsensing process faces several challenges, including fact discovery, knowledge quality, and estimation quality using sensing data \citep{capponi2019survey}.

To overcome these limitation, the Blockchain technology is introduced in crowdsensing that can support the joining of the maximum number of highly qualified users in the crowdsensing method, using a rewarding mechanism that attracts and motivates skilled users to participate in the data collection process, in order to receive high rewards as an incentive for their services for the crowdsensing process. \textcolor{black}{The aim of integrating the Blockchain network with current crowdsensing systems is to use the characteristics of Blockchain, such as decentralisation, to provide a way for emerging decentralised systems to solve the issues of a single point of failure in traditional systems and to provide equal opportunity to contribute in the fair data collection. Furthermore, the distributed ledger, as a core component of Blockchain in crowdsensing, ensures the immutability and traceability of users' data and their feedback for use in different processes. Thus, the overall role of Blockchain technology in crowdsensing can be summarised in terms of achieving the following objectives: increasing worker efficiency, implementing a fair compensation system, protecting confidential data, and lowering deployment costs \citep{chen2021blockchain}. Fig.\ref{fig:Crowd} describes the Blockchain-based crowdsensing framework, consisting of different nodes such as assigners, groups of users and miners to participate and control the overall crowdsensing process.}

For instance, Wang et al. \citep{wang2018blockchain} proposed the Blockchain-based \textit{``privacy-preserving''} incentive scheme for crowdsensing applications that allows highly skilled users to participate in the sensing process publically and securely, in order to gain high incentives. In the proposed mechanism, the k-anonymity scheme is utilised to anonymise skill users’ profiles to protect privacy. Another similar approach presented by Cai et al. \citep{cai2018leveraging} to \textit{``protect the personal information''} of the crowd uses the knowledge discovery method, without disclosing personal information. In this method, the public Blockchain platform collects knowledge from different sensing users in different distributed places. To provide the \textit{``guaranteed privacy''} of mobile users and crowdsensing providers, Chatzopoulos et al. \citep{chatzopoulos2018privacy} proposed the Blockchain-based crowdsensing scheme that specifies the smart contracts to ensure the secure relationship ? between them. The proposed scheme uses the secure multi-party computation algorithm in conjunction with smart contracts to protect users’ privacy and incentive payments.

\subsubsection{E-Commerce}

In modern times, electronic commerce (eCommerce) business has been widely accepted as a leading trading platform for the purchase and sale of online goods or services to promote their business via the Internet. With the widespread adoption of Blockchain technology in every field, traditional eCommerce platforms have been shifted to Blockchain technology to allow customers to carry out fair transactions without having a trusted party between them. However, one of the key issues identified in these systems is that it does not protect customer transactional privacy, such as content, addresses, account details and trading information. While several Blockchain-based privacy security mechanisms have been put in place to protect financial transactions, there is still a trade-off between privacy and the speed at which transactions are processed. To resolve this ongoing challenge, Li and Wang \citep{Li2018RZKPBAP} proposed RZKPB – a Blockchain-based \textit{``Privacy Preservation''} scheme that does not allow financial information to be stored in a plain-text format on the Blockchain. Multiple cryptography primitives, such as hashing and signatures, are used in the proposed methodology to verify transactions to establish a secure relationship between trading partners.

\section{Security and Privacy Techniques for Blockchain-based Industry 4.0 Applications}\label{sec:6}

\begin{figure*}[!ht]
    \centering
  \includegraphics[width=16cm,height=10cm]{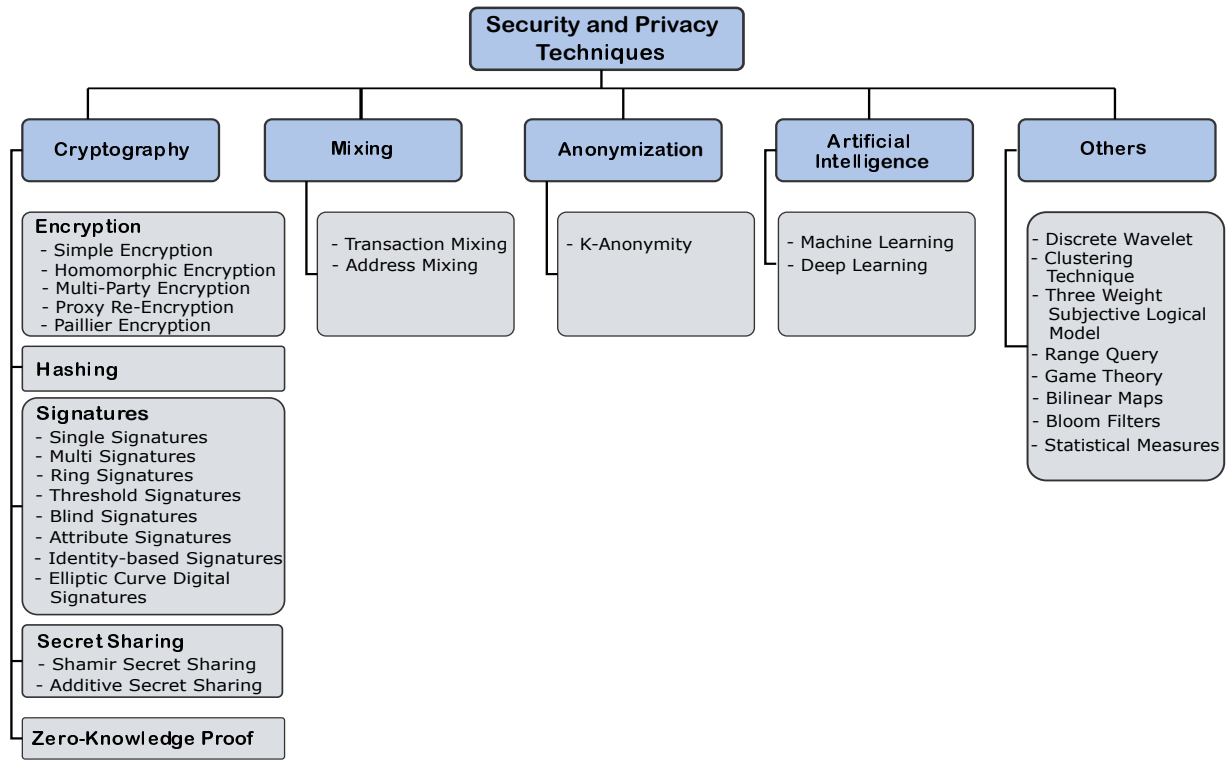}
  \caption{\colorbox{white}{A Taxonomy of Security and Privacy Techniques for Blockchain-based Industry 4.0 Applications}}
    \label{fig:solutiontaxonomy}
\end{figure*}

This section provides a detailed description of security and privacy preserving methods used in different Blockchain-based Industry 4.0 applications. We categorise them as cryptography, mixing, anonymisation, artificial intelligence and others, such as discrete wavelet, clustering and bloom filters. These categories are further divided into sub-categories, giving a better understanding of security and privacy methods and their usage in Blockchain-based Industry 4.0 applications. A taxonomy of security and privacy techniques is illustrated in Fig. \ref{fig:solutiontaxonomy}.

\subsection{Cryptography}

Cryptography is a security technique that consists of a mathematical set of rules and logic used to secure the communication between parties. In cryptography, a plaintext is converted into some hidden text using a logic known only to both the sender and the receiver. The objective of the cryptography technique is to send the information securely only to those who are authorised to see and read the information. Cryptography techniques are used in many ways, such as protecting information from theft or modification, providing authentication and ensuring user and data access \citep{wang2019cryptographic}.

\subsubsection{Encryption}
In cryptography, encryption is the most frequently used technique to transform simple text or plain data into an encoded format that can only be read by those persons who have access to decode it. It is important to describe here that this encoding/decoding process would not be possible without a shared key, which two or more parties mutually set for secure communication. The key is an important encryption element that provides data security between end-to-end components and should not be disclosed to other network components.

There have been several other techniques proposed that are purely based on the encryption used to solve the security and privacy challenges in different Blockchain application domains. Here, we divide the encryption technique into further sub-extensions of encryption.

\begin{itemize}
\item \textbf{Simple Encryption:} As stated above, encryption is a method to convert the simple text format into another format, which others cannot easily understand. There are two types of encryption methods, that is symmetric and asymmetric. In the symmetric method, one secret key is shared between two parties to encrypt and decrypt the data. However, in the asymmetric approach, each party has two types of keys, such as public and private, in order to encode and decode the data, respectively. In the Blockchain, each transaction is encrypted in other forms, which hide content details.

\item \textbf{Homomorphic Encryption:} This is an extended version of encryption that allows the user to perform computations, such as addition and exponents of encrypted text. In contrast, simple encryption does not allow users to perform any useful computations on the ciphertext. Therefore, the main advantage of homomorphic encryption is two-fold: (i) to perform complex mathematical functions on encrypted data and (ii) to analyse encrypted data without losing the original data. Additionally, homomorphic encryption is the most common technique used in cloud computing, in which different organisations are involved in-service analysis of their encrypted data stored in the public cloud. In Blockchain-based applications \citep{Zheng2018, wang2018designing}, homomorphic encryption is used to preserve users’ privacy without revealing personal and sensitive information to others.

\item \textbf{Multi-Party Encryption:} Multi-party encryption or multiparty computation is another encryption form with which multiple users jointly perform the encryption of data. However, data remain private during the multi-party process and are not disclosed to others in the group. The advantage of using multi-party encryption in secure computations is that it can prevent the attacker from obtaining the secret information of any targeted users in the network. To illustrate this concept, \citep{Zyskind2015EnigmaDC} utilise multi-party computation to provide users’ guaranteed security and privacy in the decentralised Blockchain network.

\item \textbf{Proxy Re-Encryption:} In cryptosystems, proxy re-encryption is a third-party encryption technique in which the third-party (medium party) changes the plaintext into ciphertext, without knowing the actual content inside it. Generally, proxy re-encryption is considered a public-key cryptography technique that uses public and private keys to encrypt and decrypt the data. The proxy re-encryption scheme is commonly used in those applications when the different users want to exchange encrypted data without sharing their secret key. In Blockchain-based IoT applications \citep{dorri2017blockchain,aitzhan2016security,dagher2018ancile,zhang2018towards,guo2018secure,ji2018bmpls}, proxy re-encryption is used to share private contracts between users so as to control and manage the IoT devices.

\item \textbf{Paillier Encryption:} The Paillier cryptography or Paillier encryption is a key- pair based algorithm that utilises two keys, that is public and private, to encrypt and decrypt the data, respectively. It is also known as probabilistic asymmetric cryptography since it performs n-th computations on multiple residue classes. In Paillier cryptography, additive homomorphic encryption applies to the given set of messages, and each message is encoded/decoded with the key pairs of respective users. Considering the implementation of Blockchain in IoT \citep{cha2018blockchain} and cloud computing \citep{Zheng2018} domains, Paillier cryptography is used to achieve privacy and anonymity in such decentralised applications.

\end{itemize}

\subsubsection{Hashing}

A hashing technique is used to compact or digest any arbitrary size input into a fixed-size output. The input can be given in any sizes and formats such as integer, character, string and binary. Hash functions work according to a specific data format, which is often called a hash table. This is used to map the input data values on stored values to produce the output. A hash function’s strength is that it is designed to be a one-way function, which means that a user cannot change the input back from a given output value.

In cryptography, the hash functions are utilised to achieve the integrity of data messages because any single change in data value can be detected easily by a change in output value. Apart from data integrity, the cryptography hash functions are also used in digital signatures and different checksum protocols to achieve users’ authentication. The best cryptography hash function must have the following important key features. Firstly, it should be collision-resistant which means that two of the same inputs must produce different output values. Secondly, it should be impossible for everyone to regenerate the same input from the output values. Finally, the hash function should easily detect any small modification in data values. Many families of hash functions in cryptography have been proposed but the most commonly used hash families are MD5 and SHA-0 to SHA-3 with different output sizes.

In Blockchain, cryptography hash functions are extensively used to link and maintain the integrity of blocks. The blocks are linked with other blocks so that the hash of the previous block is stored in the header of the next block to form a complete hash chain. Moreover, hash functions are used in different Blockchain applications such as financial systems \citep{bhaskaran2018double,narula2018zkledger}, eHealth \citep{zhou2018distributed}, IoT \citep{cha2018blockchain,le2018capchain}, IoT \citep{dorri2017blockchain,knirsch2018privacy,lu2018bars} and crowdsensing \citep{vos2018defend} to achieve the security and privacy of transactions and users.

\subsubsection{Signatures}

In the past, the signature method was utilised as the simplest method to authenticate any document by placing hand-written signatures at the bottom of documents \citep{rivest1978method}. In the digital world, the signature method is used to protect software ownership and digital communication in a well-defined way. Also, this is an essential tool in information security to achieve the authenticity, integrity and non-repudiation of messages \citep{lamport1979constructing}.

We classify the signatures into the following sub-categories, such as digital signatures, multi-signatures, threshold signatures, ring signatures, blind signatures, attribute-based signatures, identity-based signatures and elliptic curve digital signatures algorithms.

\begin{itemize}

\item \textbf{Digital Signature:} This is a public-key cryptography technique that binds the identity of users to their digital data using a signature mechanism. In digital signatures, a private key (or secret key) is only known to the specific user and is utilised to sign the messages \citep{pointcheval2000security}.. The digital signature method confirms the authenticity of the text from the sender, to the receiver and he/she cannot repudiate the origin of the message. In addition, digital signatures are also used to check the integrity of a message and ensure that an adversary in the communication does not modify it \citep{bellare1996exact}.

\item \textbf{Multi-Signature:} Similar to digital signatures, multi-signature is also a digital signature technique used to prove the authenticity of digital documents. However, multi-signature allows people to sign one single document instead of a single user per document. For example, in government organisations, one document is passed on from many people having different ranks (bottom-up) to authenticate and prove it. Typically, a multi-signature scheme generates one single joint signature from a group of people, rather than individual signatures \citep{harn1994group}. In the Blockchain, multi-signatures are used to sign cryptocurrency transactions in order to add extra security protection to them \citep{salman2018security}. The total number of signatures required for one document is decided before the generation of addresses. In \citep{aitzhan2016security}, a multi-signature scheme is utilised in energy trading systems to protect users’ privacy and their related energy consumption data.

\item \textbf{Threshold Signature:} The working of threshold signatures is similar to a multi-signature where a group of people signs one document to ensure its authenticity. However, the only difference in threshold signatures comparing to the multi-signature is that it requires a fixed number of peoples to produce a valid signature for the document \citep{boldyreva2003threshold}. To illustrate the proof of threshold signatures, \citep{ziegeldorf2018secure,bao2018privacy} proposed the Blockchain-based E-voting scheme to achieve the security and privacy of voters.

\item \textbf{Ring Signature:}

Another  most important type of digital signature is a ring signature that works in a group pattern arranged in a ring shape. Ring signatures enhance the idea of a group signature to provide better security and privacy to group users. In a group, the signature can be generated by any group member who is assigned valid cryptography keys. In this way, it is challenging for others to determine the group’s actual user who generated the signature with their public key \citep{zhang2002id}. In Blockchain-based applications, ring signatures are employed to protect the input transactions signatures  with the public key of any node. Therefore, it is complicated for adversaries to find the correct identity  of the group \citep{kopp2017design,liu2018unlinkable}.

\item \textbf{Blind Signature:} In this signature algorithm, the user acts as a blind person in the overall signature process and generates a signature without knowing the actual content; thus, it is called a blind signature. However, the result of the blind signature is publicly available to everyone to verify the original contents. It is essential to describe here that blind signatures are mostly used in privacy-preserving protocols to achieve the anonymity of users (or signers) belonging to different parties \citep{chaum1983blind}. In the Blockchain scenario, blind signatures are widely used in e-voting applications to achieve the security and privacy of voters and candidates \citep{bhaskaran2018double,Owiyo2018DecentralizedPP}.

\item \textbf{Attribute-Based Signature (ABS):} This is a modern technique in digital signature methods that allows signature parties to sign documents with users’ information on fine-grained access control policies. This known information is formally called attributes given by the central authority in attribute-based signatures \citep{shanqing2008attribute}. Each user in the attribute-based signature comes with different types of attributes that are not identical to others. Therefore, the changing nature of attributes can also generate different signatures. The ABS is mostly employed in Blockchain-based E-health applications in which concern for users’ privacy is essential \citep{sun2018decentralizing, Lin2018BSeInAB}.

\item \textbf{Identity-Based Signature (IBS):} Certificates follow the idea of a public-key certificate to the user for signing documents. In this way, no one can go beyond their credential limits as defined by certification authorities. Compared with other digital signature schemes, the IBS scheme has some advantages in terms of implementation and computation \citep{shamir1984identity}. However, the drawback of the IBS scheme is that it can increase the length of the signatures by combining the two different signatures, that is one from the user and the other from the certification authority. Similarly, it needs two verifiers to verify and prove the generated signatures. In the Blockchain, the IBS scheme is frequently used in authentication systems in which users require authentication before using system resources \citep{malik2018blockchain}.

\item \textbf{Elliptic Curve Digital Signature Algorithm (ECDSA):} This algorithm combines the elliptic curve cryptography and digital signature algorithms to generate the signature of data contents. Therefore, it is considered the most powerful digital signature algorithm and mostly used in IoT based-applications \citep{khalique2010implementation}. In Blockchain-based applications, the ECDSA is utilised to ensure the integrity and authenticity of transactions \citep{zyskind2015decentralizing}. However, the drawback of ECDSA is that the key size needs to be double that of the other available cryptography algorithms.

\end{itemize}

\subsubsection{Secret Sharing}

In cryptography, secret sharing is a common technique used in distributed computing when one secret is shared equally among all of the group participants. In particular, this secret sharing scheme builds the trust of group participants by fulfilling the following criteria: a sufficient number of participants and the conditions and types of shares to reconstruct the share later. To build the secret, an (n, m) - threshold method is used, which is often called the (n, m) - threshold scheme \citep{yang2004t}. The advantage of using secret sharing with Blockchain-based applications is that it can reduce the communication and storage costs for sharing and storing data on distributed ledgers.

In cryptography, secret sharing is a common technique used in distributed computing where one secret is shared equally among all of the group participants. Especially, the secret sharing scheme builds the trust of group participants by fulfilling the following criteria, such as a sufficient number of participants, conditions and types of shares to reconstruct the share later. To build the secret, an (n, m) - threshold method is used, which is often called (n, m)- threshold scheme . The advantage of using secret sharing with Blockchain-based applications is that it can reduce the communication and storage costs for sharing and storing data on distributed ledgers.

\begin{itemize}
\item \textbf{Shamir Secret Sharing:} This is a security scheme that uses the concept of encryption to provide evidence securely from the majority of community participants. The Shamir secret scheme works in the form of a hierarchy in which most participants trust each other to maintain the trusted relationship between them \citep{feldman1987practical}. For instance, one participant must be reliable in distributing the private keys to other group members. In the Blockchain, the secret Shamir scheme requires a number (or secret) to define the threshold value to reconstruct the secret used by the miners \citep{chen2019light}.

\item \textbf{Additive Secret Sharing:} Additive secret sharing is another cryptography primitive used to achieve privacy by employing the multi-party computation method. Similar to the Shamir secret sharing scheme, no one can recover the total value of a secret using their shared secret \citep{doganay2008distributed}. The advantage of using the additive secret sharing scheme in Blockchain applications is that it follows homomorphic encryption for bitwise transactions processing.

\end{itemize}

\subsubsection{Zero-Knowledge Proof}

Zero-Knowledge Proof (ZKP) is one of the fundamental concepts of applied cryptography used to ensure security properties, for example anonymity, privacy and verification of transactions. In ZKP, a verifier party verifies the proof of the claimant and provides proof of knowledge without disclosing personal information to others \citep{rackoff1991non}. The advantage of ZKP in the Blockchain is that each verifier (or miner) can prove a shared secret challenge if the claimant does not provide any information (zero-knowledge). ZKP is most commonly used in business applications in which parties exchange confidential information without fear of personal information leakage \citep{goldreich1996construct}.

\subsection{Mixing}

Mixing service plays a vital role in privacy to conceal the senders’ and receivers’ transactions in such a way that no one can know what is actually inside them. In the mixing technique, both incoming and outgoing transactions are mixed up with the same type of transactions of others. This approach aims to separate the transaction detail from the identities of the sender/receiver. Generally, there are two types of mixing techniques used in Blockchain applications, that is transaction mixing and address mixing.
\subsubsection{Transaction Mixing}

In the Blockchain, transactions are stored in the distributed ledger and are open to everyone for graph analysis and other purposes. An adversary can easily track individuals by knowing the information stored in the Blockchain. To protect the information, especially financial information (bitcoins) from adversaries, transaction mixing services help users to mix their financial transactions so that it would be difficult for others to trace the original user who is involved in the transactions. According to defined criteria, these services accept the transactions as inputs from different users, mix or shuffle them according to defined criteria, and send them as outputs at different addresses. The mixing services are mostly charge-based services that charge the users to mix their transactions. However, the main issue identified in existing mixing services is that it depends on third parties that hold the transactions (coins) for some time. These services are often attacked or operated by malicious parties that use attacking techniques such as DDOS and ransomware to steal coins from both parties’ accounts. To resolve such issues, many Blockchain-based mixing services, such as CoinJoin and CoinShuffle, have been proposed to remove the third party’s requirement to mix financial transactions between the sender and the receiver.

\subsubsection{Address Mixing}

Address mixing (or address shuffling) is also a mixing service in which an input address of the transaction is related to the transaction’s output address. The address mixing service can be implemented in two ways, that is explicit address shuffling and implicit address shuffling. In explicit address shuffling, the mixing party explicitly knows the senders’ and their addresses for which the shuffling service is performing. CloakCoin \citep{cloakcoin} is the cryptocurrency that used the explicit address shuffling scheme to relate the input address to the output address. Simultaneously, the senders’ and senders’ addresses are completely hidden in implicit (or hidden) shuffling services. In an implicit shuffling service, the mixing server cannot relate the address of the sender with the address of the receiver. An example of implicit address shuffling is Maxwell’s CoinJoin \citep{Gmaxwell} cryptocurrency that used blind signatures to implement this strategy.

\subsection{Anonymisation}

Anonymisation is a commonly used data hiding technique that ensures that users are anonymous throughout the process. The anonymisation technique is designed to make it impossible for others to identify the specific user and their data stored in the database. The main aim of the anonymisation process is to protect users’ privacy with cryptography or generalisation methods.

\subsubsection{K-Anonymity}

K-anonymity is the most popular technique used to achieve the anonymisation of users’ data. The goal of the k-anonymity technique is to protect users’ privacy while performing complex operations on data, for instance, data required to fulfil the k-anonymity property if it is not distinguished from the at-least remaining k-1 users involved in the computation process. In this way, the k-anonymity property ensures that the probability of identifying the users in a given data set must not exceed 1/k \citep{sweeney2002k}. To achieve k-anonymity, the two most commonly used approaches are generalisation and data suppression. In the Blockchain, k-anonymity is the basic method used to achieve users’ privacy in the public environment  \citep{li2007t}.

\subsection{Artificial Intelligence}
Artificial Intelligence (AI) is a combination of different intelligence features and practices  for using native hardware in meaningful ways. At present, this technology can increase the level of thinking using the concepts of the neural network, machine learning and deep learning \citep{maloof2006machine}. Blockchain technology is designed to store the data on the immutable ledger by using different cryptography algorithms. In addition, AI algorithms have also been used to examine user activities in the Blockchain network and practise  different heuristic approaches in a deterministic way \citep{zeng2010social}.

\subsubsection{Machine Learning}
This is a basic and more frequently used technique in artificial intelligence that enables the system to learn from experience and make an automatic decision to improve it. The inputs to machine learning techniques are usually in learning data, collected heuristic observations and experiences. Thus, the only objective of machine learning technology is to allow the system to take automatic action without human intervention.

Blockchain technology, with a combination of machine learning methods, plays a significant role in developing Blockchain-based applications. In fact, it changes the way of thinking about building decentralised applications \citep{bishop2006pattern}. In addition, machine learning approaches also improve the Blockchain system’s security, using analytical learning methods, and provide a way to design new privacy-preserving models for decentralised applications \citep{maloof2006machine}.

\subsubsection{Deep Learning}
Deep Learning is another widely used technique of artificial intelligence (sometimes referred to as machine learning sub-field) that has a significant impact on some different areas capable of performing multiple tasks to produce accurate results \citep{lecun2015deep}. Similar to the machine learning technique, deep learning also gives instructions to the computer in some available datasets, such as text, images, audio and videos \citep{shokri2015privacy}. As the deep learning method integrates and works with large data sets to produce high-quality results, data security is an important requirement that requires significant solutions in terms of security and privacy.

Building on Blockchain technology, deep learning features can ensure user data security and can also meet the privacy needs of different applications. In addition, decentralised deep learning approaches are used with Blockchain applications to ensure consistency and transparency of Blockchain data \citep{mcmahan2016communication}.

\subsection{Others}

Many other approaches are available to respond to the security and privacy problems in the Blockchain-based application. For example, a solution to privacy problems in the E-Health application \citep{hussein2018medical} is a discrete wavelength transformation method in which a conversion strategy is utilised to convert the wavelets into discrete sampling, in order to achieve accurate frequency and timing information of stored data. In Blockchain security, the discrete wavelength transformation is used with cryptography functions to generate the key pairs for encryption/decryption. In \citep{dorri2017blockchain}, a clustering technique, in which each vehicle is linked and tracked by the respective controlling unit, called RSU (Road-Side Unit), is used to protect vehicle users’ privacy. The unit authenticates the vehicles using some asymmetric cryptography primitives. Another privacy preserving approach, called the three-weight subjective logical \citep{kang2018blockchain} , is employed in the vehicular network to protect vehicles’ data. This approach is purely based on a probabilistic logic model in which data is assigned a different weight, in order to calculate the subjective logic in the decision-making process. There are also a few other methods, such as Range Query \citep{li2018efficient}, Game Theory \citep{chatzopoulos2018privacy}, Bilinear Maps \citep{zhang2018towards}, Bloom Filter \citep{kopp2017design} and Statistical Measures \citep{Rawat2018iShareBP} which are used to deal with the problem of security and privacy in different Blockchain-based applications.

\section{Security and Privacy Attacks on  Blockchain-based Industry 4.0 Applications}\label{sec:7}

This section describes the various security and privacy attacks on Blockchain-based Industry 4.0 Applications, in which the attacker uses different approaches to obtain data and information. Since our survey paper aims to integrate Blockchain technology with various industrial applications, it is important to mention here that the attack surface includes attacks on both platforms, such as the Blockchain network and the industrial applications. 

We categorise the attacks based on their layers as follows: data layer, network layer, consensus layer, incentive layer, smart contract layer and application layer. We also categorise the security breaches into three different primary branches: (i) breach of confidentiality, (ii) breach of integrity and (iii) breach of availability. In breach of confidentiality, the attacker tries to listen to the communication between two parties without the consent of the owner of the data rights. In breach of integrity, the attacker aims to change or modify the original data into another form, after listening to the communication channel. Undoubtedly, breach of availability is one of the most severe breaches because the attacker’s intention is to disrupt the network or data services using some malicious attacks, such as a denial of services, to make these services unavailable to legitimate users. Moreover, we explain each attack with the attacker’s goals and objectives to expose vulnerabilities and threats in the system. We also sort attacks and targeted applications whereby some attacks, such as 51\% attacks, double spending attacks and selfish mining attacks, are specially designed for Bitcoin and Ethereum applications. However, most of the attacks can also be targeted generally to the other domains, including IoT, SG, medical and vehicles. We also present state-of-the-art solutions and techniques used to protect the applications and their underlying systems against a subset of such malicious attacks.

Table \ref{table:attacks} presents a summary of the reviewed attacks on both the Blockchain network and industrial applications, along with their layer categories, attacker goals and objectives, security breaches and vulnerabilities, as exploited in the system or network. In addition, we also include the targeted applications suffering from these potential threats and vulnerabilities, and prevention methods and security approaches in the table.

\renewcommand{\arraystretch}{0.3}

\begin{sidewaystable*}
\footnotesize	    
    \centering
    \caption{Security and Privacy Attacks on Blockchain-based Industry 4.0 Applications}
    \label{table:attacks}
    
        \begin{tabular}{P{0.09\textwidth}|P{0.08\textwidth}|P{0.15\textwidth}|P{0.12\textwidth}|P{0.15\textwidth}|P{0.1\textwidth}|P{0.2\textwidth}}
    
      \toprule 
    \multicolumn{1}{c|}{ \textbf{Layer}} & \multicolumn{1}{c|}{\textbf{\thead{Attack} }}& \multicolumn{1}{c|}{\textbf{\thead{Attacker \\ Objectives}}} & \multicolumn{1}{c|}{\textbf{\thead{Security Breaches \\ Occurred}}} & \multicolumn{1}{c|}{\textbf{\thead{Vulnerabilities \\  Exploited}}}& \multicolumn{1}{c|}{\textbf{\thead{Target \\ Applications}}} & \multicolumn{1}{c}{\textbf{Countermeasures}} \\\midrule

            & Malleability Attack& Duplication of signatures to create impact of double spending problem  & Breach of Integrity
& Change few bytes of signatures & Bitcoin and other cryptocurrencies & \thead{Segregated Witness, \\ Modification of the bitcoin specification\\, Time commitment method } \\\cline{2-7} 

& Time Hijacking Attack& To modify the time stamps of both the network and a node& Breach of Integrity
& Incorrect time stamps are broadcast to other network nodes & Bitcoin and other cryptocurrencies& \thead{Hardware oriented systems, \\ Network time protocol, \\ time constraints}  \\\cline{2-7} 

\multirow{7}{*}{Data}& Quantum Attack& utilise quantum computing to solve the cryptographic portion of Blockchain & \thead{Breach of Integrity, \\ Breach of Availability}
& Performed hash collisions& All Applications & Quantum post-signature scheme  \\\cline{2-7}

  &Replay Attack& To hold the valid  transactions not to be  added in the   Blockchain  & Breach of Integrity,  Breach of Availability & Delay caused in a   P2P communication  network  &   Online Digital  Platform,  Vehicle & \thead{Add nonce in each transaction, \\  Mixing techniques,\\  Digital signatures can be used \\ to prevent replay attack}   
 \\\cline{2-7}

& Modification  Attack &  To perform  modification in  transmitted data&   Breach of  Confidentiality,  Breach of Integrity&   Caused the harmful  activities in the system  or network &   SG, IoT,  Medical & \thead{Consensus algorithms, \\ cryptography techniques} \\\cline{2-7} 

&Fault Injection  Attack & To impersonate the  Blockchain by adding  fake data or block in existing Blockchain   
&Breach of Integrity& The reputation of the  system is getting worst when  performance is the prime factor   &Vehicle& Digital Signatures,  Hashing \\ \cline{2-7}

& Upgraded  Attack & To make modifications in  the threshold values to impersonate the miners   
&Breach of Integrity& An illegal decision made by the fake  miners to perform  malicious activities  &Vehicle&Digital Signatures\\ \cline{2-7}

           \hline

            &51\% attack & To take control of the overall network by colluding with more than 50\% devices in the  network &Breach of Integrity& Make wrong decisions even when there are honest miners, Transferring of  bitcoins into targeted  accounts
& Bitcoin, Ethereum &  Random selection of miners \\\cline{2-7} 

& DoS Attack& To disrupt the  network resources provided by the host &Breach of Availability&  Unavailability of legitimate services such as network and security& All Applications & \thead{Decentralised networks, \\ Consensus algorithms} \\\cline{2-7} 

 \multirow{10}{*}{Network} & DDoS Attack& To disrupt or overload  the network resources  as collectively&Breach of Availability&Unavailability of  legitimate services  such as network, and  security  & All Applications&  Consensus algorithms\\\cline{2-7}

&Eclipse Attack & To target the specific  node in P2P network rather than  attacking the whole  network to perform malicious activities&Breach of Integrity& A single targeted  node hijack and  bypass the whole  network communication at once, A node can also govern their own rules in the network &Bitcoin& \thead{A number of possible solutions  can\\  be applied including the usage\\ of a private network,\\ randomly  selection of miners\\ and limiting  the number of \\incoming/outgoing connections}    
 \\\cline{2-7}

\hline

        \end{tabular}
\end{sidewaystable*}

\begin{sidewaystable*}
\footnotesize	    
    \centering

        \begin{tabular}{P{0.09\textwidth}|P{0.08\textwidth}|P{0.15\textwidth}|P{0.12\textwidth}|P{0.15\textwidth}|P{0.1\textwidth}|P{0.2\textwidth}}
    
      \toprule 
    \multicolumn{1}{c|}{ \textbf{Layer}} & \multicolumn{1}{c|}{\textbf{\thead{Attack} }}& \multicolumn{1}{c|}{\textbf{\thead{Attacker \\ Objectives}}} & \multicolumn{1}{c|}{\textbf{\thead{Security Breaches \\ Occurred}}} & \multicolumn{1}{c|}{\textbf{\thead{Vulnerabilities \\  Exploited}}}& \multicolumn{1}{c|}{\textbf{\thead{Target \\ Applications}}} & \multicolumn{1}{c}{\textbf{Countermeasures}} \\\midrule
        
            \multirow{7}{*}{}

&Sybil Attack&   Creation of fake  accounts to exploit the  network 
&Breach of Integrity& Leakage of personal  information of users 
& Bitcoin,  Online Digital   Platforms,  IoT  Medical,  Vehicle,  Cloud  
& Limit the creation of identities in  the network,  Assign reputations to the users
 \\\cline{2-7}

& BGP Hijacking Attack & To determine the data path routes in the  network from source to destination   
&Breach of Integrity&All network traffic  directed to the  malicious server  &Bitcoin& \thead{ Verify and monitoring the network traffic, \\ BGP sec protocol can be used to avoid \\ malicious traffic}    \\\cline{2-7} 

&Phishing  Attack & Obtaining the confidential details of a network user 
&Breach of Confidentiality& Use the obtained personal details in inappropriate ways   &All applications& \thead{Anti-spyware softwares \\ Upgrade firewalls} \\ \cline{2-7}

&  Liveness Attack& To hold the  transaction longer  than their  confirmation time&Breach of Availability& A valid verification of a transaction is not  possible on time   & Bitcoin,  Ethereum &Round trip time \\ \cline{2-7}

&Routing Attack& To perform malicious modifications in data packet before  transmitting to other  nodes in the network
&Breach of Integrity& All traffic of targeted server is routed  towards the malicious one\textquotesingle s  
&All applications& \thead{Audit and verification protocols\\ to  check the integrity of the message,\\  Round trip time to check\\ the  propagation time,\\ cryptography techniques}  \\ \cline{2-7}

& Man-in-the- middle (MITM)  Attack & To bypass the  communication   channel to obtain  secret information  between parties  
& Breach of  Confidentiality,  Breach of Integrity   
& Different secrets such   as private key and  personal identities are  disclosed to the  attackers   
& Online Digital  Platforms,  Medical,  Vehicle,  Finance  IoT  
&  Authentication schemes can be  used to prevent a MITM attack 
 \\\cline{2-7}

           \hline
           
        \multirow{1}{*}{Consensus} 
            
            &Double spending  Attack& To send the same  money (bitcoins) to  different users&Breach of Integrity&Duplicate and reproduction of digital  transactions& Bitcoin,  Digital Currencies & \thead{Consensus protocols, \\  digital signatures} \\\cline{2-7}
            
            &Stake Bleeding Attack &To gain ownership of the blocks that will be added to the Blockchain&Breach of Integrity&Increase the number of stakes&Incentives-based applications& Validity of context-oblivious transactions \\\cline{2-7}

\hline

           & Selfish mining Attack&To hold the valid block  without broadcasting to other nodes in the  network& Breach of Integrity, Breach of Availability & Resource consumption is high,  Claim the more rewards compared to  other nodes&Bitcoin,  Ethereum & Define the scale and height to detect the miner behaviour \\\cline{2-7}

           &Bribery Attack &To create a fake mining capacity&Breach of Availability&To gain the more power on the network&Incentive-based applications& PoW consensus mechanism \\\cline{2-7}

       \multirow{10}{*}{Incentive}    &Refund Attack &To reclaim payments sent to users&Breach of Availability&To reimburse payments made to users &Bitcoin and digital currencies-based applications& Request that the user submit a payment request with sufficient evidence \\\cline{2-7}
           
           &Block Withholding Attack &Disguising the hash of the puzzle rather than returning it to the mining pool &Breach of Availability, Breach of Integrity&Waste computing resources and reduces the overall mining pool’s income&Incentives-based Applications& \thead{silent  time  stamps,\\  zero  determinant methods,\\ contribution of smaller pools, \\ and a game model based on the \\ consensus protocol and Nash equilibrium}  \\\cline{2-7}

\hline

        \end{tabular}
\end{sidewaystable*}

\begin{sidewaystable*}
\footnotesize	    
    \centering

        \begin{tabular}{P{0.09\textwidth}|P{0.08\textwidth}|P{0.15\textwidth}|P{0.12\textwidth}|P{0.15\textwidth}|P{0.1\textwidth}|P{0.2\textwidth}}
    
      \toprule 
    \multicolumn{1}{c|}{ \textbf{Layer}} & \multicolumn{1}{c|}{\textbf{\thead{Attack} }}& \multicolumn{1}{c|}{\textbf{\thead{Attacker \\ Objectives}}} & \multicolumn{1}{c|}{\textbf{\thead{Security Breaches \\ Occurred}}} & \multicolumn{1}{c|}{\textbf{\thead{Vulnerabilities \\  Exploited}}}& \multicolumn{1}{c|}{\textbf{\thead{Target \\ Applications}}} & \multicolumn{1}{c}{\textbf{Countermeasures}} \\\midrule
        
            \multirow{7}{*}{}

&Balance Attack& To delay the Communication  among nodes having  the same mining power(balance) in the  Blockchain network &Breach of Integrity& Selected miners can  increase their balance  by disrupting the  communication of  others  
& Bitcoin,  Ethereum & Limiting the numbers of   miners with more balance  in the network
 \\\cline{2-7}

           \hline

        &Integer Overflow Attack& To overflow the defined limits &Breach of Integrity&This resulted in memory overflow and system halting problems&Ethereum& Carefully Code analysis, rewriting, and testing \\\cline{2-7}
        
      \multirow{1}{*}{Smart Contract}    &Re-Entrancy Attack&To create malicious smart contracts that call re-entrance functions&Breach of Integrity&Take ethers from other people's wallets&Ethereum-based applications& \thead{Dynamic taint tracking of smart contract \\ data flows, fuzzing testing} \\\cline{2-7}
         
          &Short Address Attack&To enter a short address in order to execute malicious code modifications&Breach of Integrity, Breach of Availability&Exploitation of smart contract code&Ethereum-based applications& Correct code verification and synthesis \\\cline{2-7}

\hline

&Location  cheating Attack &  To send the wrong  location to the road-  side unit 
&Breach of Integrity&   Consume the   maximum resources of the system &Vehicle& Calculate the new location from  existing stored locations  \\\cline{2-7}

&Ballot stuffing  Attack & To do the multiple  entries instead of  permitted one entry in  the system   
&Breach of Integrity& Perform malicious actions by taking  control of the overall  network  & eCommerce,  E-Voting 
& Digital Signatures \\\cline{2-7}

&Badmouthing Attack&   To destroy someone \textquotesingle s reputation by giving  negative feedback 
& Breach of  Confidentiality  Breach of Integrity& Malicious activities  can occur in the  system by single or  group of attackers to  take unlawful actions   
&eCommerce&Self-organizing maps \\ \cline{2-7}

&Guess Attack& To guess the keyword   by performing brute forcing or matching  techniques  & Breach of  Confidentiality,  Breach of Integrity& Disclose the system  secrets such as private  keys or personal  information   
&Medical& Encryption,  Hashing 
 \\ \cline{2-7}

\multirow{10}{*}{Application} &Chosen Ciphertext  Attack & To obtain the secret  key to recover the  original data from the ciphertext  & Breach of Confidentiality,  Breach of Integrity& Obtain the personal data  or information 
&Medical& Elliptic Curve Digital Signature  Algorithm (ECDSA) 
 \\ \cline{2-7}

&Impersonation   Attack & To create a fake profile or mimic the  behaviour of others  nodes to gain  maximum benefit   
&Breach of Integrity& Control the overall  network by  maximising the benefits from others  
& Vehicle,  Crowd Sensing 
& \thead{Verification techniques,\\  Cryptography} 
\\ \cline{2-7}

&Linking Attack&   To map the stored  data using various linking algorithms & Breach of  Confidentiality,  Breach of Integrity &   Extract some useful  information to harm  the privacy of users  & Bitcoin,  Online Digital  Platforms,  IoT,  Vehicle,   
&  Generate the new key every time  to encrypt the data  \\\cline{2-7}

&Collusion Attack& Tries to mix the  different inputs to  reveal the secrets  about the node in the  Blockchain network & Breach of  Confidentiality,  Breach of Integrity& Personal or secret  information is  revealed to malicious users   & Medical,  eCommerce
& The pseudo-random method can be   used to prevent collusion attack 
\\ \cline{2-7}

\hline

        \end{tabular}
\end{sidewaystable*}

\subsection{Data Layer}

The data layer is the last layer of the Blockchain framework; it defines the physical structure and properties of a block and encapsulates the data and chain of connecting blocks to the Blockchain. The data layer is responsible for handling data that is stored on the Blockchain (on-chain) and in the database (off-chain). The attacks on the data layer of Blockchain architecture, which include malleability attack, time hijacking attack, quantum attack, replay attack, modification attack, fault injection attack and upgraded attack, are listed in detail.

\subsubsection{Malleability Attack}
The malleability attack is a specific type of double-spending attack that often happens in networks due to the malleability of signatures. In this attack, the attacker broadcasts the two transactions to the Blockchain network, resulting in the appearance of double-spending. For instance, attackers can monitor transactions on the Bitcoin network and change their signatures while still authenticating the transactions. The signature uses the secure sockets layer protocol, which means that even though an attacker modifies any bytes, the signature remains valid. Upon modifying the signature of a transaction, a new transaction identifier is generated \citep{decker2014bitcoin}.

Numerous solutions have been proposed to this attack in order to prevent the network from being malleable, including segregated witness \citep{sgwitness}, modification of the bitcoin specification \citep{andrychowicz2014fair} and time commitment methods \citep{andrychowicz2013deal}.

\subsubsection{Time Hijacking Attack}
Time hijacking attacks occur as a result of a loophole discovered in the time stamp protocol of Bitcoin and their related cryptocurrencies. In this attack, the attacker's goal is to change the time counters of both the node and the network.

One strategy for resolving this problem is implementing hardware-oriented systems, which benefit from replacing older network time technologies. Time hijacking attacks can also be mitigated by using tolerance range constraints and improved network time protocols \citep{ma2020achieving}. 
\subsubsection{Quantum Attack}

Quantum computers are explicitly developed to solve cryptographic problems based on complex mathematical problems. Quantum attacks are generally aimed at the Blockchain's cryptographic component, with the primary objective of resolving the mathematical problem of cryptographic dependency. For instance, attackers may conduct quantum attacks against Blockchain in order to perform hash collisions against consensus protocols such as PoW.

Khalifa et al. \citep{khalifa2019quantum} suggested general security steps against quantum elastic Blockchain using the quantum post-signature scheme to minimise the problem of quantum attacks on Blockchain.

\subsubsection{Replay Attack}
In Blockchain systems, the replay attack is the most common issue faced by Blockchain transactions and can cause a long delay in the communication between the two parties. In response to the attack, the opponent holds certain transactions in the network and does not send them to the miners for verification. As a result, each node has to wait a long time to acknowledge its transactions and results. Some significant works have shown that this problem is addressed by using multiple approaches, such as adding a nonce to each transaction, mixing techniques and digital signature methods \citep{androulaki2019resisting}.

\subsubsection{Modification Attack}
An adversary always tries to change the broadcasted transactions in the Blockchain network before sending them to the miners for verification. Furthermore, it can also involve the modification of acknowledgment messages received from miners. As a result, the attacker breaches the system’s integrity to launch harmful activities and can take complete control of the underlying system. To overcome this problem, some approaches \citep{kuo2018modelchain,Zyskind2015EnigmaDC,Lin2018BSeInAB} use cryptography operations as attribute-based signatures and consensus algorithms utilised to provide resistance to modification attacks in Blockchain systems.

\subsubsection{Fault Injection Attack}
In a fault injection attack, the adversary can change the programme’s execution by inserting malicious or fake code inside the program. The purpose of the fault injection is to either stop the execution of specific instructions or disrupt the complete code. This vulnerability usually occurs in the system due to the improper use of code validations in the program. In the Blockchain, the adversary attempts to impersonate the Blockchain by adding fake data or blocks to the existing Blockchain, using fault injection methods. To give the idea of a fault injection attack, \citep{li2018efficient} proposed the Blockchain-based vehicular system in which different cryptography methods, such as hashing and digital signatures, were used to protect the system from injection attacks.

\subsubsection{Upgraded Attack} 
The upgraded attack is a type of data attack in which the adversary involves changing the trust values defined as threshold values of the system’s participant amount. The system’s trust level can reach higher than the current level with many users in the Blockchain network. In an upgraded attack, the Blockchain miner controls the overall network to launch the network’s upgraded attack. Multiple fake miners can change the current threshold value to perform malicious activity in the system. A possible solution to overcome this problem was proposed in \citep{li2018creditcoin}, in which digital signatures were utilised to verify the defined threshold values in the vehicular network. In this way, the modification in the threshold values can easily be detected and mitigated.

\subsection{Network Layer}

The network layer is the fifth layer in the Blockchain layered architecture and it is primarily responsible for information transmission between Blockchain nodes. As we all know, Blockchain operates on a network known as a P2P network, in which peers exchange knowledge about the state of the network. For example, any node in the public Blockchain may enter the network. That node can be any ordinary home computer or mobile device; therefore, network layer protection must be implemented to prevent further network attacks. Therefore, security and privacy are also important components of the network layer. The network layer is subject to the following attacks such as 51\% attack, denial-of-service attack, distributed denial-of-service attack, eclipse attack, Sybil attack, BGP Hijacking attack, phishing attack, liveness attack, routing attack and man-in-the-middle attack. These attacks are described in detail below.

\subsubsection{51\% Attack}

One of the common vulnerabilities found in the Blockchain network, especially in Bitcoin and Ethereum applications, is the 51\% attack. A group of miners wants to control the network with more than 50\% mining (or computing) power \citep{hajdarbegovic2014bitcoin}. The mining group prevents the newly created transactions and would not allow them to go to other miners to pass the confirmation successfully. In this case, the 51\% controlling group takes control of the overall Blockchain and creates wrong decisions to dispute the network’s reputation. Moreover, 51\% of miners would be able to transfer all of the bitcoins from the user account to their targeted accounts. One approach to solve this problem involves selecting random miners and restricting them from recycling their bitcoins to participate in the consensus process \citep{bastiaan2015preventing}.

\subsubsection{Denial of Services (DoS) Attack}
Denial of Services is a common attack on centralised systems; this prevents the host’s network communication or services from performing some legitimate actions. In this way, the particular node or system cannot provide legal services to others until it recovers so as to provide the same services again. In the Blockchain systems, the DoS attack could restrict one particular node from sending and receiving updates from other nodes in the P2P network. Generally, applications that are specifically based on centralised architectures may suffer most from a DoS attack. The standard solutions to this problem are decentralised networks and optimal consensus protocols that better protect the systems from DoS attack \citep{vasek2014empirical,back2002hashcash}.

\subsubsection{Distributed Denial of Services (DDoS) Attack}

Compared with the DoS attack, the attacker’s main aim in a DDoS attack is to interrupt the complete services of the integrated network, instead of an attack on a specific node in the network \citep{karami2013understanding}. In most cases, this attack happened in the network due to poorly managed cache records of Domain Name Services (DDoS), in which the nodes could not receive the updates from other nodes on time. This attack is most dangerous for network applications when all legal services go down at once; therefore, it requires significant solutions to overcome this problem. In Blockchain-based cryptocurrencies, the consensus mechanism plays a vital role in a decentralised and distributed environment to make an optimal decision at some common points \citep{feder2018impact}.

\subsubsection{Eclipse Attack}
In an eclipse attack, the attacker aims to target a single specific node rather than capture all the P2P network nodes. In this way, the attacker could stop the targeted node from receiving the new updates from the other nodes. Indeed, the attacker wants to connect the target node with the other malicious captured nodes in the network. The major difference between the eclipse and Sybil attack is that the eclipse attack only targets the specific node. In contrast, the Sybil attack captures and takes control of all network nodes at once. If the attacker successfully launches the eclipse attack, he can govern his own rules in the network. Several possible solutions can resist the eclipse attack by applying the following methodologies, such as using the private network, a random selection of miners, static IP address and limiting the number of incoming/outgoing connections \citep{singh2006eclipse,kendler2015eclipse}.

\subsubsection{Sybil Attack}
Sybil attack refers to the most important issue of the P2P network. An adversary exploits the performance of the network by creating multiple fake identities of the same user. Similarly, a malicious node in the Blockchain can cover a large portion of the network by creating multiple fake profiles of the same node. In this scenario, the honest nodes in the same network cannot detect fraudulent behaviour and seem to receive the transactions from other honest nodes in the network. This malicious behaviour of the network nodes infers that the attacker aims to control the overall Blockchain network. There are many solutions available to reduce the risk of Sybil attack in Blockchain-based applications such as eHealth \citep{kuo2018modelchain},smart vehicles \citep{li2018creditcoin}, trusted computing \citep{zyskind2015decentralizing} and online digital platforms \citep{friebe2018decentid}. However, one simple technique that can restrict the access of a malicious user is to apply some identity-based mechanisms in the systems. Moreover, consensus algorithms, such as PoW, are also used in many cryptocurrencies to protect the Sybil attack. Each node requires solving the expensive puzzle problem to participate in the mining process.

\subsubsection{BGP Hijacking Attack}
In a BGP (Border Gateway Protocol) hijacking attack, the adversary takes advantage of a vulnerability found in network operators to intercept and manipulate the network traffic routing through gateways \citep{zhang2007practical}. In Blockchain systems, the BGP attack controls miners’ mining power (or mining pool server) by splitting them into different groups to cause the propagation delay of blocks in the network. Thus, all traffic from the Blockchain nodes is directed towards the malicious server to gain others’ bitcoins. The common strategy used to tackle the BGP hijacking is the BGPsec protocol that prevents the malicious traffic from gaining access to the system \citep{lepinskibgpsec}. Moreover, monitoring and verifying network traffic after some intervals can also be a useful solution.

\subsubsection{Phishing Attack}

A phishing attack is a type of social engineering activity that is often used to obtain financial benefits by stealing user personal information, login credentials and banking information such as credit card numbers. This attack is typically carried out when an attacker attempts to pose a trustworthy party and persuades a victim to act in various ways, such as opening an email or responding to a text message.

The most successful way to prevent phishing attacks on the system is to install anti-spyware software and periodically upgrade firewall settings. Furthermore, firewall security can prevent unauthorised file access by blocking malicious attempts \citep{hong2012state}.

\subsubsection{Liveness Attack}
This attack happens in Bitcoin and Ethereum applications, when the attacker can hold the broadcasted transactions longer than their confirmation time, in order to cause a delay in the network. As a result, an attacker can build a single chain consisting of transactions that are not transferred to honest miners on the network. The size of the private chain is longer than the public chain maintained by honest nodes in the network \citep{kiayias2017trees}. The round-trip time (RTT) is used to solve the liveness attacks in Blockchain-based applications in order to overcome this problem.

\subsubsection{Routing Attack}
In a routing attack, an attacker compromises and intercepts the network channel to perform malicious modifications to the data packets. An attacker in a routing attack aims to temper with  the data values inside the packet before transmitting to other nodes in the network. The adversary in the routing attack routes the data traffic towards the malicious server by changing the packet header’s destination addresses. This attack is a common attack on client-server-based applications, and especially on those applications based on a P2P network. As Blockchain technology follows the idea of a P2P network, the attacker can change the destination address of broadcasted transactions to get the maximum reward from the system \citep{apostolaki2017hijacking,song2018blockchain,banerjee2018blockchain}. In the Blockchain, the standard solution used to detect the routing attack is simply discarding those updates that do not match the other received updates. Moreover, the network parameters, such as round-trip time (RTT) and irregular patterns, can also help users identify and detect the routing attacks in the network.

\subsubsection{Man-in-the-Middle (MITM) Attack}

This is one of the common vulnerabilities found in network systems in which the attacker plays the role of a middleman to bypass the network traffic to obtain users’ personal information or secrets. Once the attacker successfully captures the communication channel’s data, he can use this data in further malevolent activities. The MITM attack also takes advantage of vulnerabilities found in key-agreement protocols and storage systems to retrieve the secret keys from them. In Blockchain-based cryptocurrency systems, the attacker uses the MITM attack to steal money from the victim’s wallet by changing the destination address with their fake wallet address. The most significant way to overcome the MITM attack is using an advanced authentication mechanism which does not allow the adversary to enter into the system.

\subsection{Consensus Layer}
The consensus layer is regarded as the foundation layer of the layered architecture of Blockchain. Further, it contains numerous consensus algorithms essential to the operation of all Blockchain networks; for example, they allow Blockchain nodes to agree on the validity of newly generated data blocks. The security of the Blockchain is based on the participation of each node in the network. The security of Bitcoin, for example, is dependent on the high hash power of the nodes that participated in the PoW. There are several distinct consensus protocols, for example, PoW, PoS, PBFT and DPoS. The attacks on consensus layers are described below, including the double-spending attack and the stake bleeding attack.

\subsubsection{Double Spending Attack}

In Blockchain-based cryptocurrencies, especially in Bitcoin, an attacker with some bitcoins tries to collude with the network by sending a transaction of already consumed bitcoins to others, with the intention of a newly generated transaction. In this case, the attacker can use and spend the double bitcoins to collude with someone in the network. This type of vulnerability is only found in digital cryptocurrencies due to the limitation of miners’ abilities during the verification process. Therefore, recently proposed cryptocurrencies and different Blockchain systems \citep{aitzhan2016security,zhong2019secure,gao2018blockchain} are trying to overcome the double-spending attack by using the latest consensus protocols and cryptography mechanisms, such as digital signatures.

\subsubsection{Stake Bleeding Attack}

As the name implies, a stake-bleeding attack is a type of attack against the PoS consensus mechanism. The attacker used transaction fees and processed transactions out of context in this attack, enabling attackers to track the newly added block to the Blockchain. Stake-bleeding attacks on Blockchain networks grew in probability as the number of legitimate transactions and adversarial shares increased.

To address this problem in Blockchain networks, Gazi et al.\citep{gavzi2018stake} suggested a protocol focused on perspective transactions for validating low-growth chains in order to avoid stake-bleeding attacks.

\subsection{Incentive Layer}

The incentive layer in the Blockchain architecture is intended to provide rewards to nodes for participating in the mining process in order to ensure security and verification of blocks added to the Blockchain. The security of the Blockchain is determined by a few factors such as the number of miner nodes, the consensus protocol and the mining method. This layer, however, is vulnerable to the following attacks, as described below: selfish mining attack, bribery attack, refund attack, block withholding attack and balance attack.

\subsubsection{Selfish Mining Attack}
In a selfish mining attack, an adversary acting as a miner shows selfish behaviour by holding the confirmed blocks, without broadcasting to the other miners in the pool network. More than one miner is involved in selfish mining behaviour in order to govern their own rules and policies. In this way, the selfish miners collecting the validated blocks can demonstrate and claim more reward against their PoW (hashing power) than other honest miners in the mining pool. Recently, one solution was proposed to tackle the selfish mining attack when a fair mining mechanism is adapted to determine the scale and height of the block. It also allows the network to block the selfish miners in the event of a discrepancy in the blocks \citep{saad2019countering}.

\subsubsection{Bribery Attack}
In a bribery attack, the attackers attempt to obtain a temporary majority of miners to increase mining ability on the network. This attack is made by renting it from the nominal owners and mining on the fork that does not require this transaction. Consequently, the attackers may execute a transaction first and the supplier will then await transaction confirmation. Attackers used various methods to bribe miners and increase mining power, including direct payment, fraudulent mining pools and inside payout via tokens.

An efficient strategy to alleviate this problem is using the PoW consensus mechanism as attackers have to pay considerable costs to discredit the miner \citep{judmayer2019pay}.

\subsubsection{Refund Attack}
In a refund attack, the attacker aims to refund the transactions (or payments) made to illegal users by honest users and then fairly deny them to participate in the refund transaction phase. In such an attack, the intruder impersonates unauthorised traders in order to exploit the entire network by rejecting all sent transactions.

McCorry et al. \citep{mccorry2016refund} suggest an effective solution to this problem in which users are asked to include payment request message along with a few verifiable pieces of evidence such as a delivery address, which reduces the attacker's incentive for profitable attacks.

\subsubsection{Block Withholding Attack}

A block withholding attack is a form of resource squandering attack in which miners violate the mining rules by disguising the hash of the puzzle rather than returning it to the mining pool for a greater self-reward. In a real-world example, as miners engage in mining activities through a mining pool, all miners share the reward for successfully solving the computing puzzle based on their computing power quota. However, block withholding attacks waste computing resources and reduces the overall mining pool's income, as only malicious miners profit from this attack and collect additional rewards.

A wide range of solutions to this problem have been suggested, including silent time stamps \citep{chang2019silent}, zero determinant methods \citep{hu2019game}, contribution of smaller pools \citep{elliott2019nash} and a game model based on the consensus protocol and Nash equilibrium \citep{li2020mining}.

\subsubsection{Balance Attack}

As the name suggests, a balance attack is an attack on some consensus mechanism to increase the balance by using some unfair means. Balance Attack is a special type of attack on a PoW-based consensus protocol. An attacker with low balance or power attempts to delay communication between those subgroups of miners who have the same hashing power (or balance) \citep{natoli2016balance}. In this way, the attacker captures some of the information from their communications and allows for other types of attacks, such as double-spending. This type of attack is most common in Bitcoin and Ethereum-based applications that have coins and ethers to spend. Generally speaking, this problem is resolved by limiting the number of more balanced miners in the network.

\subsection{Smart Contract Layer}
The contract layer is the fifth layer of the Blockchain network and comprises three major components: the smart contract itself, scripting code and an algorithm or logic. These three elements represent the key logic and conditions in the executed contract. These logics are generally written in Solidity, a programming language. The smart contract layer attacks are as follows: integer overflow attack, re-entrancy attack and short address attack, all of which are described further below.

\subsubsection{Integer Overflow Attack}
Integer overflow is a common security vulnerability in many applications, especially ethereum smart contracts on the Blockchain, which occurred primarily due to a lack of code validations. Smart contracts are a series of programme codes in which unique numbers determine the upper and lower limits of an integer. When the value executed reaches their prescribed limits, an integer overflow problem occurs, causing the machine to halt for specific errors.

A few solutions have been suggested to mitigate the risk of integer overflow attack in smart contracts; however, most of them focus on careful analysis, rewriting, verification of codes writing \citep{torres2018osiris, gao2018blockchain}. 

\subsubsection{Re-Entrancy Attack}
Reentrancy attacks are normally triggered by those functions that are not meant to be re-entered by developers. In this attack, attackers can create malicious contracts that call these functions reentrantly with the intent of stealing Ether from an honest user's account, causing the user to lose his credentials and all Ethers. An example of this type of attack is the DAO attack on smart contracts, which occurred in 2016 and resulted in the loss of 60 million Ether.

Many solutions have been suggested to overcome the re-entrancy attack on smart contracts. For example, one approach called Sereum \citep{rodler2018sereum} is proposed to solve the re-entrance attack, allowing for dynamic taint tracking of smart contract data flows. ReGuard \citep{liu2018reguard} is an automatic detection system that conducts fuzzing tests in order to fix the issue of re-entry attacks.

\subsubsection{Short Address Attack}

A short address attack is actually a bug in developer-side code that causes users to enter a short address instead of the full address. For example, if a user uses the transfer method to withdraw coins and is needed to enter a short address. If the size of the address entered by the user is not checked due to a lack of validation measures, a short address attack may occur.

A solution to this problem is suggested by a technique called SmartScopy \citep{feng2019precise}—automatically synthesising adversarial contracts to achieve smart contract stability.

\subsection{Application Layer}

The application layer is responsible for the execution of applications used by end-users to communicate with the Blockchain network. Application layer security refers to the protection of this layer and the users who communicate with it. Since this layer is a combination of different Blockchain components and third-party technologies to develop an application, it is vulnerable to a wide range of attacks, including location cheating attack, ballot stuffing attack, badmouthing attack, guess attack, chosen ciphertext attack, impersonation attack, linking attack and collusion attack. These attacks are detailed below.

\subsubsection{Location Cheating Attack}

A location cheating attack is a common attack on most vehicular networks when both passengers and drivers are involved in the location cheating activities by sending false locations to a central authority called a road-side unit (RSU). In the case of passengers, an adversary launches the location cheating attack by sending a false and long-distance location to an RSU to pick up the passengers from some point. On the other hand, the driver can also be involved in the location cheating attack by spoofing the identity of an RSU to cheat passengers. To rectify the issue of location cheating in vehicular networks, \citep{li2018fppb} proposed a robust and secure Blockchain-based method to calculate drivers’ and passengers’ authenticity.

\subsubsection{Ballot Stuffing Attack}

Ballot stuffing (or ballot-box stuffing) is an attack on the electronic voting system. An adversary tries to carry out this illegal activity by casting several votes (ballots) favouring a target. This fraudulent behaviour triggers the system’s breach of integrity to increase the number of votes for one candidate, thus reducing another candidate’s credibility. In this case, the attacker may be able to exploit the system’s vulnerability by taking control of the overall election process. One way to overcome this problem is to use digital signatures in the voting process to validate voters’ and candidates’ authenticity. In addition, there are a variety of Blockchain-based e-voting applications designed to detect ballot stuffing attacks in the systems \citep{Owiyo2018DecentralizedPP, dagher2018broncovote, heiberg2018trade}.

\subsubsection{Badmouthing Attack}
A badmouthing attack is a widespread attack on rating or feedback systems in which the opponent is trying to respond negatively to the target party. In this attack, the opponent’s main objectives are twofold: (i) to degrade the reputation of a particular user in the system by making the scale system negative or (ii) to increase the rating of a favourable user. As a result, a badmouth attack can significantly reduce a network’s reputation and , or corrupt the entire system. To solve this issue, the self-organising maps technique \citep{bankovic2011detecting} is used to detect and prevent users’ malicious behaviour.

\subsubsection{Guess Attack}
In this attack, the attacker’s goal is to discover a specific user’s personal information from some stored data using brute force or other matching techniques. The guess attack can occur during the searching process, when an attacker tries to match the randomly typed keywords with the stored information. If the attacker is successful in his actions, he can easily steal the record,  that is, the private keys and personal information of any user. To solve these issues, \citep{zhang2018towards} proposed the Blockchain-based E-health scheme in which cryptography primitives, such as encryption and hashing, are employed to protect the personal information of both patients and doctors.

\subsubsection{Chosen Ciphertext Attack}
The chosen ciphertext attack aims to obtain the target user’s private key by analysing the different chosen ciphertexts obtained from the communication channel \citep{biryukov2011chosen}. In this attack, an adversary uses some packet sniffing tools to access the network’s ciphertexts and attempts to retrieve the secret keys for decryption. This problem can overcome by using advanced cryptography methods, such as order-preserving encryption and ECDSA in different systems, to transform the secret key to others \citep{ji2018bmpls}.

\subsubsection{Impersonation Attack}
This is an illegal attempt against an individual or a group to retrieve personal information from databases. In this attack, the attacker first creates a fake profile of the legitimate user in the network and then uses some social engineering methods, such as email and links, to reach a targeted system. In Blockchain systems, the miners involved in the consensus process gain some incentives to reward their computations. Thus, the attacker uses some common techniques to access miner’s systems to maximise profit. To reduce the chances of impersonation attacks on Blockchain systems, that is transport and crowdsensing applications, \citep{malik2018blockchain,wang2018blockchain,Lin2018BSeInAB} proposed the security mechanism that prevents the miners from impersonation attack and only validates those miners who have the valid attributes to log in to the system.

\subsubsection{Linking Attack}

In a linking attack, the adversary’s goal is to create the link between the external data and stored data by using some de-anonymisation techniques to expose the personal information of users. The linking attack is a severe type of attack that is applied over many Blockchain systems such as Bitcoin \citep{liu2018unlinkable}, online digital platforms \citep{friebe2018decentid}, IoT and vehicular networks \citep{dorri2017blockchain}, in order to extract the secret data from stored transactions. This problem has been widely discussed and several approaches have been proposed to resolve a link attack but the most accurate approach is to create a new key pair each time so as to encrypt the data.

\subsubsection{Collusion Attack}
In a collusion attack, the adversary aims to retrieve users’ secrets by combining several different copies of data obtained from the network. For this purpose, the attacker examines the network traffic using some packet sniffing tools to obtain data packets containing users’ personal information. The same issue could occur in the Blockchain network if multiple miners try to collude with the network to get the maximum reward for their mining activities. To address collusion activity that occurred in Blockchain-based eHealth and eCommerce applications, \citep{Owiyo2018DecentralizedPP,guo2018secure} utilise pseudo-random methods in which the random seed is shared between two users in such a way that it can resist N-1 other fake users in the system.

\subsection{Analysis and Discussion on Attacks}

\begin{figure*}[t]     
\centering
\subcaptionbox{Percentages of Blockchain Layers Impacted by Attacks\label{subfig:1}}{\includegraphics[width=8cm,height=6cm]{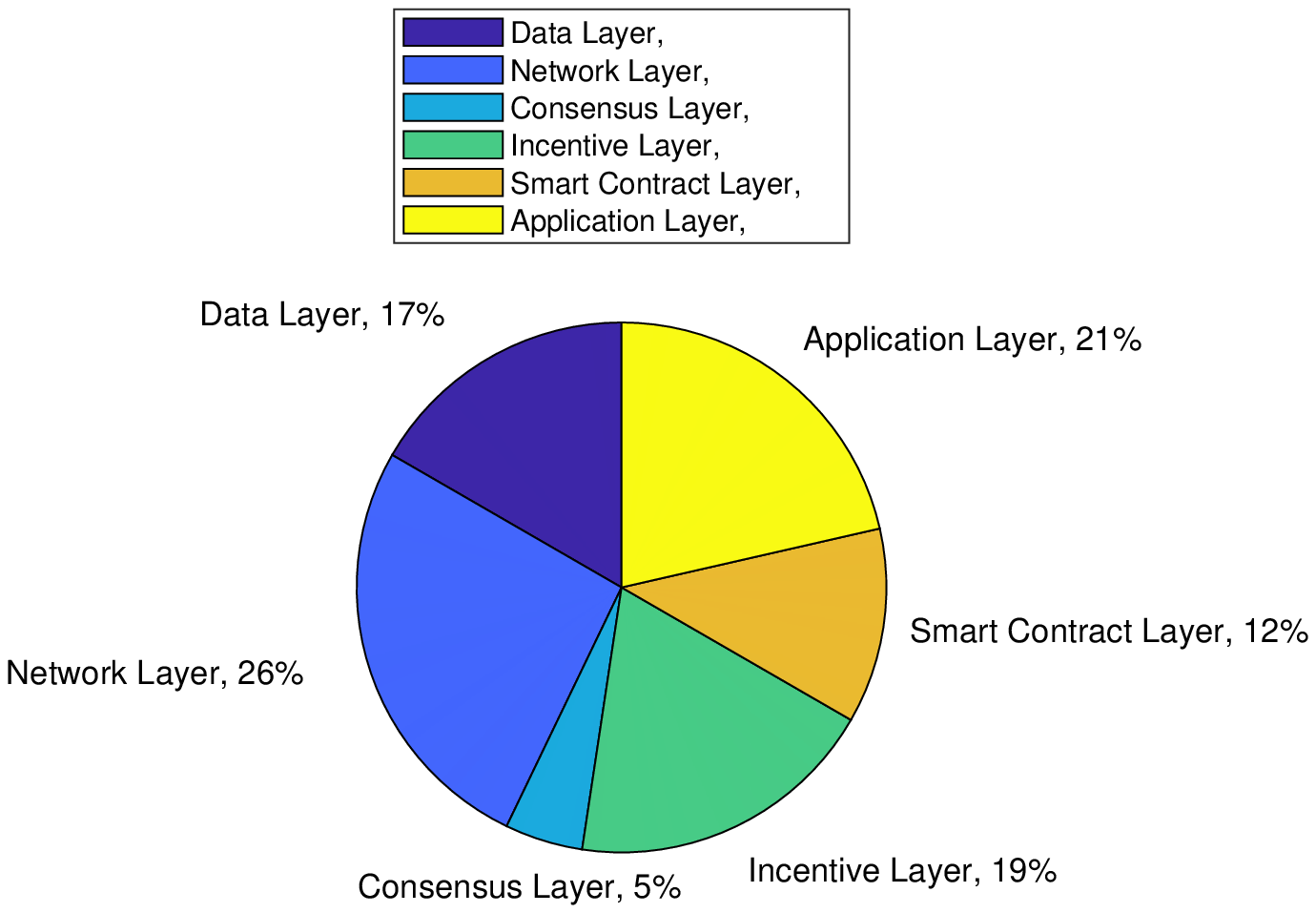}}
\subcaptionbox{Percentages of Security Breaches Cause of Attacks\label{subfig:2}}{\includegraphics[width=8cm,height=6cm]{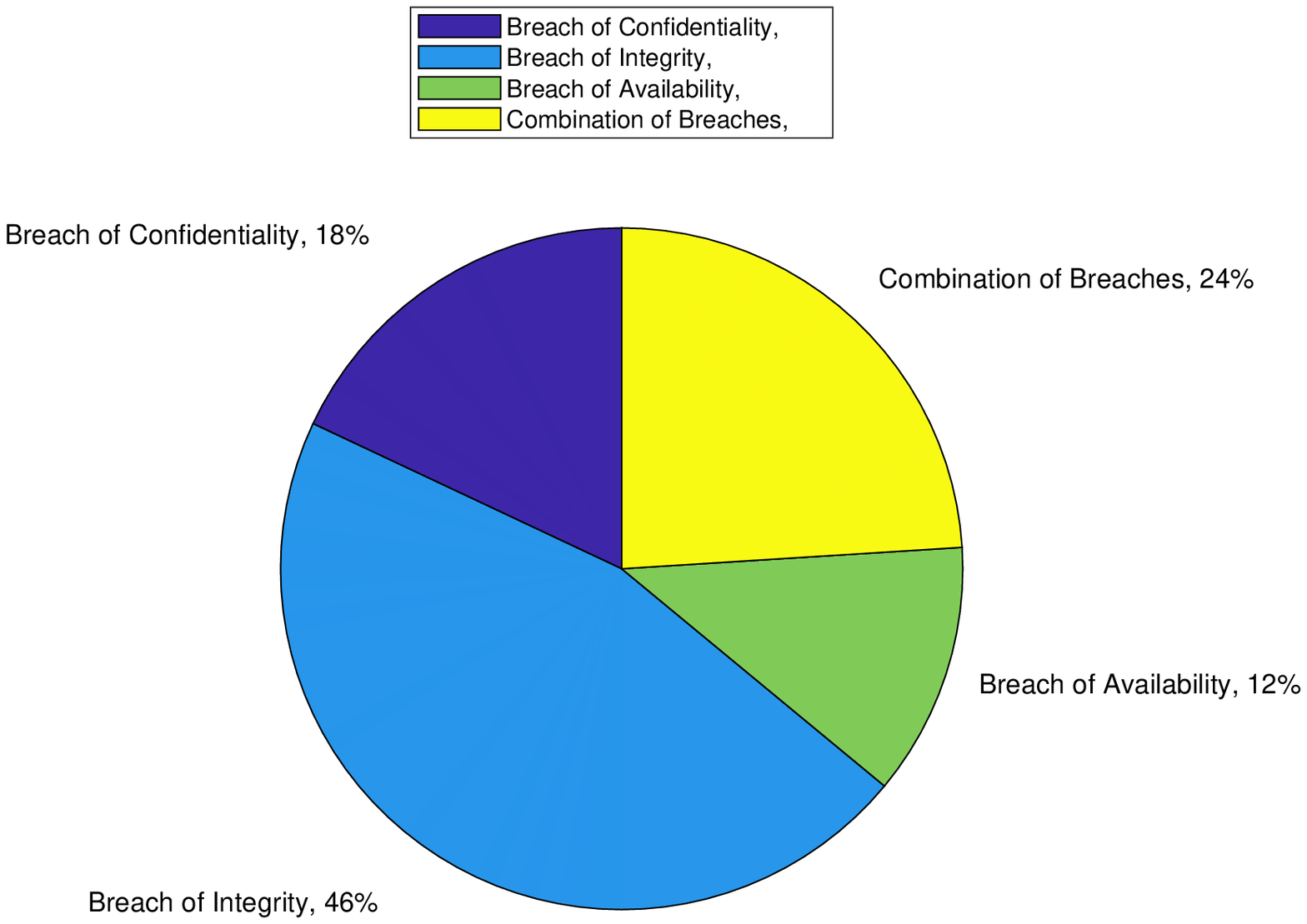}}

\subcaptionbox{Percentages of Industrial Applications Impacted by Attacks\label{subfig:3}}{\includegraphics[width=8cm,height=6cm]{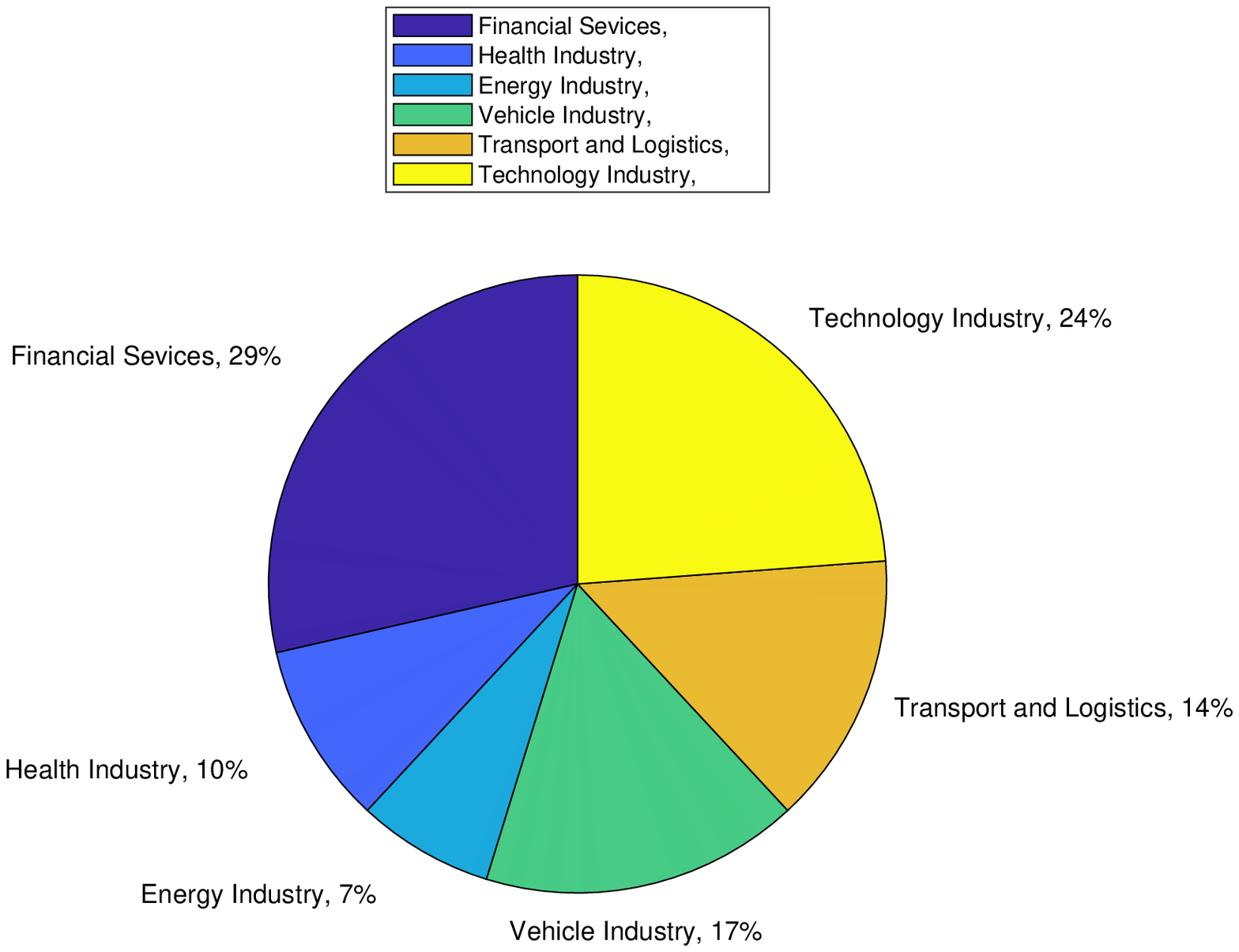}}
\subcaptionbox{Percentages of Security Solutions Used to Mitigate Attacks \label{subfig:4}}{\includegraphics[width=8cm,height=6cm]{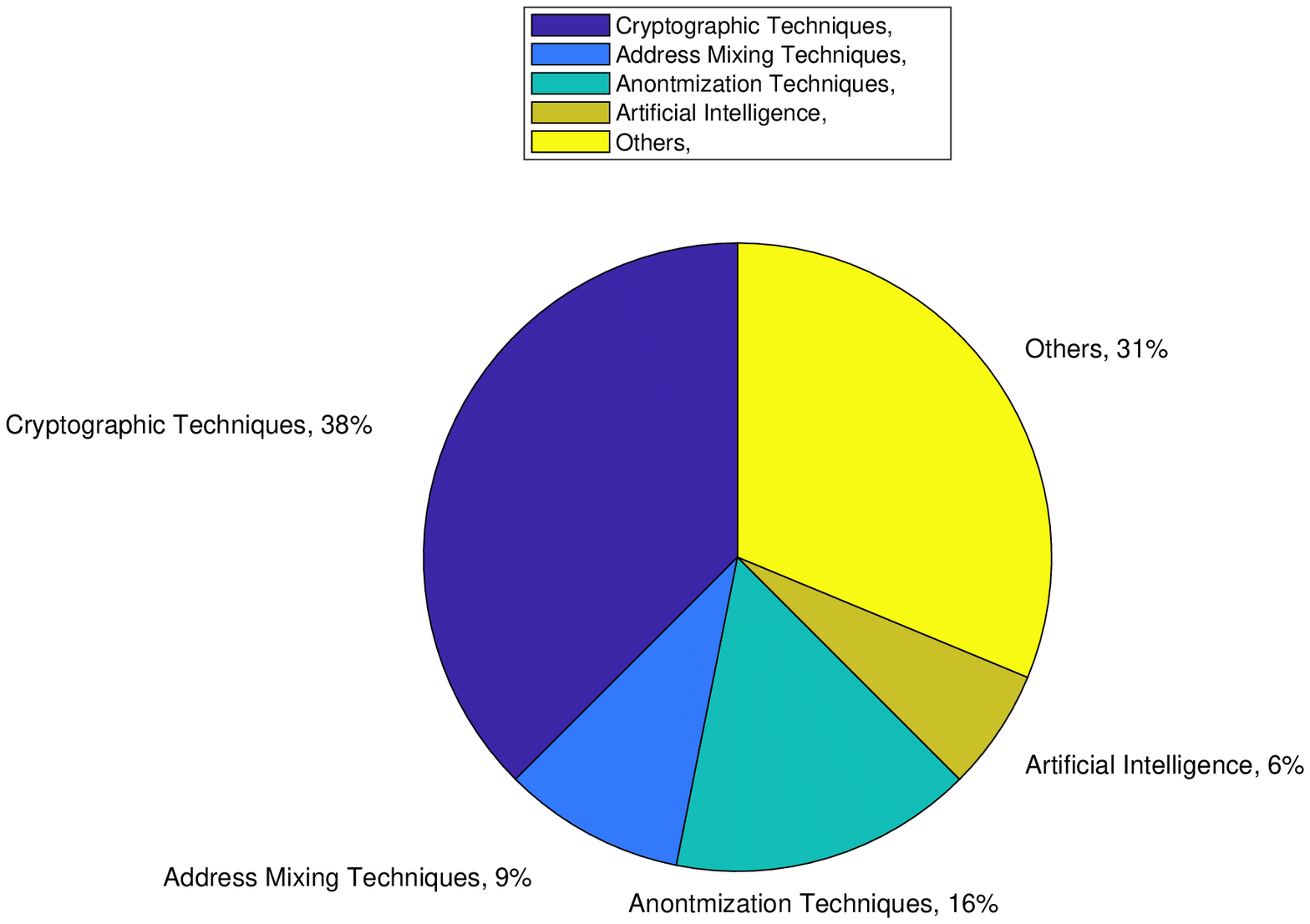}}

\caption{A Graphical Representations of Performed Analysis of Security and Privacy Attacks in terms of Layers, Security Breaches, Applications and Security Solutions}
\label{MainFigure}

\end{figure*}

\textcolor{black}{We provide an in-depth analysis of the security and privacy attacks covered in this survey paper. For analysis, we compared the attacks with the various parameters listed in the table, such as layers, security breaches, applications and solutions. Figure \ref{MainFigure} shows various graphical representations of security and privacy attacks in percentages, with several parameters such as layers, security breaches, impacted applications and security solutions.}

\textcolor{black}{Fig. \ref{subfig:1} shows the percentage of each Blockchain layer affected by security and privacy attacks included in the paper. The Blockchain layers are the network layer, data layer, consensus layer, incentive layer, smart contract layer and application layer.
Overall, it is clear that the network layer of Blockchain architecture is the target of a large number of attacks, accounting for 26\% of all attacks. It is also worth noting that the application layer and the incentive layer are the most affected layers by these attacks, with just a 2\% gap between the two. Attacks on these layers account for 21\% and 19\% of all attacks, respectively. In comparison, the data layer is impacted by these attacks by 17\%, while the smart contract layer is only impacted by 12\%. Finally, only a small percentage of attacks are countered at the consensus layer, approximately 5\%.}

\textcolor{black}{Figure \ref{subfig:2} depicts the percentage of different security breaches that cause the security and privacy attacks discussed in our paper. In our work, we mainly considered three types of security breaches: breach of confidentiality, breach of integrity and breach of availability. Additionally, attacks may occur as a result of a combination of those security breaches.
Overall, it is clear that a significant proportion of attacks on various Blockchain and industrial applications are the result of a breach of integrity, which is regarded as the primary security breach in this analysis part. Furthermore, it is worth noting that nearly half of all attacks for breach of integrity result from a combination of these security breaches. A combination of security breaches may come in the form of a breach of confidentiality and a breach of integrity, or a breach of confidentiality and a breach of availability. On the other side, there is just a 6\% difference between the remaining two breaches (breach of confidentiality and breach of integrity).}

\textcolor{black}{Fig. \ref{subfig:3} illustrates the percentage of various industrial applications susceptible to security and privacy attacks explored in our work. Financial services, health industry, energy industry, transport and logistics and technology industry are among the industrial applications covered by our research.
At first glance, it appears that financial services are the most affected application by these security attacks. The primary goal of the attacker is to obtain financial benefits in the form of Bitcoins, Ether and other related cryptocurrencies. The percentage of financial applications affected is 29\%, which is less than a third of the overall percentages. However, it is clear that the technology industry is also affected by these attacks, with just a 5\% difference from the financial industry. Furthermore, the vehicle industry and transport and logistics have been exposed to these attacks, with just a 3\% gap between the two. Finally, the same comparison applied to the other two industrial applications, such as the energy industry and the health industry, with just a 2\% disparity between the two.}

\textcolor{black}{Fig. \ref{subfig:4} represents the percentages of various types of security solutions used to mitigate security and privacy threats on Blockchain-based Industry 4.0 applications. These solutions are classified as cryptographic methods, address mixing, anonymization, artificial intelligence and a few others. Overall, it is clear that the most effective solution for securing industrial applications from attacks is the use of various cryptographic techniques such as encryption, hashes, digital signatures, which accounts for nearly two-fifths of all other security solutions. By leading this, nearly a third used a combination of security solutions such as bilinear maps, three weight models, time ranks and consensus mechanisms to minimise the risks of attacks on Blockchain-based applications. Anonymization techniques, which account for 16\% of the total, also play an important role in securing those applications from attacks. Address mixing and artificial intelligence-based solutions are also used to solve security issues, but these are in the minority.}

\section{Open Issues} \label{sec:8}
\begin{figure*}[t]
    \centering
\includegraphics[width=16cm,height=12cm]{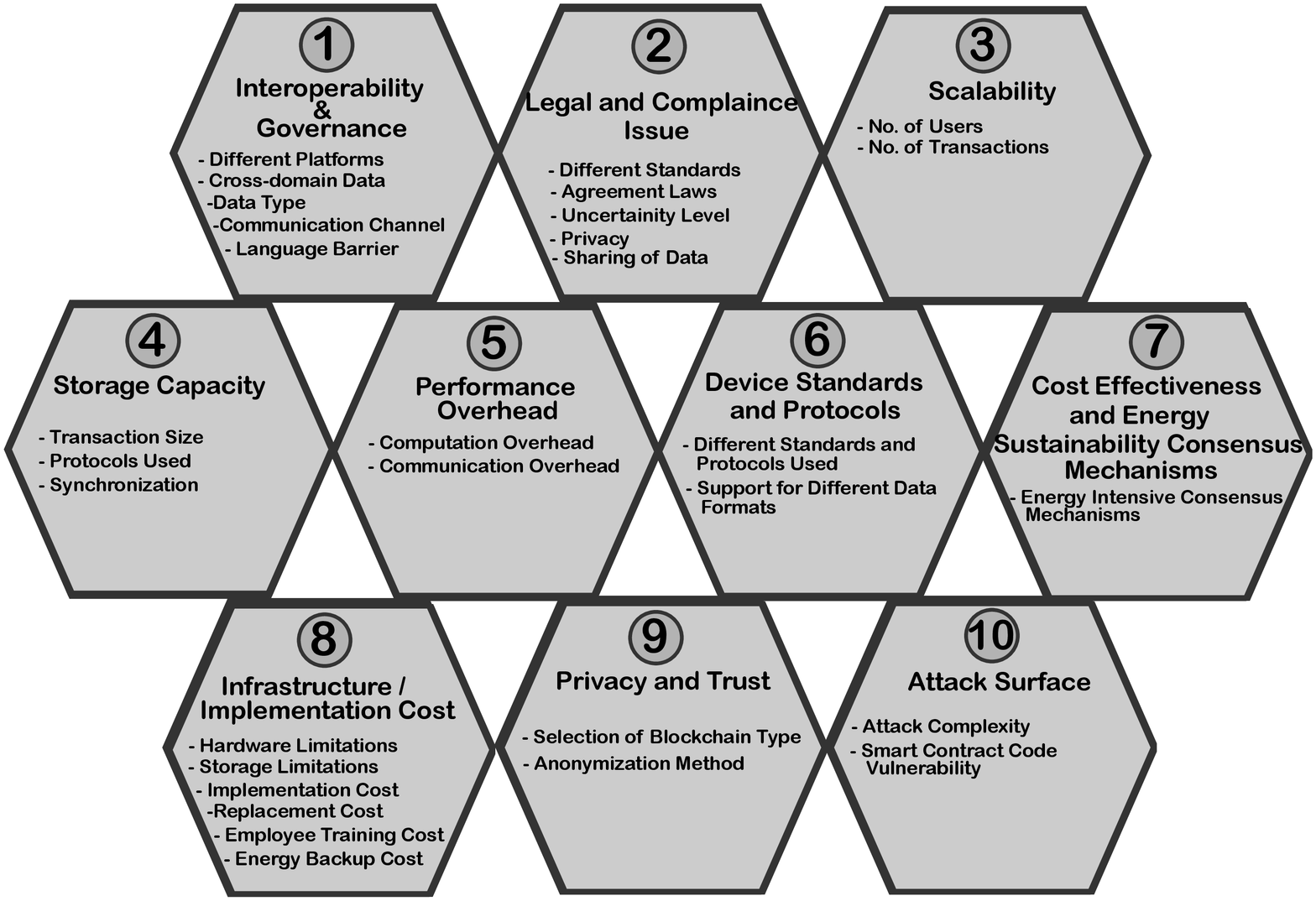}
  \caption{\colorbox{white}{Open Issues of Blockchain Adoption in Industry 4.0-based Applications}}
   \label{fig:openissues}
\end{figure*}

Despite the fact that potential Blockchain features offer enormous benefits to industry users, some issues must be addressed in order to broaden the scope of industry and business applications. In this section, we highlight open issues that arise from integrating Blockchain technology into industrial applications, limiting its applicability to Blockchain technology adoption at a broader industry level. The open issues of Blockchain adoption in Industry 4.0-based applications are summarised in Fig. \ref{fig:openissues}.

\subsection{Interoperability and Governance}
As the scope of Industry 4.0 is not limited to some applications, it includes a wide range of applications and businesses, in order to interact and share valuable data or assets with others. To achieve this interaction between different industrial applications, interoperability is defined as a network infrastructure or capacity that enables the industry partner to exchange information over the network. Similarly, Blockchain interoperability allows different Blockchain systems to interact by sending messages and sharing trusted values with others \citep{saxena2021blockchain}. However, the issues arising during the interoperability of Blockchain systems are the security of the exchanged data and the appropriate industry guidelines for the regulation and control of applications \citep{madine2021application}. Thus, there is a need to design a mechanism that can ensure interaction between different Blockchain platforms and define the appropriate rules and regulations, in line with industry principles and guidelines \citep{belchior2020survey}.

\subsection{Legal and Compliance Issues}

\textcolor{black}{Some of the significant obstacles to the deployment of Blockchain technology in the industrial context is using a different standard, agreement laws and some uncertainty around government regulatory bodies, which prevents this technology from being deployed in larger industrial domains. For example, stakeholders in the manufacturing industry use Blockchain tools to ensure that their internal processes and goods adhere to legal and regulatory frameworks \citep{zachariadis2019governance}. According to commercial law, there must be specific contracts between industry users and government regulations relating to negotiation, execution, administration and Blockchain management. Furthermore, liability must be acknowledged if a contract were miscoded and is not carried out as intended \citep{amato2021model}.}

\textcolor{black}{Another crucial problem between the parties’ legal and compliance arrangements is to adhere to substantive law, effective governance, jurisdiction and settlement, as well as ensuring the privacy of both consumers and the product. Additionally, the sharing of manufacturing data across various platforms can pose challenges for manufacturers and their products. Therefore, when developing industry platforms, it is imperative to protect both the users’ privacy and their data \citep{yeoh2017regulatory}. In regard to the public interest, when making new laws, rules, creating guidelines and applying laws in the industries, governments should hold to account a specific government obligation. One way to achieve this aim is to use a private Blockchain to stop illicit activities such as money laundering or bypassing regulations. Additionally, there should be some mechanism to prevent new blocks from being produced by fake miners.}

\subsection{Scalability}

Blockchain technology has been proven to be better due to its adherent features, such as decentralisation, in reforming current centralised systems. It provides a dynamic environment for P2P network nodes to mine blocks over a specified time and frequently updates transactions to other network nodes. For example, Bitcoin and Ethereum Blockchain systems have both their capabilities and their limitations in handling users and transactions. However, the issues of scalability in existing Blockchain technology discourage broad adoption and implementation. One such example is that VISA \citep{vermeulen2017bitcoin} can process about 2,000 transactions per second, while Bitcoin \citep{alladi2019blockchain} can process just seven transactions per second. With the increased adoption and applicability of Blockchain technology in Industry 4.0, the length of a particular system increases, and thousands of nodes are needed to join the network for block creation and mining. Thus, the scalability of Blockchain is questioned; this is a significant challenge for network security systems and their related applications \citep{zhou2020solutions,shukla2020systematization}.

\subsection{Storage Capacity}

\textcolor{black}{Storage capacity has been questioned in many Blockchain-based applications as part of the storage constraint for storing data on a secure distributed ledger. As in Bitcoin, the chain is increasing by one megabyte every 10 minutes and each node in the network has a copy of the full chain. A full node can store the entire blocks but total storage requirements expand exponentially with transaction size, thereby growing the entire system’s capacity. Since the amount of manufacturing data in an industrial setup is enormous, incorporating Blockchain technology with many industrial processes is a difficult problem \citep{tian2016agri}. Furthermore, the underlying Blockchain protocols cause significant traffic congestion in the system, increasing the system’s need for overall Blockchain storage space \citep{dorri2016blockchain}. A further problem with oversized chains is that they put unnecessary synchronisation overload on new users. For example, in the industrial IoT context, with the growing number of sensors and produced data, the problem is greatly amplified \citep{azeem2021big}. Even though Ethereum-inspired frameworks have recently been developed, resource constrained IoT solutions are still in their infancy and are perhaps incapable of supporting more industrial applications at this time \citep{farahani2021convergence}.}

\subsection{Performance Overhead}

Most industrial applications, such as IoT, SG, eHealth, and so on, are simple in nature and have minimal computing, storage and energy capabilities. With the integration of Blockchain with those applications, the most challenging problems arise in node computation limitations \citep{rathee2021secure}. They require significant computation to mine blocks and perform extensive cryptography operations, including hashing, encryption/decryption and digital signatures. Many solutions have been proposed that distinguish mining nodes and simple nodes as full nodes and simple nodes in Blockchain-based IoT applications, respectively \citep{yang2019integrated}. Another performance issue arising in industrial applications is communication overhead, in which each mining node is responsible for performing multiple functions, such as mining and updating Blockchain, along with sending updated blocks to other peer nodes over the network \citep{kang2020scalable}. This issue creates an extra network overhead that affects the overall bandwidth and performance of the network. Thus, computation and communication overheads in Blockchain-based Industry 4.0 applications create barriers to their adoption at a wider industry level \citep{lao2020survey}.

\subsection{Device Standards and Protocols}

\textcolor{black}{The key part of any industrial setup is the use of various heterogeneous IoT devices that are involved in continuous monitoring and controlling of the environment in order to undertake some future measures. However, device standards and protocols have become another barrier for IoT and Industry 4.0, in which a vast range of smart devices are deployed. Integrating them is expensive and challenging since they all record data in different formats and use different protocols \citep{majeed2021blockchain}. Manufacturing companies such as Bosch and the Eclipse Foundation aim to create more common data formats and communication protocols for data sharing, such as MQTT \citep{mishra2020use}. The main objective is to assist smart devices in providing popular data formats that enable them to interact seamlessly with one another; however, more data formats mean more complexity in developing a single data model \citep{rubi2021iot}.}

\subsection{Cost-Effectiveness and Energy Sustainable Consensus Mechanisms}

\textcolor{black}{Consensus algorithms such as POW and POS are considered very energy-intensive, as they tend to use more energy and computational resources \citep{mendling2018blockchains, bozic2016tutorial}. As the Blockchain’s size grows over time, more successful miners, as well as some energy-efficient consensus protocols, such as DPoS \citep{yang2019delegated} and Proof of Trust (PoT) \citep{zou2018proof}, have been proposed recently to execute this process in cost-effective ways. The key advantage of using these protocols is that they can only store the most recent transaction data on the Blockchain. However, since energy and resource-scarce industrial IoT devices are surrounded by enormous amounts of data, more efficient consensus algorithms are required \citep{bouraga2021taxonomy}.}

\subsection{Infrastructure / Implementation Cost}

\textcolor{black}{Blockchain technology generally requires specialised infrastructure, such as additional storage and computationally intensive hardware resources to store the Blockchain. Blockchain storage is built on Distributed Ledger Technology (DLT) and serves as a shared database of knowledge regarding transactions between various parties. The DLT must be able to store an increasing number of blocks indefinitely and immutably. According to one estimation, between April 2019 and March 2021, Bitcoin is seen a growth rate of 340GB; however, this growth varies with the discovery rate of new blocks \citep{blockchainchart}. Furthermore, as the number and size of Bitcoin miners have grown dramatically, mining the blocks requires more sophisticated hardware and computational power \citep{de2021bitcoin}.}

\textcolor{black}{As a result, using Bitcoin’s infrastructure cost as an example, one must consider various cost scenarios when implementing Blockchain technology in industrial settings. These scenarios can include: (i) the cost of implementing and deploying the Blockchain setup, (ii) the cost of replacing the current industry infrastructure, (iii) the cost of training employees to be familiar with Blockchain technology, and (iv) the cost of energy running resources as a backup \citep{lee2019blockchain}.}

\subsection{Privacy and Trust}

One of the most appealing features and benefits of Blockchain technology is to achieve anonymity of user identity and transactions, using pseudo-anonymity methods \citep{zhang2019security}. However, the selection of public Blockchain type and the use of different pseudo-anonymous techniques in Blockchain applications may connect the identity of users to transactions such as public keys, thus increasing the chances of disclosing personal information to others \citep{swarnkar2021security}. Therefore, there is a substantial need to implement pseudo-anonymity methods that must be fully anonymous and must achieve a higher level of privacy and trust among Blockchain users \citep{ma2020survey}.

\subsection{Attack Surface}

With an increasing number of industrial applications that have adopted Blockchain technology, many attack surfaces have been targeted. These attacks take advantage of various methods and vulnerabilities found in applications to get into the environment and they have made some malicious changes to user data \citep{latif2021blockchain}. For example, the Bitcoin Double Spending Attack comes in various ways that sometimes combine with other attacks, such as a Sybil attack, to obtain details of user wallets, balance and private keys \citep{zhang2019double}. In addition, the vulnerabilities identified in the smart contract code and, in some cases, open source applications, have made Blockchain systems more vulnerable to other malicious users \citep{hu2021transaction}. Therefore, carefully designed Blockchain applications with proper guidelines and secure cryptography mechanisms across different layers can minimise the chances of attacks \citep{wen2021attacks}.

\section{Conclusion and Future Work}\label{sec:9}
The integration of Blockchain technology with various industrial applications such as IoT, finance, SG, eHealth, transport and logistics and the cloud is growing at a rapid pace, and it has changed human lives in a beneficial way. However, the integration of Blockchain into existing applications has increased many security and privacy challenges for the users and their data. Considering the design of secure Blockchain-based Industry 4.0 applications, we present a detailed study achieving a significant number of contributions regarding security and privacy for Blockchain-based Industry 4.0 applications. Firstly, we provide an overview of Blockchain features and properties that map and initiate the design requirements for secure Blockchain applications. Secondly, we categorise the design phase of such applications into different design, security and privacy requirements. Thirdly, we extend our survey with various types of security and privacy attacks on Blockchain applications, along with the categories of attacks, the attacker’s objectives, exploited vulnerabilities and targeted applications. Furthermore, we provide an in-depth review of Blockchain-based Industry 4.0 applications in terms of defined requirements. Then we explore different security and privacy enhancement solutions used to achieve security and privacy requirements in Blockchain-based Industry 4.0 applications. Finally, we discuss open design issues which provide the fuel to researchers and developers for the advancement of secure applications.

In the latter part of the work, we present several future recommendations for design and security requirements, as well as open issues, providing the directions to both researchers and developers for designing secure, scalable, efficient and flexible Blockchain-based applications. Currently, most of the existing Blockchain-based schemes designed for applications are lacking in the achievement of the following requirements: scalability, interoperability, usability, flexibility, modularity and transparency. Scalability is the most critical requirement among all others for designing the Blockchain applications, especially in the industrial domain, in which systems perform poorly with the increasing number of users and their real-time transactions. However, many systems (such as Ethereum \citep{wood2014ethereum} and Hyperledger \citep{Hyperledger}) are subject to their restrictions regarding mining and validating transactions that influence the system’s scalability. Therefore, researchers and developers need to address the scalability challenge during the development of Blockchain applications. Another design requirement issue found in the Blockchain applications is the linking gap between the system internal modules, which is also referred to as an interoperability issue. In practice, industrial processes are dependent on each other and linked together to obtain recent updates about the system activities. Moreover, the interoperability of system components also helps the administrator to take instant security actions against malicious activities occurring in the system. Thus, the interoperability challenge needs to be addressed in the design course of secure Blockchain applications.

Blockchain consists of many supporting and underlying features which provide ease of use for the completion of multiple tasks. However, not all ordinary users  understand the technical detail and working of Blockchain technology. Thus, it stops them from adopting the Blockchain applications in different businesses and industries. As a result, a Blockchain model must endorse usability and flexibility requirements, while building the Industry 4.0 applications and business domains. Moving forward to the modularity design requirement, developers need to write the application code that achieves greater applicability among different Blockchain applications, and it should support the wide variety of services that offer network resources efficiently. As Blockchain applications are gaining popularity within the community, it becomes essential for researchers and developers to provide a secure environment in which users can share resources efficiently and communicate transparently with each other.

\section*{Declarations of Interest}
The authors declare that they have no known competing financial interests or personal relationships that could have appeared to influence the work reported in this paper.

\bibliography{main.bib}

\end{document}